\documentclass[rmp,nopacs,twocolumn,longbibliography,nofootinbib]{revtex4-1}

\usepackage[utf8]{inputenc}
\usepackage{epsfig}
\usepackage{epstopdf}
\usepackage{graphicx,color,xspace}
\setkeys{Gin}{draft=false} %%% used by \includegraphics
\usepackage{bm,amssymb,amsmath}
\usepackage{dcolumn,xfrac}
\usepackage{natbib}
\usepackage[colorlinks=true,citecolor=blue,linkcolor=black]%[colorlinks] %%% sometimes colour links are not seen!
{hyperref}
\usepackage[english]{babel}
\newcommand{\fref}[1]{Fig.~\ref{#1}}
\newcommand{\Fref}[1]{Figure~\ref{#1}}
\newcommand{\sref}[1]{Sec.~\ref{#1}}
\newcommand{\Eref}[1]{Eq.~(\ref{#1})}
\newcommand{\tref}[1]{Table~\ref{#1}}

\newcommand{\cm}{cm$^{-1}$}
\def\mr{\mathrm}

\def\fm#1{\ifmmode #1 \else $#1$\fi}
\def\ket#1{{%
  \ifmmode |\,#1\,\rangle \else $|\,#1\,\rangle$\fi}}
\def\DfB{\fm{\Delta f_\mathrm{B}}\xspace}
\def\Al{\fm{\mathrm{Al}^{+}}\xspace}

\def\Artp{\fm{\mathrm{Ar}^{13+}}\xspace}

\def\Sr{\fm{\mathrm{Sr}^{+}}\xspace}
\def\Ca{\fm{\mathrm{Ca}^{+}}\xspace}
\def\Yb{\fm{\mathrm{Yb}^{+}}\xspace}

\def\Be{\fm{\mathrm{Be}^{+}}\xspace}
\def\Ben{\fm{^{9}\mathrm{Be}^{+}}\xspace}

\def\tpz{\fm{{}^3\mathrm{P}_{0}}\xspace}
\def\tpo{\fm{{}^3\mathrm{P}_{1}}\xspace}

\def\dsoh{\fm{{}^2\mathrm{S}_{1/2}}\xspace}
\def\dpoh{\fm{{}^2\mathrm{P}_{1/2}}\xspace}
\def\dpth{\fm{{}^2\mathrm{P}_{3/2}}\xspace}

\def\Erf{\fm{E_\mathrm{rf}}\xspace}

\def\Ekin{\fm{E_\mathrm{kin}}\xspace}
\def\Orf{\fm{\Omega_\mathrm{rf}}\xspace}
\def\hci#1#2#3{\fm{{}^{#1}\mathrm{#2}^{#3}}}

\begin{document}
\title{Highly charged ions: optical clocks and applications in fundamental physics}

\author{M. G. Kozlov$^{1,2}$}
\author{M. S. Safronova$^{3,4}$}
\author{J. R. Crespo L\'opez-Urrutia$^{5}$}
% crespojr@mpi-hd.mpg.de
% jose.crespo@mpi-hd.mpg.de
\author{P. O. Schmidt$^{6,7}$}

\affiliation{$^1$Petersburg Nuclear Physics Institute of NRC
``Kurchatov Institute'', Gatchina 188300, Russia}

\affiliation{$^2$St.~Petersburg Electrotechnical University
``LETI'', Prof. Popov Str. 5, St.~Petersburg, 197376, Russia}

\affiliation{$^{3}$Department of Physics and Astronomy, University
of Delaware, Newark, Delaware 19716, USA}

\affiliation{$^{4}$Joint Quantum Institute, National Institute of
Standards and Technology and the University of Maryland,
Gaithersburg, Maryland 20742, USA}

\affiliation{$^5$Max-Planck-Institut f\"{u}r Kernphysik,
Saupfercheckweg 1, 69117 Heidelberg, Germany}

\affiliation{$^6$Physikalisch-Technische Bundesanstalt Braunschweig,
38116 Braunschweig, Germany}

\affiliation{$^7$Institut f\"{u}r Quantenoptik, Leibniz
Universit\"{a}t Hannover, 30167 Hannover, Germany}

\date{\today}

\begin{abstract}
Recent developments in frequency metrology and optical clocks have
been based on electronic transitions in atoms and singly charged
ions as references. The control over all relevant degrees of freedom
in these atoms has enabled relative frequency uncertainties at a
level of a few parts in $10^{-18}$. This accomplishment not only
allows for extremely accurate time and frequency measurements, but
also to probe our understanding of fundamental physics, such as a
possible variation of fundamental constants, a violation of the
local Lorentz invariance, and the existence of forces beyond the
Standard Model of Physics. In addition, novel clocks are driving the
development of sophisticated technical applications. Crucial for
applications of clocks in fundamental physics are a high sensitivity
to effects beyond the Standard Model and Einstein's Theory of
Relativity and a small frequency uncertainty of the clock. Highly
charged ions offer both. They have been proposed as highly accurate
clocks, since they possess optical transitions which can be
extremely narrow and less sensitive to external perturbations
compared to current atomic clock species. The large selection of
highly charged ions in different charge states offers narrow
transitions that are among the most sensitive ones for a change in
the fine-structure constant and the electron-to-proton mass ratio,
as well as other new physics effects.  Recent experimental advances
in trapping and sympathetic cooling of highly charged ions will in
the future enable advanced quantum logic techniques for controlling
motional and internal degrees of freedom and thus enable high
accuracy optical spectroscopy. Theoretical progress in calculating
the properties of selected highly charged ions has allowed the
evaluation of systematic shifts and the prediction of the
sensitivity to the ``new physics'' effects. New theoretical
challenges and opportunities emerge from relativistic, quantum
electrodynamics, and nuclear size contributions that become
comparable with interelectronic correlations. This article reviews
the current status of theory and experiment in the field, addresses
specific electronic configurations and systems which show the most
promising properties for research, their potential limitations, and
the techniques for their study.

\end{abstract}

\maketitle

\tableofcontents

%%%%%%%%%%%%%%%%%%%%%%%%%%%%%%%%%%
%\input{introduction_v14}
%%%%%%%%%%%%%%%%%%%%%%%%%%%%%%%%%%

% !TeX spellcheck = en_US
\section{Introduction}
\label{sec:Introduction} In this review we cover current and new
research directions that arise from high precision spectroscopy and
novel optical clocks using trapped highly charged ions (HCI). The
recent interest in high resolution optical spectroscopy of HCI has
been triggered by their high sensitivity to a change in the
fine-structure constant $\alpha$ (\sref{sec:variation}). Since the
required HCI physics is generally less well known compared to
neutral and singly-charged atoms, \sref{sec:EST-QED} recapitulates
the current status of theory, including quantum electrodynamics
(QED), mostly dealing with few-electron HCI. In \sref{theory2} we
present recent theory on more complicated electronic systems that
are particularly interesting for frequency metrology, and discuss
them in detail.

We present the current status of experimental methods in
\sref{sec:ExpMethHCI}, with a particular emphasis on the field of
optical spectroscopy (\sref{Sec:Lab_OS_HCI}). Novel methods needed
for high resolution optical spectroscopy with HCI are introduced in
\sref{sec:CompactEBITs} and \ref{sec:prep_cold_HCI}, and an analysis
of clock shift systematics with HCI candidates follows in
\sref{sec:hires}. Future directions of research are discussed in
\sref{sec:OtherApps}.

\subsection{Atomic clocks for exploring old and new physics}

Celestial mechanics, based on the (for their time) exceptionally
precise measurements of astronomy, became one of the most fruitful
disciplines in theoretical physics and was the key for many
discoveries in other fields. In the same way, atomic and optical
clocks \cite{Ludlow2015}, aided by advances in atomic structure
theory are evolving into an exceptional driving force for the search
for new physics. Frequency measurements of narrow transitions in
atoms, with their universal reproducibility that is paradigmatic in
science, serve as benchmarks for the most subtle deviations from the
theoretical framework of physics, while high-energy physics aims at
testing the boundaries of our knowledge in huge facilities. These
complementary approaches eventually aim at answering the same
questions.

In atomic clocks, physics interactions appear as contributions to
the electronic binding energies or the transition rates in a
hierarchy now stretching over 18 orders of magnitude. Novel clocks
are capable of addressing those in more and more depth, and their
rapid development in the last decades heralds further advances.
Progress in optical clocks is based on an improved control over all
degrees of freedom, including the internal state, the motion, and
external fields affecting the transition frequency, paired with
increased accuracy of electronic structure calculations of atomic
properties. This has been accomplished through the rapid development of
suitable laser cooling and trapping techniques based on the pioneering work of 
\citet{Metcalf2007,Hansch1975,phillips_laser_1998,Wineland1978,Neuhauser1978,Ashkin1978,wineland_laser_1979,Bollinger1985,Phillips1985,Aspect1986,Stenholm1986,Lett1988}.

However, progress in
atoms (such as most HCI), that do not posses the required properties
for cooling and trapping could not profit from these advances. This
situation has recently changed with the development of quantum logic techniques for
spectroscopy in which a cooling ion, trapped together with the
spectroscopy ion, provides sympathetic cooling as well as control
and readout of the internal state of the spectroscopy ion. This
arrangement forms a compound system combining the advantages of both
species and makes previously intractable atoms, such as HCI,
accessible for high resolution spectroscopy.

Similarly to the progress in experimental techniques, advances in
atomic structure theory are required for exploiting the present and
future experimental precision in order to improve our knowledge of
the various fundamental interactions that are intertwined within the
electronic shell. For the present status of optical clocks, we refer
to the recent review of \citet{Ludlow2015}, and for a general work
on atomic physics methods for probing fundamental physics to
\citet{ClockMasters2017}. Within this review we will be concerned
with the specific aspects related to the applications of HCI as
novel atomic frequency references.

\subsection{Atomic physics at the extremes: Highly charged ions}

Since the epoch of reionization -- roughly 13 billion years ago --
atomic matter in the universe mostly appears as ions
\cite{Shull2012}. Now, the majority of the chemical elements can be
found as highly charged ions. Following Big-Bang nucleosynthesis and
400 million years of expansion, reionization was driven by strong
radiation sources, stars coming into being by the gravitational
collapse of cold neutral gas. Supernovae later infused space with
heavy elements, heating interstellar matter in shocks to very high
temperatures \cite{hitomi2017}. Furthermore, as a consequence of
energy virialization in deep gravitational potentials, the
translational temperature of the diffuse intergalactic medium
containing most of the baryonic matter is also very high
($10^{5...7}$\,K) \cite{Reimers2002}. Here, ions reign absolute, no
matter how low the radiation temperature of the medium around them
might be. Galaxy clusters, galaxies with their active galactic
nuclei, binary x-ray systems and stars are extremely hot
environments; hot winds and relativistic jets expel highly ionized
material out of the galactic cores \cite{hitomi2016}. X-ray
astronomy missions such as the present \textit{Chandra} and
\textit{XMM-Newton} allow us to observe such environments, and they
as well as the future ones \textit{XARM} \cite{hitomi2016} and
\textit{Athena}  \cite{athena2017} will produce more and more
quantitative results. This emphasizes the need for laboratory data
and advanced atomic structure calculations for their interpretation;
therefore, studying HCI is essential for astrophysics diagnostics
\cite{Beiersdorfer2003e}. In addition, HCI play a central role in
plasma physics \cite{Beiersdorfer2015Applications}, and in various
aspects of fundamental research such as tests of quantum
electrodynamics (QED) \cite{Bei10}, developing relativistic atomic
structure theory, and other applications
\cite{Martinson1989,zou2016handbook,BeyerKlugeShevelko1997,Gillaspy2001}.
In the present review we address novel avenues of research which
have recently been proposed, and recapitulate about their foundation
in the exciting atomic physics properties of HCI.

Due to their fundamental role in atomic structure theory and quantum
mechanics, the most investigated isoelectronic sequences have been
the hydrogenlike and lithiumlike ones, with one single $ns$ electron
being amenable to accurate theoretical calculations of the dominant
terms of the energy Hamiltonian. The heliumlike sequence has also
seen a wealth of calculations (see, e.g.
\cite{Indelicato1987,Lindgren1995,Cheng2000,Lindgren2001,Johnson1995,Andreev2001,Chen1993,Cheng1994,Plante1994,Drake1979,Drake1988,Drake1995,Drake2002,ASYP05}),
since understanding the two-electron correlations  is believed to be
the gateway to the study of more complex quantum systems.

Relativistic fine-structure, QED and nuclear size effects (see
\sref{sec:EST-QED}) show along any given isoelectronic sequence a
much stronger dependence on the nuclear charge than electronic
correlations. Advantageously, this tunability of the various terms
of the Hamiltonian can be used to tailor studies capable of
separating their relative contributions.

Owing to the reduction of the number of bound electrons, electronic
structure theory becomes -- in principle -- less complicated.
Nonetheless, in astrophysics and plasma applications (see, e.g.,
reviews by \citet{Beiersdorfer2003e,Beiersdorfer2015Applications}),
which often deal with ions having more complex electronic
structures, an accurate treatment of electronic correlations is as
important as that of relativistic effects and QED. This is crucial
in order to improve the quantitative understanding of plasmas.

In most cases, the strongest transitions observed in HCI are in the
X-ray region. Nonetheless, they also display observable transitions
in every spectral range, and in particular also in the optical
regime, around which this review is centered. There are several
possibilities for optical transitions arising in HCI
\cite{Crespo2008}. As a function of increasing charge state, fine-
and even hyperfine structure splittings reach the optical regime.
Furthermore,  level crossings between different electronic
configurations can result in transitions in the optical wavelength
range.

Production of HCI in a laboratory environment is a long-standing
problem. Few groups could routinely generate those ions, since the
techniques were complicated and expensive. For these reasons, there
is still an enormous scarcity of data and the majority of possible
ions in this ``spectral desert'' remains unexplored. As for
high-accuracy measurements, the best results in the field of HCI
spectroscopy lag by more than ten orders of magnitude behind of what
is now possible in atomic physics with neutral atoms, or singly
charged ions. Compounding the lack of experimental data, theory
encounters a rugged terrain when scouting beyond its usual range.
Complex electronic configurations with several shell vacancies in
the ground state configuration are still very hard challenges for
advanced atomic structure theory.

Further development of both HCI theory and experiment became crucial
in the past several years after a completely new avenue of research
blossomed following the pioneering work by \citet{BDF10}, which
identified HCI optical transitions between states with different
electronic configurations and proposed their applications for
optical metrology and tests of  variation of the fine-structure
constant. This work and subsequent studies
\cite{BerDzuFla11b,BerDzuFla11a,BerDzuFla12,DzuDerFla12,DzuDerFla12a,DerDzuFla12,BerDzuFla12a,DzuDerFla13E,
KozDzuFla13,SDFS14,SDFS14a,SDFS14b,YTD14,DzuFlaKat15,DSSF15}
demonstrated that despite very large ionization energies, certain
HCI have very narrow transitions that lie in the optical range and
can be used for the development of ultra high-precision clocks and
tests of fundamental physics.

This new  dimension in the parameter space of precision atomic
physics opens a wide field of research opportunities. HCI are less
sensitive to external perturbations than either neutral atoms or
singly charged ions due to their more compact size. Recent studies
of systematic uncertainties
\cite{DerDzuFla12,DzuDerFla12a,DzuDerFla13E} have shown that the
achievable fractional inaccuracy of the transition frequency in the
clocks based on HCI may be smaller than 10$^{-19}$ using shift
cancelation schemes. At the same time, the clock transitions in HCI
are more sensitive to the variation of $\alpha$ than those of
neutral atoms \cite{BDF10}. Therefore, HCI-based clocks may allow
significant improvement of the current experimental limit on
$\alpha$ variation on the level $\dot{\alpha}/\alpha \lesssim
10^{-17}$~yr$^{-1}$ \cite{Ros08,GodNisJon14,HunLipTam14}. Moreover,
optical clocks are sensitive not only to a linear variation of
$\alpha$, but also to hypothetical oscillations and occasional jumps
of this parameter. Such effects can be caused by cosmological fields
\cite{SF15c,StaFla16} and topological defects \cite{DePo14}, which
are often considered as candidates for dark matter
\cite{Der16,AHT15}. In all these cases sensitivity to $\alpha$
variation is given by the same sensitivity coefficients. Therefore,
HCI based clocks are also more sensitive to all these effects than
state-of-the-art atomic clocks.

Two major obstacles on the way toward the realization of such
proposals in 2010 were the lack of accurate theoretical descriptions of
potential clock candidates and, even more importantly, lack of
experimental techniques to decelerate, trap, cool, and control HCI to observe
such weak transitions and support the development of the frequency
standards.

At the present time, theoretical  studies have identified a list of HCI
candidates for optical clock development and provided extensive calculations
of relevant atomic properties. In 2015, crucial experimental steps were achieved
with a breakthrough demonstration of sympathetic cooling of Ar$^{13+}$ with a
laser-cooled Be$^+$ Coulomb crystal in a cryogenic Paul trap
\cite{SchVerSch15}. This experiment heralded the start of a new era in
exploration of HCI with the techniques previously reserved for neutral and
singly-charged systems. Combined with quantum logic spectroscopy, in which another ion is used for cooling and readout of
the clock transitions, as it is done in the Al$^+$ clock \cite{schmidt_spectroscopy_2005}, cold trapped HCI, suddenly become an experimentally-accessible resource for precision fundamental
studies. These developments represented a turning point for  HCI studies and this review aims to
summarize both theoretical and experimental
findings and to discuss possible directions of further research.

Both theory and experiment have to be improved in order to achieve the
best possible scientific harvest from these investigations.
On the other hand, this is a worthwhile endeavor given the plethora of available ground state configurations with laser-accessible transitions that HCI offer.
Furthermore, studies along isoelectronic sequences in HCI afford a large degree
of tunability concerning the systematics of the various binding energy
contributions, aiding the identification of yet unassigned HCI spectra, as shown in \cite{WinCreBel15}.
New measurements will provide benchmarks of the theoretical predictions and help to further
improve the theory.

%In this review we discuss recent theoretical development in the description of
%HCI and identification of the optical transitions suitable for development of
%atomic clocks and subsequent tests of fundamental physics. The present
%paper is complementary to the recent review in atomic clocks based on
%neutral and singly-charged systems \cite{Ludlow2015}. We will systematize
%various HCI clock proposals that will aid identifying the best systems for
%experimental developments. Production of the HCI, cooling, trapping, and quantum logic spectroscopy schemes, systematic uncertainties that will arise in the %development of HCI frequency standards, and a perspective on the coming decade of research will be described. Finally we will discuss other potential applications %of cold HCI.

\section{Background: variation of fundamental constants}
\label{sec:variation}

Variation of fundamental constants is predicted by many extensions
of the Standard Model (SM). Experimental searches for
$\alpha$-variation allow testing of these models and search for the
new physics beyond the SM. Light scalar fields appear very naturally in
cosmological models, affecting parameters of the SM including
$\alpha$. Space-time variations of these scalar fields are expected
because of the evolution of the Universe's composition. Theories
unifying gravity and other interactions predict spatial and temporal
variation of physical ``constants'' in the Universe
\cite{Mar84,DSW08,CalKel15}. Moreover, all coupling constants and
masses of elementary particles in such models can be both space and
time-dependent, and influenced by the local environment
\cite{Uza11}.

Searches for variation of fundamental constants are conducted in a
very wide range of systems including astrophysical studies of quasar
spectra and observation of the H I 21 cm line, atomic clocks, the
Oklo natural nuclear reactor, meteorite dating, stellar physics,
cosmic microwave background, and big bang nucleosynthesis (see e.g.
reviews \cite{Uza11,CFK09,KoLe13}). Here, we only briefly discuss
the laboratory limits set on the variation of the fundamental
constants by atomic clocks and focus on the HCI proposals.

The transition frequency $\nu$ between two electronic energy levels
in an atom is only dependent on the fine-structure constant:
\begin{equation}
\nu \simeq c R_{\infty} A F(\alpha),
\label{eq3-FC}
\end{equation}
where $R_{\infty}$ is the Rydberg constant, $c$ is the speed of
light in vacuum, $A$ is a numerical factor depending on the atomic
number $Z$ and $F(\alpha)$ is a function which depends upon the
particular transition. Based on their respective frequency ranges,
current atomic clocks based on such electronic transitions are
referred to as optical clocks, while clocks based on hyperfine
transitions are referred to as microwave clocks. The frequency of
the electromagnetic radiation associated with transitions between
the hyperfine levels, such as the Cs transition defining the SI
second, may be expressed as

\begin{equation}
 \nu_\textrm{hfs} \simeq c R_{\infty} A_\textrm{hfs}
 \times g_i \times \frac{m_\mr{e}}{m_\mr{p}}
 \times \alpha^2 F_\textrm{hfs}(\alpha),
 \label{eq2-FC}
\end{equation}
where $m_e$ and $m_p$  are electron and proton masses, respectively,
$A_\textrm{hfs}$ is a numerical quantity depending on the particular
atom, and $F_\textrm{hfs}(\alpha)$ is a relativistic correction
specific to each hyperfine transition. The dimensionless quantity
$g_i=\mu_i/\mu_\textrm{N}$   is the   g-factor associated with the
nuclear magnetic moment $\mu_i$, where
$\mu_\textrm{N}=e\hbar/2m_\mr{p}$ is the nuclear magneton. The
potential variation of $g$-factors may be reduced to more
fundamental quantities, such as $X_q=m_q/\Lambda_{\textrm{QCD}}$,
where $m_q$ is the average light-quark mass and
$\Lambda_{\textrm{QCD}}$ is the  QCD energy scale.
 As a result, the hyperfine transition frequencies are sensitive to
the variation in $\alpha$, $\mu=m_\mr{p}/m_\mr{e}$, and $X_q$.

Measuring the ratios $R=\nu_1/ \nu_2$ of optical to hyperfine clocks
over time  sets limits on the variation of $\alpha$, the
proton-to-electron mass ratio $\mu$, and nuclear $g$ factors,
specifically $g_\textrm{Cs}$ and $g_\textrm{Rb}$, as these
correspond to the two  microwave clocks with the smallest
uncertainties.

The ratio of frequencies of any two optical clocks depends only upon
$\alpha$. The sensitivity of the particular optical atomic clock to
$\alpha$-variation depends on the parameter $q$ that links variation
of the transition energy $E_0$, and hence the atomic frequency
$\nu=E_0/h$, to the variation of $\alpha$
\begin{equation}
 \frac{\delta E}{E_0} = \frac{2q}{E_0}\frac{\delta
 \alpha}{\alpha_0} \equiv K  \frac{\delta \alpha}{\alpha_0},
 \label{eq:K}
\end{equation}
where
\begin{equation}
 K=\frac{2q}{E_0}
 \label{eq4-FC}
\end{equation}
is a dimensionless sensitivity factor.

The relationship between the ratio of two clock frequencies and the
variation of $\alpha$  is then given by the difference in their
respective $K$ values for each clock transition, i.e. $\Delta
K=|K_2-K_1|$. The larger the value of $K$, the more sensitive is a
particular atomic energy level to the variation of $\alpha$.
Therefore, it is advantageous to select transitions with
significantly different values of $K$, preferably of the opposite
sign.  These $K$ factors allow comparison of the sensitivity to
$\alpha$-variation between transitions with significantly different
frequencies.

The $K$ factors are small for most clocks currently in development,
Al$^+$ (0.008), Ca$^+$ (0.15), Sr$^+$ (0.4) , Sr (0.06), Yb (0.3),
Yb$^+$ quadrupole transition (0.88) \cite{DzuFla09,
dzuba_relativistic_2003}. The $K$ factors for Hg$^+$ and Yb$^+$
octupole clock transitions are $-3$ and $-6$, respectively,  making
them the best candidates for one member of a clock-comparison pair,
with the other member taken from the previous group. A particular
attraction of HCI is the availability of transitions with much
larger $K$ factors.

The most accurate single laboratory
test of $\alpha$-variation comes from the  Al$^+$/Hg$^+$ optical clock comparison \cite{Ros08}, setting the limit
\begin{equation}
\frac{\dot{\alpha}}{\alpha}=(-1.6\pm2.3) \times 10^{-17} \,
\textrm{yr}^{-1}. \label{eq6-FC}
\end{equation}
The global limits to the present day variation of
$\alpha$ and $\mu$ from all present clock comparisons and Dy measurements \cite{LeeWebCin13} are given by \citet{GodNisJon14, HunLipTam14}:
\begin{eqnarray}
\label{Eq_alpha_var}
\frac{\dot{\alpha}}{\alpha}&=&(-2.0 \pm 2.0) \times 10^{-17}\,  \textrm{yr}^{-1}\\
\label{Eq_mu_var}
\frac{\dot{\mu}}{\mu}&=&(0.2 \pm 1.1) \times 10^{-16} \,
\textrm{yr}^{-1}.
\end{eqnarray}

A number of optical and near-optical transitions in various HCI were
shown to be very sensitive to the possible variation of the
fine-structure constant
\cite{BDF10,BerDzuFla11b,BerDzuFla11a,BerDzuFla12,DzuDerFla12,
DzuDerFla12a,DerDzuFla12,BerDzuFla12a,DzuDerFla13E,
KozDzuFla13,YTD14,DzuFlaKat15,SDFS14,SDFS14a,SDFS14b,DSSF15}. The
energy scale for HCI levels is large ($\sim
(Z_a+1)^2$~$R_{\infty}$), where $Z_a$ is the ionization charge.
Optical transitions arising due to a level crossing take place
between states with accidental near degeneracy. If the orbital
angular momenta of the corresponding  energy levels differ by two
units or more, these energies have a different dependence on
$\alpha$. Relativistic corrections for HCI scale as $\alpha^2 Z^2
(Z_a+1)^2$~$R_{\infty}$, where $Z$ is the nuclear charge, while
optical frequencies are of the order of a fraction of $R_{\infty}$.
As a result, the frequencies of optical transitions depend on
$\alpha$ as \cite{BDF10}:
\begin{align}\nonumber
 \frac{\delta \nu}{\nu} \sim \alpha^2 Z^2 (Z_a+1)^2,
 \quad \Rightarrow\quad K\sim 2\alpha^2 Z^2 (Z_a+1)^2
 \,.
\end{align}
Therefore, the sensitivity of optical transitions in HCI to
$\alpha$-variation is enhanced by the factor $(Z_a+1)^2\sim 10^2$
compared to similar transitions in neutral atoms. Brief reviews of
$\alpha$-variation in HCI were recently published by
\citet{OngBerFla14} and \citet{DzFl15}.

In 2011, a very large analysis of quasar absorption spectra that
combined data taken by the Keck telescope in Hawaii and the Very Large
Telescope (VLT) in Chile indicated $4\sigma$ spatial gradient in the
value of $\alpha$ \cite{alpha-dipole}. A 2015 study of systematic
distortions in the wavelengths scales of high-resolution
spectrographs \cite{WhitMur15} showed that instrumental error may
weaken the spatial variation result, but can not
explain all of the observed $\alpha$-variation.
 Calculated sensitivity coefficients of the optical
transitions in HCI to $\alpha$-variation are indeed much higher than
in neutral atoms or singly-charged ions. This opens the possibility to
drastically improve present laboratory limits on $\alpha$-variation,
or find such variation and explore new physics beyond the Standard
Model.

Up to now, most laboratory searches of $\alpha$-variation were
focused on  smooth drifts during the whole duration of the
experiment \cite{Ros08,HunLipTam14,GodNisJon14}, or on annual
variations, which can be linked to the variations of the
gravitational potential of the sun (see
\cite{BlaLudCam08,LeeWebCin13} and references therein). However,
some modern models of dark matter predict oscillations of the
fundamental constants at the Compton frequency of the cosmological
field, or even random jumps when the Earth passes domain walls, or
other topological defects, associated with such a field
\cite{DePo14,SF15c,AHT15,Der16}. In the first approximation the
sensitivity coefficient of a given optical transition to
$\alpha$-variation is the same for gradual variation and for
periodic variation if its frequency is much smaller than transition
frequency. Because of that, HCI can be used to search for such
variations and to test predictions of this class of theoretical
models of dark matter. Recently, HCI have also been identified as
potential candidates for significant improvement for tests of
Lorentz symmetry \cite{ShaOzeSaf17}. New experimental techniques for
HCI described in this review combined with an improved theoretical
description  will allow for more stringent QED tests.

%%%%%%%%%%%%%%%%%%%%%%%%%%%%%%%%%%
%\input{theory_v11}
%%%%%%%%%%%%%%%%%%%%%%%%%%%%%%%%%%

\section{Electronic structure theory and tests of QED}
\label{sec:EST-QED}

\subsection{Theoretical treatment of QED in HCI}
\label{sec:QEDtreatment}

QED laid the foundation of the modern formalism of the Standard
Model as the first relativistic quantum field theory
\cite{BjoDre64,PeskinSchroeder1995QFTbook,alb1965}. It is arguably
the most stringently tested part of the Standard Model. Highly
charged ions are extremely relativistic systems and an accurate
prediction of their electronic structure has to include large QED
contributions. The understanding of QED contributions is crucial for
a number of precision tests of physics beyond the SM, including
those described in this review. QED contributions are also needed
for determining fundamental constants. Therefore, we start the
discussion of HCI electronic structure with the chapter on QED
calculations in HCI and briefly review recent bound-state QED
(BSQED) tests. We refer the reader to reviews by
\citet{VolGlaPlu13,StuWerBla13,ShaGlaPlu15,StuVogKoh17,Indelicato2017,Eides2001,DraYan08,Bei10,Karsh2005}
for detailed discussion of QED calculations and tests of QED. The
electronic structure of HCI specific to the metrology applications
and searches for the variation of fundamental constants is discussed
in Section~\ref{theory2}.

While in heavy atoms the QED contributions to the binding energy of
the inner-shell electrons are equally strong to those in HCI, their
investigation is hindered by the presence of filled outer shells.
This causes, on one side, noticeable energy shifts, and, on the
other side, reduces the lifetime of excited states through Auger
decay coupling the initial state to the ionization continuum
\cite{Zimmerer1991}. The resulting broadening of the electronic
transitions of interest reduces spectral accuracy. In addition,
theoretical methods are best developed for few-electron systems, and
therefore the research quest was primarily the study of hydrogenlike
and lithiumlike heavy HCI, and experimental efforts have also
focused on such systems \cite{Bei10}.

Since the expansion parameter $Z\alpha$ in perturbative theory would
approach a value close to 1 for heavy elements, it was not clear how
far the usual expansion-based approximations would remain valid, and
if contributions from two-loop QED \cite{Yerokhin2003} could be
appropriately be accounted for. To address this, non-perturbative,
all-order methods have been developed, e.g.,
\citet{Shabaev2002,Lindgren2001}, and two-loop calculations carried
out by e.g., \citet{ASYP05,Yerokhin2003}.

A consistent QED approach is possible only within $1/Z$ perturbation
theory for systems with up to three, or four electrons
\cite{YIS03,ASYP05,YAS07,SC11,YeSha15a}. For many-electron ions,
which are of most interest to this review, the use of mean-field
approximations (such as Dirac-Fock approximation) is necessary, and
correlations are treated within relativistic quantum mechanics. In
this case, QED corrections can be included by means of model
potentials
\cite{Blu92,CJS93,FlaGin05,TuBe13,STY13,RDF13a,STY15,GiBe16}. In
such calculations, the electron-electron interaction is usually
treated within the Coulomb-Breit approximation, see
e.g.~\cite{SPPB15,KonKoz15}.

Electron-electron correlation effects in  many-electron ions, can be
included with the help of several methods: (i) many-body
perturbation theory (MBPT) \cite{Johnson07}; (ii) configuration
interaction (CI) \cite{KT87,FFG02,FTGG07,YeSu12}, or
multiconfiguration Dirac-Hartree-Fock (MCDHF) \cite{FTGG07}; (iii)
coupled cluster (CC) method \cite{HN94,EKI94}; (iv) combinations of
CI and MBPT (CI+MBPT) \cite{DFK96b,SJ02,KPST15}, or CI and all-order
(CI+AO) methods \cite{Koz04,SKJJ09}. Recently \citet{TKSSD16}
incorporated four of the most popular effective QED potentials into
the CI+AO method, and tested them for several HCI. Recent
developments of these theoretical methods have considerably improved
the accuracy and reliability of newer calculations for HCI, and
raised the predictive power of theory. We discuss the QED studies in
HCI with one to three valence electrons (i.e.\ H-like to Li-like) in
Section \ref{Sec:QED} below and return to the subject of
many-electron HCI in Section~\ref{theory2}.

\subsection{Tests of QED effects in HCI}
\label{Sec:QED}
%{\bf Notes: \\}

\subsubsection{Lamb-shift studies in the x-ray region}
\label{Sec:QEDX-ray} In first order, the Lamb shift -- understood as
the difference between the Dirac binding energy and the actual one
-- of a hydrogenic ion scales with $(\alpha Z)^4/n^3$, and rises for
the $1s$ electron of U$^{92+}$ to more than 425\,eV
\cite{Mohr1974,Johnson1985}. For this reason, very soon after the
development of experimental methods for the production of HCI
\cite{Mokler1985}, they were seen as potentially very interesting
probes of QED calculations
\cite{Mohr1985,Mohr1992,Mohr1998,Persson1997,Blu92,Shabaev2002,Johnson2004,Volotka2013}
and, accordingly, studied in many experiments
\cite{Briand1983a,Briand1983,Deslattes1984,Richard1984,Tavernier1985,Marmar1986,Beyer1985,Briand1990,Beyer1991,Beyer1995,Gumberidze2004,GumStoBan05}.
In fact, one of the strongest drives for research with heavy HCI in
large facilities such as Bevalac at Lawrence Berkeley National
Laboratory, GSI in Darmstadt, Germany, and GANIL in Caen, France,
was testing QED in the non-perturbative regime. The HITRAP facility
\cite{Rodriguez2010} at GSI continues pursuing this type of
research.

For the ground state of hydrogenlike uranium (U, $Z=92)$, the most
recent Lamb-shift measurement by \citet{GumStoBan05} has yielded
$460.2\pm 4.6$ eV, to be compared with predicted 463.99(39)~eV.
Radiative corrections from QED contribute 265.2~eV, and a shift of
-1.26(33)~eV results from 2nd order QED
\cite{Yerokhin2003,Beiersdorfer2005TwoLoop}. A comparably large
correction arises from finite nuclear-size effects, with a total of
198.54(19)~eV \cite{KozAndSha08}. These results confirm QED theory
predictions at the 2\% level in the strongest stationary
electromagnetic fields that nature provides, an extreme regime where
the rest mass of the electron is only four times larger than its
binding energy to the nucleus. Given the high accuracy of theory,
hydrogenic systems have been proposed as calculable x-ray standards
\cite{Flowers2001}.

For two-electron systems, and specifically for heliumlike ions,
there is abundant theoretical literature
\cite{Indelicato1987,Lindgren1995,Cheng2000,Lindgren2001,Johnson1995,Andreev2001}.
At medium $Z$, relativistic configuration interaction and
perturbative many-body methods
\cite{Chen1993,Cheng1994,Cheng2000,Plante1994} as well as
unified-method calculations
\cite{Drake1979,Drake1988,Drake1995,Drake2002} have yielded
reasonably accurate results in close agreement with measurements of
the $1s-2p$ x-ray transitions
\cite{Briand1984,Indelicato1986,Widmann1996,BruBraKub07,Kubicek2014,KubBraBru12,AmaSchGue12,Rudolph2013,Epp2015,Beiersdorfer2015,msa2017}.

A controversy arising from a claimed $Z$-dependent divergence
between earlier experimental data and calculations
\cite{ChaKinGil12,ChaKinGil13,Gil14}, soon disputed by
\citet{Epp13}, has been settled, with all newer results agreeing
very well both with older calculations and advanced theory
\cite{ASYP05}. Nonetheless, better measurements will be needed to
test higher-order QED as well as interelectronic contributions to
the binding energy in more detail.

In lithiumlike systems, the $2s_{1/2} \rightarrow 2p_{1/2,3/2}$
transition energies display the largest relative QED contributions
(up to 15\%), and have been studied in detail both experimentally
\cite{Schweppe1991,Staude1998,Bosselmann1999,Feili2000,Brandau2002,Brandau2003,Epp2007,Zhang2008,Epp2010,Andreev2001,Andreev2012}
and theoretically among others by
\citet{Indelicato1990,Kim1991,Artemyev1999,Cheng2000b,SC11,YeSu12,Artemyev2003b}.
A prominent example, the $2s_{1/2} \rightarrow 2p_{1/2}$ transition
energy in lithiumlike U$^{89+}$, 280.645(15)\,eV, was measured with
an 0.005\% uncertainty, and agrees perfectly with the theoretical
value of 280.71(10)~eV by \citet{KozAndSha08}. These results tested
second-order (in $\alpha$) QED effects to 6\% \cite{VolGlaPlu13}.
Currently, the theoretical accuracy of HCI QED tests is limited by
the nuclear polarization correction.

At this point, there is a general consensus that significant
contributions from nuclear-size effects
\cite{Johnson1985,Mohr1993,Beier1998,Shabaev1993,Shabaev2000},
nuclear polarization corrections \cite{Plunien1989} and also
nuclear-recoil
corrections\cite{Shabaev1985,Palmer1987,Artemyev1995a,Artemyev1995b,Shabaev1998a,Shabaev1998b}
which are not sufficiently well known, compromise the ability to
extract information on high-order QED contributions. There have been
proposals \cite{ShaArtYer01} based on the different scaling of the
QED and nuclear size contributions with $Z$ and $n$ that could help
solving this conundrum; however, due to the lack of experimental
data, their application has only been possible in a few cases until
now \cite{Beiersdorfer2005TwoLoop}.

\subsubsection{Fine and hyperfine structure transitions}
\label{Sec:QED-FS-HFS}

Scaling laws push up the fine and hyperfine-structure (HFS) energy
splittings in HCI by several orders of magnitude. Among other
things, this gives rise to intra-configuration transitions due to
changes in total angular momentum $J$ \cite{Crespo2008} at energies
well below the x-ray and soft x-ray region that have been mentioned
above. Rearrangements of configurations giving rise to level
crossings are another possibility for optical transitions
\cite{BDF10,OngBerFla14,WinCreBel15}.

Experimentally, the accuracy that can be obtained  is much higher in
the optical range than in the x-ray region: e.g., an uncertainty of
only 0.6~ppm was achieved for $1s^{2} 2s^{2} 2p$~$^2P_{3/2}$ -
$^2P_{1/2}$ transition in boronlike Ar$^{13+}$ ions at 441.25568(26)
nm \cite{Draganic2003,Maeckel2011}. Theory is two orders of
magnitude less accurate \cite{Tupitsyn2003,artemyev2007}. Since the
non-relativistic energies of $p_{1/2}$  and $p_{3/2}$ states are the
same, this line and many other intra-configuration electric-dipole
(E1) forbidden transitions in a large number of isoelectronic
sequences that  contain similarly large relativistic and QED
contributions are excellent candidates for future precision tests of
many-body BSQED, in particular by means of sympathetically cooled
HCI \cite{SchVerSch15}.

The first direct observation of a hyperfine splitting (HFS) in the
optical range was achieved by resonant laser excitation of the M1
transition coupling the two hyperfine levels of the ground state of
hydrogenlike $^{209}$Bi$^{82+}$ ions circulating at relativistic
velocities in the GSI heavy-ion storage ring ESR \cite{Klaft1994},
followed by spontaneous-emission measurements of $^{187}$Ho$^{66+}$
(holmium, $Z=67$) (\cite{Crespo1996}), $^{185,187}$Re$^{74+}$
(rhenium, $Z=75$) (\cite{Crespo1998}), and $^{203,205}$Tl$^{80+}$
(thallium, $Z=81$) (\citet{Beiersdorfer2001}) ions trapped in an
EBIT, and a further experiment on $^{207}$Pb$^{81+}$
\cite{Seelig1998} at ESR. In all those experiments, systematic
effects, low resolution and statistics limited the relative
wavelength accuracies to about $1\times 10^{-4}$.

On the theoretical side, the one-loop self-energy correction to the
first-order hyperfine interaction in hydrogenic ions for various
nuclear charges was theoretically studied by \citet{Persson1996}.
Vacuum-polarization corrections to the HFS of Bi HCI were analyzed
by \citet{Labzowsky1997}, and leading non-recoil radiative
corrections to the HFS including effects of extended nuclear
magnetization calculated by \cite{Sunnergren1998}. As for the
nuclear recoil effect,
\citet{Shabaev1998a,Shabaev1998b,Shabaev2000,ShaArtYer01} performed
a sophisticated analysis of its influence on the various
transitions.

\subsubsection{Nuclear-size effects: Charge radius and magnetization distribution}

The uncertainty of the leading radiative terms in all mentioned
above calculations seems to be small compared with that of finite
nuclear-size effects (NSE) appearing at the few \% level in the HFS
splitting \cite{Shabaev1997}. Since the finite charge radius can be
independently measured in scattering experiments, its contribution
(at the 10\% level of the transition energy!) to the HFS could be
reasonably inferred. However, the nuclear magnetization
distribution, or Bohr-Weisskopf (BW) effect (at the level of 1\% of
the total HFS), is extremely difficult to determine independently.
Basically, the only other method that can measure this quantity is
$\gamma$ spectroscopy, and more recently also laser spectroscopy, on
muonic atoms. Therefore, in most cases the BW effect is accounted
for based on uncertain models of the nuclear magnetic structure. The
situation in the BSQED tests with HCI is akin to that of the laser
spectroscopy of hydrogen and the proton-size puzzle
\cite{BeyMaiMat17,PohlJPSJ,PohAntNez10,Pohl2013,Pohl2016}: our
limited knowledge of the nucleus is the frontier for the most
stringent tests of QED.

In order to suppress the uncertainties stemming from finite NSE,
\citet{ShaArtYer01} introduced the concept of the specific
isonuclear difference between the ground-state HFS of the Li-like
ion, $\Delta E(2s)$, and the H-like ion, $\Delta E(1s)$:
\begin{equation}
 \Delta^{\prime}E=\Delta E(2s)-\xi \Delta E(1s)\,,
 \label{gg}
\end{equation}
where the parameter $\xi$ is theoretically chosen to cancel the NSE in this
difference.

By scaling with $1/n^3$ and applying relativistic corrections, $\xi$
can be calculated with high precision. For Bi (bismuth, $Z=83$)
ions, the method achieves now a relative uncertainty of $\approx
10^{-4}$ following better calculations ~\cite{VolGlaAnd12} of the
two-photon exchange corrections. If HFS were measured at the
$10^{-6}$ level, many-body QED effects could be benchmarked at a few
percent level \cite{VolGlaPlu13}. Experiments are already
approaching this level of accuracy, e.g.\ for the $2s_{1/2}
\rightarrow 2p_{1/2}$ EUV hyperfine transitions in Li-like and
Be-like ions of $^{141}$Pr \cite{BeiTraBro14}.

\citet{Karpeshin2015} have proposed the opposite approach, namely to
investigate the nuclear magnetization distribution based on the HFS
experimental data for various isoelectronic sequences, as had been
demonstrated by \citet{Crespo1998,Beiersdorfer2001} for the cases of
Re and Tl HCI.

In order to solve some open questions, \citet{Lochmann2014} repeated
the HFS measurements in hydrogenlike and lithiumlike Bi, and
obtained values in disagreement with earlier experimental work by
\citet{Klaft1994}. Recently, \citet{UllAndBra17}re-measured the HFS
transitions in $^{209}$Bi$^{82+}$ and $^{209}$Bi$^{80+}$, and
obtained $\xi$ with more than an order of magnitude improvement in
precision. Its theoretical value of $\xi=0.16886$ would allow to
cancel the BW correction for $^{209}$Bi \cite{ShaArtYer01}. However,
the experimental result $\Delta^{\prime}E=-61.012(5)(21)$~meV
(statistical and systematic uncertainties given in parentheses)
disagreed by 7$\sigma$ with the predicted value of
$-61.320(4)(5)$~meV \cite{VolGlaAnd12} (uncertainties in first and
second parentheses arise from uncalculated higher-order terms and
the uncertainty of the complete cancelation of all nuclear effects,
respectively). This result was considered a challenge to bound-state
strong-field QED theory \cite{UllAndBra17}. However, the explanation
which was found soon after was rather mundane: the value of the
nuclear magnetic moment for $^{209}$Bi that was used to analyze the
data was simply wrong, as recently found out by
\citet{Skripnikov2018}. Chemical shifts which are difficult to be
taken into account have now been properly included. This was a
suspicion that \citet{Gustavsson1998,Gustavsson1998b} had expressed
in their analyses of the results of \citet{Klaft1994}.

%A measure of caution is therefore always in order:
In this context, it is important to mention that the ubiquitous
atomic diamagnetism modifies the magnetic field experienced by the
nucleus in every determination of the nuclear gyromagnetic ratio.
This intrinsic effect is always present when bound electrons
surround the nucleus. Calculations of the diamagnetic shielding
factors that result from this effect have theoretical uncertainties.
Even more problematic are chemical shifts that appear in molecules
imbedded in chemical samples. Therefore, the accuracy of the derived
'corrected' nuclear magnetic moments is reduced, and the
interpretation of experiments becomes problematic. As an example,
the nuclear magnetic shielding factors by the single bound electron
in hydrogenic systems \cite{Yerokhin2011} has to be calculated to
all orders of QED expansion in the nuclear binding strength
parameter. In principle, solving these issues is a prerequisite for
the intended high-level QED tests, as pointed out by
\citet{Gustavsson1998,Gustavsson1998b}.

\subsubsection{Microwave studies of the bound-electron $g$ factor}
\label{sec:MW_g-factor}

Hitherto, the most stringent benchmark of QED calculations, and thus
of the Standard Model, comes from the very good agreement of the
measurements of the fine-structure constant $\alpha$ (a parameter of
the model that cannot be calculated from the first principles) by
strictly different methods. In the first method, the experimental
value of the free-electron magnetic-moment anomaly $a_\mathrm{e}$
\cite{CODATA2010,CODATA2014} is measured in a Penning trap
\cite{HanFogGab08} and combined with exact QED calculations that
include electroweak and hadronic contributions using expansions in
powers series of $\alpha/\pi$ with calculable coefficients. The
second approach is based on the measured Rydberg constant
$R_{\infty}$ \cite{WicHenSar02,CadMirCla08,BouClaGue11} obtained by
atom interferometry of recoiling atoms upon photon absorption.

For BSQED, the steep scaling laws governing the spin-orbit
interaction make trapped HCI extremely sensitive probes. The $g$
factor of the bound electron (for a theory review see
\citet{ShaGlaPlu15}) is determined to a very high precision also in
Penning traps \cite{StuWerBla13,StuVogKoh17}, to 10 significant
digits in the case of H-like $^{28}$Si$^{13+}$, as demonstrated by
\citet{StuWagSch11,Schabinger2012,StuWagKre13}. Here, the
experimental relative uncertainty is only $4\times 10^{-11}$,
leaving theoretical, uncalculated two-loop QED corrections of order
$\alpha^2(\alpha Z)^5$ and higher \cite{PacCzaJen05} as the largest
source of uncertainty. These results could be further improved by
combining theoretical and experimental values for two different
H-like ions \cite{StuKohZat14}. This idea of combining precise
$g$-factor measurements and QED calculations
\cite{StuWagKre13,YerHar13,Czarnecki2016}, has recently yielded a
13-fold improvement on the electron mass determination
\cite{StuKohZat14,KohStuKra15,ZatSikKar17}.

The most stringent BSQED experimental test of the $g$ factor for a
three-electron system was carried out for Li-like $^{28}$Si$^{11+}$
(silicon, $Z=14$) by \citet{WagStuKoh13,VolGlaSha14}, and is in
excellent agreement with theory that rigorously treats two-photon
exchange corrections \cite{VolGlaSha14}. With the works by
\citet{WagStuKoh13}, relativistic interelectronic interactions,
one-electron BSQED in magnetic fields and  screened bound-state QED
are tested at a level of precision corresponding to 0.001\%, 0.7\%
and 3\%, respectively.

Using cancelations of BSQED contributions between different
isotopes, e.g. for $^{40}$Ca$^{17+}$ and $^{48}$Ca$^{17+}$ ions,
nuclear effects can be tested \cite{KohBlaBlo16}. This is a
complementary approach to the specific difference scheme
(Eq.~(\ref{gg})) used to eliminate the nuclear effects in the
$g$-factor calculations \cite{ShaGlaSha02,VolPlu14}. This
methodology can be extended \citet{ShaGlaOre06} to $\alpha$
determinations using specific differences of the $g$ factors of
B-like and H-like ions with zero nuclear spin. Future experiments
with a Penning trap, ALPHATRAP \cite{StuWerBla13,StuVogKoh17},
coupled to an EBIT, and ARTEMIS \cite{LinWieGla13} at the HITRAP
\cite{Kluge2008} facility at GSI, will explore the
application of these methods to HCI in very high charge states.\\

%------------------------------------------------------------------
\begin{figure}[th]
\includegraphics[width=3in]{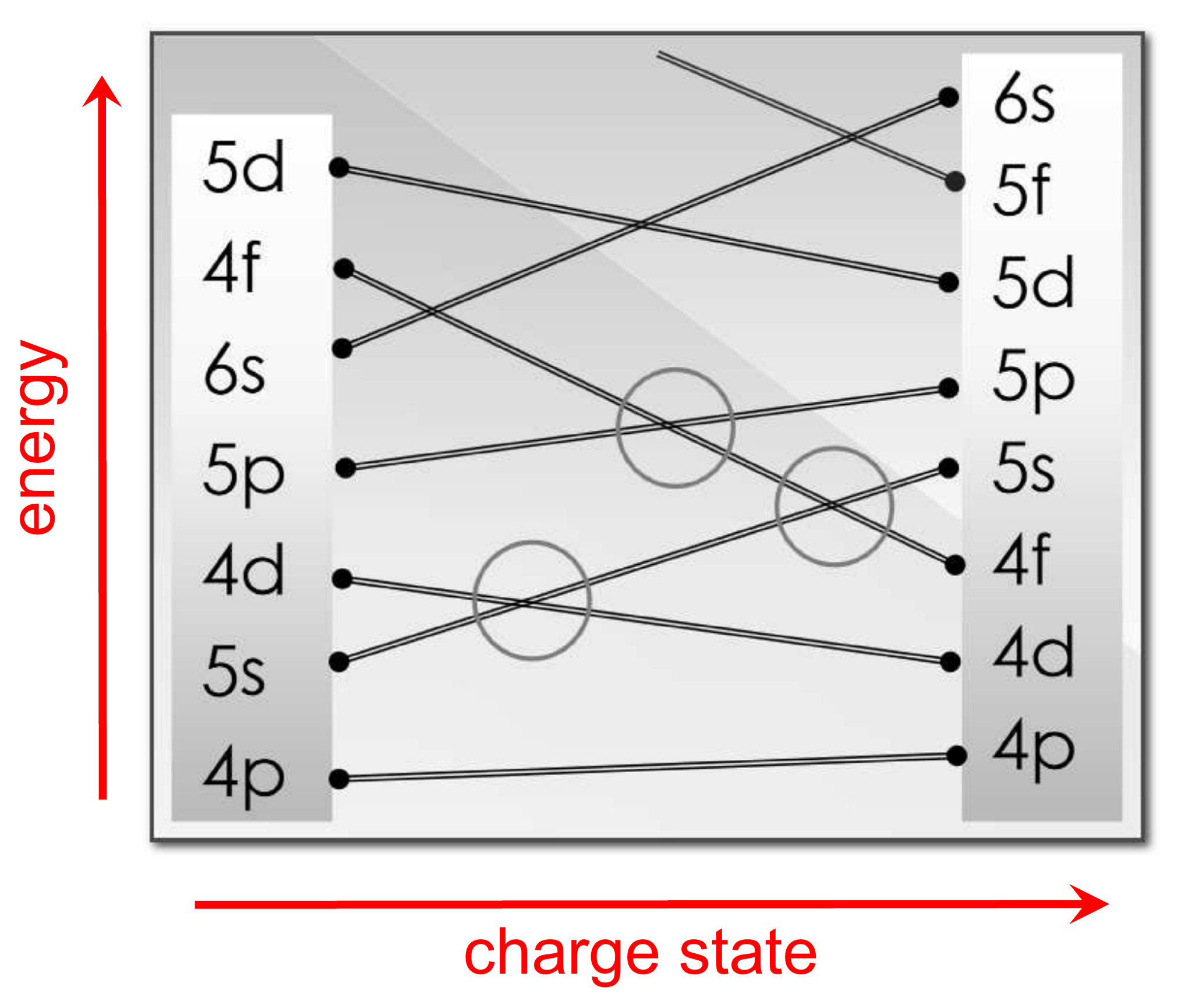}
\caption{Schematic of the shell order in neutral atoms (left) and in
hydrogen-like ions (right). One can see that the ``diving'' of $4d$ and
$4f$ shells result in level crossings in the areas marked by
circles.} \label{Fig_Shells}
\end{figure}

\section{HCI electronic structure and frequency metrology}
\label{theory2}

As we discussed in the previous section, BSQED tests and the
corresponding measurements of fundamental constants have to be
carried out in HCI with a very few valence electrons. In contrast,
the metrology applications and the search for the variation of
fundamental constants  discussed in Sec.~\ref{sec:variation}
requires HCI with rather complicated electronic structures, with the
special exception of the HFS in H-like ions. In this section, we
discuss the electronic structure of HCI relevant to these new
applications. At a first glance, the idea to use HCI for optical
clocks may appear strange. The energy scale for HCI electronic
levels is about $(Z_a+1)^2 R_\infty$, where $Z_a$ is the ion charge.
Therefore, one would expect optical transitions in HCI taking place
only between highly excited Rydberg states, being broadened by
competing X-ray transitions branches. Certainly, electric-dipole
(E1) transitions from the ground state lie in the extreme UV or
X-ray wavelength regime. Nonetheless, optical transitions of
interest to metrology in HCI occur within the ground state
configuration due to fine structure, HFS, and also near level
crossings, when the ions in the isoelectronic sequence change
between ground state configurations. We will now consider this
latter case.

First calculations of HCI systems which display such level crossings
were done by \citet{BDF10}. The general idea of that paper is as
follows: the order of shells in neutral atoms follows the $n+l$ rule
and is different from that of the hydrogen-like ions (see
\Fref{Fig_Shells}). For example, in hydrogen the $4f$ shell
immediately follows $4s$, $4p$, and $4d$ shells, while in the
periodic table it starts to fill after the $6s$ shell at $Z=58$
(cerium). The energy of the $4f$ levels decreases much faster than
energies of other levels if we move along isoelectronic sequences.
As a result, we can expect a re-ordering of electron configurations
(level crossings) near some value of $Z$. When such crossings
involve the ground state of an ion, we get low energy transitions,
which may lie in the optical range. If these transitions are of
$p-f$, or $s-f$ type (i.e. high order multipole transitions), they
are very narrow and suitable for high precision metrology. Note that
the $s-d$ crossing may also lead to narrow optical transitions, but
such crossings happen at a relatively low ionization stage. For
example, the $6s-5d$ crossing takes place in Tm-like sequence
between Ta$^{4+}$ and W$^{5+}$ \citet{BerDzuFla11a}.

Alternatively, one can consider transitions between the levels of
the ground state multiplet of an HCI. Such transition energies are
of the order of $(Z\alpha)^2 (Z_a+1)^2 R_\infty$. For moderately
heavy HCI these M1 transitions may lie in the visible range. This is
the case, for example, for Al-like ions V$^{10+}$ -- Cu$^{16+}$
\cite{YuSa16}. \citet{NaSa16} studied the fine structure transition
$^2$F$_{7/2} \rightarrow {}^2$F$_{5/2}$ in the configuration
$4f^{13}5s^2$ of the ions W$^{13+}$, Ir$^{16+}$, and Pt$^{17+}$.
\citet{Windberger2016} studied M1 optical transitions between fine
structure levels in Sn$^{11+}$ -- Sn$^{14+}$ ions. All these fine
structure transitions are of M1 type. For optical clocks we would
like to have weaker transitions to allow for significantly narrower
lines. \citet{Schiller2007} and \cite{YTD14} considered M1
transitions between hyperfine levels in heavier H-like HCI, such as
$^{171}$Yb$^{69+}$, or $^{207}$Pb$^{81+}$, where these transitions
lie in the optical range.

Below we will mostly focus on level crossings, where transitions are
of quadrupole, or even octupole type and the lines are extremely
narrow. Calculations of the ion properties in the vicinity of the
level crossing are technically very difficult. Interesting crossings
happen for degrees of ionization higher than 10 ($Z_a\gtrsim 10$),
when binding energies are of the order of 1 keV. That means that we
need atomic calculations with very high accuracy to identify optical
transitions, which correspond to energy differences of the order of
1 eV. In order to get transition energies with 10\% accuracy, one
needs fractional accuracy of the theory on the order of $10^{-4}$ or
better. This is very challenging for existing theoretical methods.
As a result, the first theoretical paper \cite{BDF10} was followed
by intensive further research
\cite{BerDzuFla11b,BerDzuFla11a,BerDzuFla12,DzuDerFla12,DzuDerFla12a,
DerDzuFla12,BerDzuFla12a,DzuDerFla13E,KozDzuFla13,YTD14,DzuFlaKat15}.
The most accurate calculations were done using a state-of-the-art
hybrid approach that combines coupled-cluster and configuration
interaction methods \cite{DSSF15,SDFS14,SDFS14a,SDFS14b}. The last
three works specifically considered all HCI which satisfied the
following criteria:
\begin{itemize}
 \item the existence of long-lived
metastable states with transition frequencies to the ground states
ranging between $(0.1\dots 1.8)\times 10^{15}$ Hz,
 \item high sensitivity
to $\alpha$-variation,
 \item the existence of stable isotopes, and
 \item relatively simple electronic structure, with one to four valence electrons.
\end{itemize}

\citet{SDFS14,SDFS14a,SDFS14b} found that only four isoelectronic
sequences satisfy the criteria above: Ag-like, Cd-like, In-like, and
Sn-like, which include ions with 46-electron core $[1s^2\dots
4d^{10}]$. \citet{BerDzuFla12a} and \citet{DSSF15} considered
heavier actinide ions which satisfy all criteria above with the
 exception of the existence of stable isotopes.

A suitable transition for laser cooling and internal state readout
is not required for the HCI when using quantum logic spectroscopy in
which a co-trapped singly charged ion provides for these features as
described in more detail in \sref{sec:QLS}. The dependence of the
clock transition on external fields and their gradients causes
systematic effects in atomic clocks. The size of HCI scales as
$1/(Z_a+1)$ and their dipole and quadrupole moments and
polarizabilities, both static and dynamic, are suppressed by an
order of magnitude and up to several orders for level-crossing and
hyperfine optical transitions, respectively. A number of papers
provided detailed investigation of systematic effects in optical
clocks based on HCI, reaching the conclusion that the next order of
magnitude improvement in the accuracy of frequency standards to
10$^{-19}$ uncertainty may also be achievable with HCI
\cite{DerDzuFla12,DzuDerFla12a,DzuDerFla13E,DzuFlaKat15,YTD14}. The
systematic effects in HCI clocks are discussed in more detail in
Sec.~\ref{unc}.

In the remaining part of this section we discuss calculations of the
spectra, lifetimes, and sensitivity to $\alpha$-variation for
particular systems. The accuracy of theoretical predictions strongly
depends on the number of valence electrons, as the valence-valence
correlations are very strong and can not be accurately treated
perturbatively. As a result, CI represents the best strategy to
include valence-valence correlations. However, the number of
configurations that has to be included into the CI calculations
grows exponentially with the number of valence electrons, limiting
accurate calculations to a few valence electrons. Systems with more
valence electrons usually also have a much denser spectrum, leading
to experimental difficulties in spectra identification, exacerbated
by large uncertainties in the theoretical predictions.

We start with the discussion of the proposals with H-like ions,
which are based on the narrow M1 hyperfine transitions. All other
HCI optical clock proposals are grouped by the number of valence
electrons, starting with the systems with one electron above the
closed shells and then discussing systems with two, three, and four
electrons. Finally, we discuss systems with one or more holes in
almost filled shells and a case with a mid-filled $4f$ shell.

\subsection{HFS of hydrogenlike ions}

As already mentioned in \sref{Sec:QED-FS-HFS}, in heavy H-like HCI
with nuclear spin $1/2$ the hyperfine transitions may lie in the
optical and near-optical range. These transitions are of the M1 type
and are very weak because they require a nuclear spin-flip.
\citet{Schiller2007} analyzed systematic effects for the cases of
$^{61}$Ni$^{27+}$ and $^{207}$Pb$^{81+}$. \citet{YTD14} discussed
hyperfine transitions in H-like  Yb, Pt, Hg, Tl, and Pb ions with
clock wavelengths below $3~\mu$m and nuclear spin $I=1/2$, listed in
Table~\ref{tab0}. For these ions, the ground-state hyperfine
structure consists of only $F=0$ and $F=1$ levels, simplifying
experimental realization since the $F=0$ level does not have Zeeman
components. The authors evaluated systematic effects due to
quadrupole shifts in inhomogeneous electric fields, Zeeman shifts,
blackbody radiation (BBR) shifts and ac-Stark shifts induced by the
clock laser. As a result, \citet{YTD14} concluded that systematic
effects can be controlled at a level below $10^{-20}$ fractional
accuracy. However, the achievable instability even for
\hci{171}{Yb}{69+}, which has the longest upper clock state lifetime
of 0.37~s,  is  $\sigma_y(\tau)\approx 5\times
10^{-15}/\sqrt{t/\mathrm{s}}$, requiring 9.5 months to reach
$10^{-18}$ fractional accuracy (see Sec.~\ref{HFS1}).

\begin{table} \caption{\label{tab0} Clock transition wavelengths $\lambda$ (in $\mu$m) and natural linewidth $\gamma/2\pi$ (in Hz)
for hyperfine transitions in H-like ions. The values are from \citet{YTD14}.}
\begin{ruledtabular}
\begin{tabular}{lcc}
\multicolumn{1}{l}{Ion}&
\multicolumn{1}{l}{$\lambda$ ($\mu$m)}&
\multicolumn{1}{c}{$\gamma/2\pi$}\\
\hline \\[-0.4pc]
$^{171}$Yb$^{69+}$ & 2.16 & 0.43 \\
$^{195}$Pt$^{77+}$ & 1.08 & 3.4 \\
$^{199}$Hg$^{79+}$ & 1.15 & 2.8 \\
$^{203}$Tl$^{80+}$ & 0.338 & 111.2 \\
$^{205}$Tl$^{80+}$ & 0.335 & 114.2 \\
$^{207}$Pb$^{81+}$ & 0.886 & 6.2 \\
\end{tabular}
\end{ruledtabular}
\end{table}

Hyperfine transitions are particularly interesting for the search
for a variation of fundamental constants because, like the Cs clock
transition, they depend on $\alpha$, the proton-to-electron mass
ratio $\mu$, and nuclear $g$ factors, which can be interpreted in
terms of the variation of $X_q=m_q/\Lambda_{\textrm{QCD}}$,
according to Eq.~(\ref{eq2-FC}). All other optical atomic clocks are
sensitive only to the variation of $\alpha$. This means that by
comparing such a HCI hyperfine clock with any other optical clock
one can significantly improve the laboratory limit on
$\dot{\mu}/\mu$, which is constrained according to
Eq.~\eqref{Eq_mu_var} at the level of $10^{-16}$ yr$^{-1}$. The
present constraint is limited by the Cs clock accuracy and the long
averaging times to reach low statistical uncertainties. Both
limitations may be overcome using optical hyperfine transitions in
HCI. Availability of different nuclei may also allow the setting of
further constraints on the variation of $X_q$. Moreover, the
sensitivity coefficients to $\alpha$-variation $K$ for heavy H-like
HCI vary monotonically from 1 for $Z=1$ to 4.3 at $Z=92$
\cite{Schiller2007}, therefore exceeding the sensitivity of the Cs
clock $K_{\rm{Cs}}=2.83$ clock for all ions listed in
Table~\ref{tab0}.

\subsection{HCI with one valence electron}
\paragraph{Ag-like ions.}

%------------------------------------------------------------------
\begin{table} \caption{\label{tab1}
Energies and $\alpha$-variation sensitivity coefficients $q$ relative to the ground state in cm$^{-1}$ for HCI with one and two valence electrons.  $K=2q/E$ is the enhancement factor. Wavelengths $\lambda$ (in nm) for transitions from the ground states and total radiative lifetimes $\tau$ (in s) are listed. Nd$^{13+}$, Sm$^{15+}$ , Nd$^{12+}$,  Sm$^{14+}$, and Es$^{17+}$  values are from CI+AO calculation \citet{SDFS14,DSSF15}. Nd$^{13+}$ and Sm$^{15+}$ wavelengths  are experimental values from \citet{ag-like-81}. Cf$^{17+}$ results are CI+MBPT calculations from
 \citet{BerDzuFla12a}. * indicates cases with no stable isotopes.}
\begin{ruledtabular}
\begin{tabular}{llrrrcc}
\multicolumn{1}{l}{Ion}&
\multicolumn{1}{l}{Level}&
\multicolumn{1}{c}{Energy}&
\multicolumn{1}{c}{$q$}&
\multicolumn{1}{c}{$K$}&
 \multicolumn{1}{c}{$\lambda$}& \multicolumn{1}{c}{$\tau$}\\
\hline   \\[-0.4pc]
Nd$^{13+}$   &$5s_{1/2}$  &     0     &            &       &        &          \\
             &$4f_{5/2}$  &   55706   &    104229  &   3.7 &179  & $1.3\times10^6$\footnote{This values includes $E3$ and $M2$ transitions. Inclusion of the hyperfine-induced E1 transition for $^{143}$Nd$^{13+}$
decreases the lifetime to $ 1.1-1.2 \times10^6$ s, depending on the hyperfine component of the transition \cite{DzuFla16}.} \\
             &$4f_{7/2}$  &   60134   &    108243  &   3.6 &166  & 0.996           \\[0.4pc]
Sm$^{15+}$   &$4f_{5/2}$  &      0     &            &       &       &         \\
             &$4f_{7/2}$  &   6444    &    5910    &   1.8 & 1526  & 0.308  \\
             &$5s_{1/2}$  &   60517   &    -134148 &   -4.4& 166 & $3.1\times10^5$ \\ [0.4pc]
*Cf$^{17+}$  &$5f_{5/2}$      &   0     &          &         &       &     \\
             &$6p_{1/2}$      & 18686   &  -449750 & -48      & 535 & \\
             &$5f_{7/2}$      &   21848    &  17900 &  1.6     &  458&     \\             [0.4pc]
Nd$^{12+}$   &$5s^2 \ ^1S_0$  &    0    &           &         &        &             \\
             &$5s4f \ ^3F_2$  & 79469   &   101461 &   2.6   & 126  &8.5$\times10^{10}$
             \\
             &$5s4f \ ^3F_3$  & 80769   &   102325 &   2.4   & 124  &19.7      \\ [0.4pc]
 Sm$^{14+}$  &$4f^2 \  ^3H_4$ &    0    &          &         &        &        \\
           &$5s4f \ ^3F_2$  &   2172  &   -127720 &   -118 &4600     &5.6$\times10^{13}$\footnote{The hyperfine quenching reduces this lifetime in $^{147}$Sm$^{14+}$ to $5\times10^7$ s -
$2\times10^9$ s  \cite{DzuFla16}.}     \\
           &$5s4f \ ^3F_3$  &   3826  &   -126746 &   -66  &2614     &8.51       \\ [0.4pc]
*Es$^{17+}$ &$5f^2 \ ^3H_4$&     0    &           &        &         &    \\
           &$5f6p \ ^3F_2$&   7445   & -46600    & -13     & 1343   & 11000\\
             \end{tabular}
\end{ruledtabular}
\end{table}

\begin{table} \caption{\label{comp} Comparison of the energies of Ag-like  Nd$^{13+}$, Sm$^{15+}$, and In-like Ce$^{9+}$ ions relative to the ground state with experiment  \cite{ag-like-81,in-like-01}. Differences with experiment are given in cm$^{-1}$ and \% in columns ``Diff.'' Adapted from \citet{SDFS14}.   }
\begin{ruledtabular}
\begin{tabular}{llrrrrr}
\multicolumn{1}{c}{Ion}&
\multicolumn{1}{c}{Level}&
 \multicolumn{1}{c}{Expt.}&
\multicolumn{1}{c}{CI+AO}&
\multicolumn{1}{c}{Diff.}&
\multicolumn{1}{c}{Diff.\%}\\
\hline   \\[-0.4pc]
Nd$^{13+}$ & $5s_{1/2}$&       0&       0&   0 &         \\
           & $4f_{5/2}$&   55870&  55706&  164& 0.29\% \\
           & $4f_{7/2}$&   60300&  60134&  166& 0.28\% \\
           & $5p_{1/2}$&  185066&  185028&   38& 0.02\% \\
           & $5p_{3/2}$&  234864&  234887&  -23&-0.01\%  \\ [0.4pc]
Sm$^{15+}$ &$4f_{5/2}$ &       0&     0  &      0  &          \\
           &$4f_{7/2}$ &    6555&  6444 &      111&   1.69\%   \\
           &$5s_{1/2}$ &   60384& 60517&     -133&  -0.22\%    \\
           &$5p_{1/2}$ &  268488& 268604&     -116&  -0.04\%    \\
           &$5p_{3/2}$ &  333203& 333385&     -182&  -0.05\%     \\ [0.4pc]
Ce$^{9+}$  &$5p_{1/2}$&      0  &      0 &    0&          \\
           &$5p_{3/2}$& 33427   &  33450 & -23&  -0.07\%   \\
            &$4f_{5/2}$& 54947   &   54683&  264&   0.48\%  \\
           &$4f_{7/2}$& 57520   &   57235&  285&   0.50\%
\end{tabular}
\end{ruledtabular}
\end{table}
The single-valent systems are simplest from the theoretical point of
view. It was pointed out already by \citet{BDF10}
that the $5s$ -- $4f$ crossing in the Ag-like isoelectronic
sequence ($N=47$) takes place near Pm$^{14+}$. The ions Nd$^{13+}$ ($Z=60$) and
Pm$^{14+}$ have the ground state configuration $[1s^2 \dots 4d^{10}]5s$
and the closely lying first excited state configuration $[1s^2 \dots
4d^{10}]4f$. These configurations exchange
places in Sm$^{15+}$.
 Energies and $\alpha$-variation sensitivity coefficients $q$ relative to the ground state for Nd$^{13+}$ and
 Sm$^{15+}$ are listed in Table~\ref{tab1}. The table also lists
 the enhancement factors $K=2q/E$,  wavelengths $\lambda$ (in nm) for transitions from the ground states and total radiative lifetimes $\tau$ (in s). The experimental wavelengths from \citet {ag-like-81} are given; the remaining values are from
  CI+AO calculation \cite{SDFS14}. Theoretical energy values are listed for consistency with the $q$ values.
  The Ag-like ions are among the very few HCI with near-optical transitions for which experimental measurements are available. The comparison of the theoretical and experimental energies relative to the ground state for Nd$^{13+}$,
 Sm$^{15+}$, and Ce$^{9+}$ is given
 in Table~\ref{comp}, adapted from \textcite{SDFS14}. The experimental values are from \citet{ag-like-81,in-like-01}. Ce$^{9+}$ has a $5s^2$ closed shell, so it can be considered as either a system
 with one or three valence electrons, as discussed below.  Excellent agreement with experiment is observed for all levels.
A detailed calculation of the Ag-like ion properties is given in \cite{SDFS14a}.
%------------------------------------------------------------------
\begin{figure}[htb]
%\begin{center}
%\hfill
\includegraphics[width=\columnwidth]{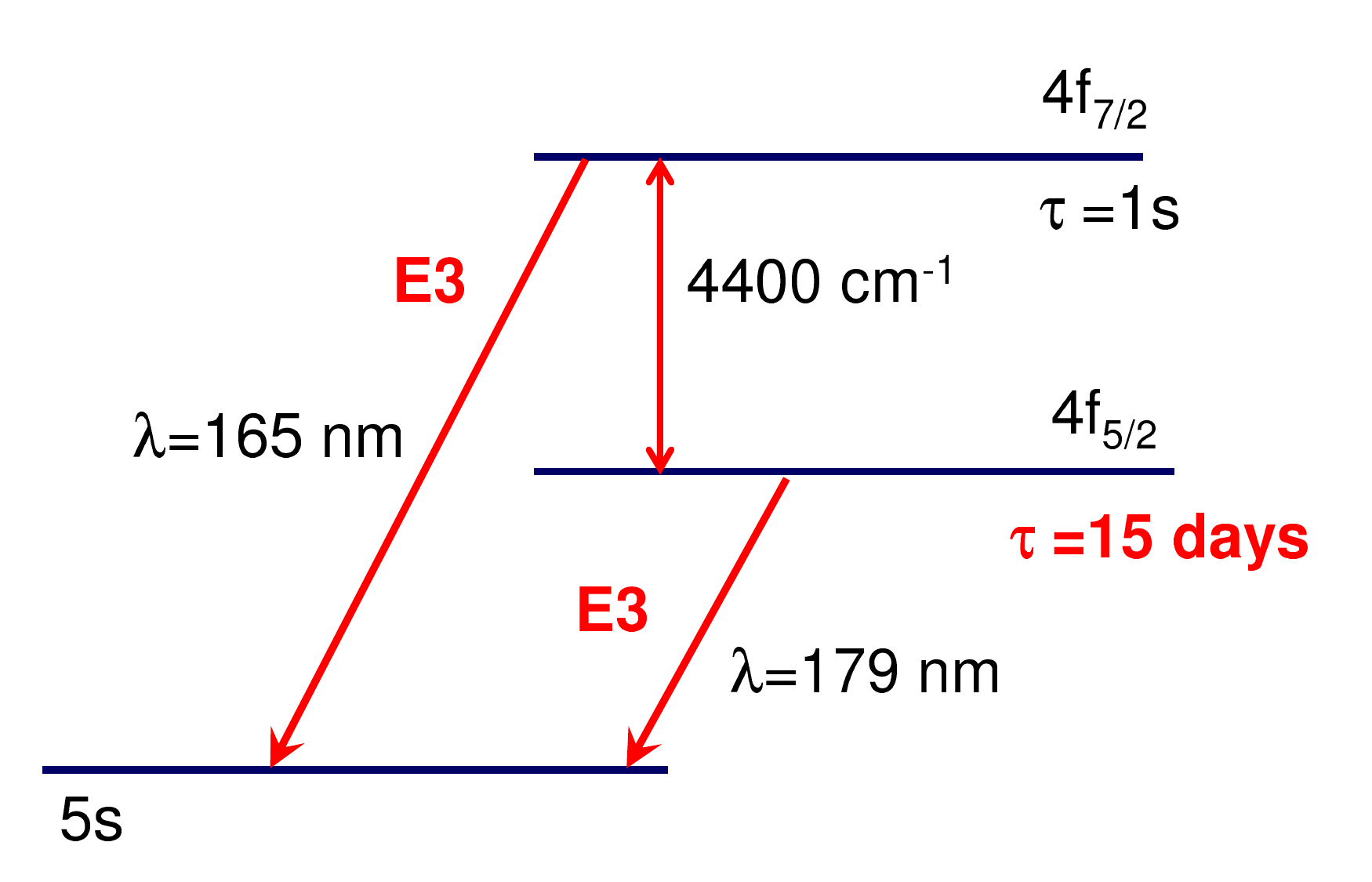}
%\hfill
%\hfill
%\end{center}
\caption{Energy levels and radiative lifetimes of low-lying levels
of Ag-like Nd$^{13+}$. Vertical intervals are not to scale.
The lowest state lifetime includes $E3$ and $M2$ transitions. Inclusion of the hyperfine-induced E1 transition for $^{143}$Nd$^{13+}$
decreases the lifetime to 12.5 -- 14 days, depending on the hyperfine component of the transition \cite{DzuFla16}.
From \citet{SDFS14a}.} \label{Fig_Ag-like_Nd}
\end{figure}
%------------------------------------------------------------------

The Pm$^{14+}$ ion has no stable isotopes and while the CI+MBPT calculation predicted
that
the $5s$ and $4f_{5/2}$
states are separated by 3973~cm$^{-1}$ \cite{BDF10}, the CI+AO calculations predicted only about $300$~cm separation \cite{SDFS14a}.

Nd$^{13+}$ represents a particularly attractive case since the
strongest transition from the metastable $4f_{5/2}$ level of this
ion is E3, resulting in the extremely long lifetime of more than 15
days (see \Fref{Fig_Ag-like_Nd}). The wavelength of the $5s-4f$
transition in Nd$^{13+}$ is in the VUV. The amplitudes
of such strongly forbidden transitions may depend on the nuclear
spin and can be significantly enhanced for odd isotopes.
\citet{DzuFla16} calculated such hyperfine-induced transitions for
$^{143}$Nd$^{13+}$, $^{149}$Pm$^{14+}$, $^{147}$Sm$^{14+}$, and
$^{147}$Sm$^{15+}$. The $^{143}$Nd$^{13+}$ lowest excited state value in Table~\ref{tab1} includes E3 and M2 transitions. Inclusion of the hyperfine-induced E1 transition for $^{143}$Nd$^{13+}$
decreases the lifetime to $ 1.1-1.2 \times10^6$~s, depending on the hyperfine component of the transition \cite{DzuFla16}.
The contribution of the hyperfine-induced E1 transition to the lowest excited state lifetime of the $^{147}$Sm$^{15+}$ is only 1\%.

\citet{DzuDerFla12} carried out a detailed assessment of the systematic uncertainties for Nd$^{13+}$ and
Sm$^{15+}$, including BBR shifts, Zeeman shifts, electric quadrupole shifts, and other perturbations,
concluding that the fractional accuracy of the clocks based on these systems may reach $10^{-19}$ if efficient shift cancelation schemes are applied.

\paragraph{Tl-like californium.}
\citet{BerDzuFla12a} considered the $5f-6p$ crossings which occur for the actinide ions. The CI+MBPT results for the
Cf$^{17+}$ ion are listed in Table~\ref{tab1}. Cf$^{17+}$  has a $6s^2$ closed shell, so it may be considered as a system with a single valence electron.
The sensitivity to $\alpha$-variation of the excited $6p_{1/2}$ state with respect to the ground state is extremely large, $q=-450000$ \cm\, with $K$ =48, and makes this system an ideal probe for that hypothesis.
Neither californium nor einsteinium, considered below, have stable isotopes.
However they have very long-lived isotopes, such as $^{249}$Cf with half life
of 351 years and $^{252}$Es with half life of 1.3 years. There are many facilities, including
Berkeley, Dubna, Darmstadt, RIKEN, etc., which produce and study unstable isotopes (see, e.g. \cite{Cf-prod,Es-prod}).

\subsection{HCI with two valence electrons}

\paragraph{Cd-like ions.}
Cd-like ions Nd$^{12+}$ and Sm$^{14+}$ have two valence electrons
and ground configurations $5s^2$ and $4f^2$, respectively. In both
cases the first excited configuration is $4f5s$. The lowest
multiplet of this configuration is ${}^3F_{2,3,4}$. In Nd$^{12+}$
the levels of this multiplet lie at approximately 79500, 80800, and
83700 \cm\ above the ground state, while for Sm$^{14+}$ they are
much lower, at 2200, 3800, and 8800 \cm\ respectively, as
illustrated in Table~\ref{tab1}. Therefore neither of the ions have
transitions in the visible range and Pm has no stable isotopes. The
lowest level ${}^3F_{2}$ is connected to the ground state by an $M2$
transition and has an extremely long lifetime, while the other
levels of this multiplet have lifetimes of the order of seconds. The
M2 lifetimes are strongly quenched by hyperfine-induced E1
transitions, by 4-6 orders of magnitude in  $^{147}$Sm$^{14+}$,
depending on the hyperfine component of the transition
\cite{DzuFla16}. Other details can be found in
\citet{SDFS14b,DzuFla16}.

\paragraph{Pb-like californium and einsteinium.} Cf$^{16+}$ has a very dense and
complex spectrum with three closely lying configurations, $5f^2$,
$5f6p$ and $6p^2$. According to the CI+MBPT calculation by
\citet{BerDzuFla12a}, the ground state is $J=3 [5f6p]$, with the
first excited state $J=0 [6p^2]$ at about 5000 \cm\ and $J=4 [5f^2]$
at roughly 10000 \cm. QED and high-order correlation corrections may
shift levels enough to even change their order. Therefore, new
studies are necessary to predict this spectrum more reliably. Note
that the sensitivity coefficients $q$ are very large and have
opposite signs: $q(J=3)\approx -371000$ \cm\ and $q(J=0)=+415000$
\cm. Es$^{17+}$ was considered by \citet{DSSF15} using the CI+AO
method; the clock transition data are listed in Table~\ref{tab1}.

\begin{table} \caption{\label{tab2} Energies and $\alpha$-variation sensitivity coefficients $q$ relative to the ground state in cm$^{-1}$; enhancement factor $K=2q/\omega$, wavelengths $\lambda$ (in nm) for transitions to the ground state, and lifetimes $\tau$ (in s) for HCI with three and four valence electron configurations. All values are obtained using the CI+AO method.
Cf$^{15+}$ and Es$^{16+}$  values are from  \citet{DSSF15}, the other data are from \citet{SDFS14}. Eu$^{14+}$ and Cf$^{15+}$
values include QED and three-electron corrections from \citet{TKSSD16,KSPT16}. * indicates cases with no stable isotopes.}
\begin{ruledtabular}
\begin{tabular}{llrrrcc}
\multicolumn{1}{l}{Ion}&
\multicolumn{1}{l}{Level}&
\multicolumn{1}{c}{Energy}&
\multicolumn{1}{c}{$q$}&
\multicolumn{1}{c}{$K$}&
 \multicolumn{1}{c}{$\lambda$}& \multicolumn{1}{c}{$\tau$}\\
\hline   \\[-0.4pc]
Ce$^{9+}$   &$5s^2 5p_{1/2}$&      0    &           &       &      &          \\
             &$5s^2 5p_{3/2}$&  33450   &   37544   &   2.2&  299&0.0030   \\
             &$5s^2 4f_{5/2}$&  54683   &   62873   &   2.3&  182&0.0812    \\
             &$5s^2 4f_{7/2}$&  57235   &   65150   &   2.3&  174&2.18     \\[0.4pc]
Pr$^{10+}$   &$5s^2 5p_{1/2}$&     0    &           &       &            &                   \\
             &$5s^2 4f_{5/2}$&  3702   &   73849   &   40  &2700&8.5$\times10^{4}$ \\
             &$5s^2 4f_{7/2}$&  7031   &   76833   &   22  & 1422&2.35 \\
             &$5s^2 5p_{3/2}$&  39141  &   44098   &   2.3 & 256&0.0018 \\[0.4pc]
Nd$^{11+}$   &$5s^2 4f_{5/2}$&    0       &          &       &          &       \\
             &$5s^2 4f_{7/2}$&  4180    &   3785    &   1.8 &  2392&1.19\\
             &$5s^2 5p_{1/2}$&  53684   &   -85692  &   -3.2& 186&0.061  \\   [0.4pc]
  Sm$^{13+}$  & $5s^2 4f~^2F_{5/2}$  &    0    &            &      &          &        \\
              & $5s^2 4f~^2F_{7/2}$ &      6203    &   5654    &   1.8& 1612&0.367 \\
              & $4f^2 5s~^4H_{7/2}$ &      20254   &   123621  &   12 &  494&0.133 \\ [0.4pc]
Eu$^{14+}$  &$4f^2 5s ~\textrm{J=7/2}$ & 0    &          &        & &\\
            &$4f^3    ~\textrm{J= 9/2}$ & 1262 &   137437  &   218 &7924 &\\
            &$4f^2 5s ~\textrm{J= 9/2}$& 2594 &   1942    &   1.5 &3855 &\\
            &$4f^3    ~\textrm{J=11/2}$ & 5388 &   141771  &   53  &1856 &\\  [0.4pc]
  *Cf$^{15+}$&$5f6p^2~  ^2F_{5/2}$&       0   &           &        & & \\
             &$5f^26p ~^4I_{9/2}$&   12898   &380000    & 59     &775 & 6900 \\
             &$5f6p^2 ~^2F_{7/2}$&   22018   &          &       &454 & 0.012 \\[0.4pc]
 *Es$^{16+}$ & $5f^26p ~^4I_{9/2} $&     0      &         &        & \\
            & $5f^26p ~^2F_{5/2} $&  6994 &  -184000    & -53&1430 &16000 \\
            & $5f^3 ~^2H_{9/2} $& 10591&            &     &944 &3.4  \\[0.4pc]
Pr$^{9+}$& $5s^2 5p^2$~$^3P_0$   &      0    &           &                 &          \\
         & $5s^2 5p4f~^3G_3$     &   20216   &   42721   &   4.2& 475& 6.6$\times10^{14}$ \\
         & $5s^2 5p4f~^3F_2$     &   22772   &   42865   &   3.8& 426& 59.0                \\
         & $5s^2 5p4f~^3F_3$     &   25362   &   47076   &   3.7& 382& 5.33                 \\[0.4pc]
Nd$^{10+}$ &$5s^2 4f^2~J=4$   &         0 &           &          &        &          \\
            &$5s^2 5p4f~\textrm{J=3}$  &   1564    &   -81052  &   -104 &         &  16000 \\
            &$5s^2 4f^2~\textrm{J=5}$  &   3059    &   3113    &   2.0  &         & 1.4  \\
            &$5s^2 5p4f~\textrm{J=2}$  &   5060    &   -60350  &   -24  &2200&  25    \\
             \end{tabular}
\end{ruledtabular}
\end{table}

\subsection{HCI with three valence electrons}

%------------------------------------------------------------------
\begin{figure}[t]
%\begin{center}
%\hfill
\includegraphics[width=\columnwidth]{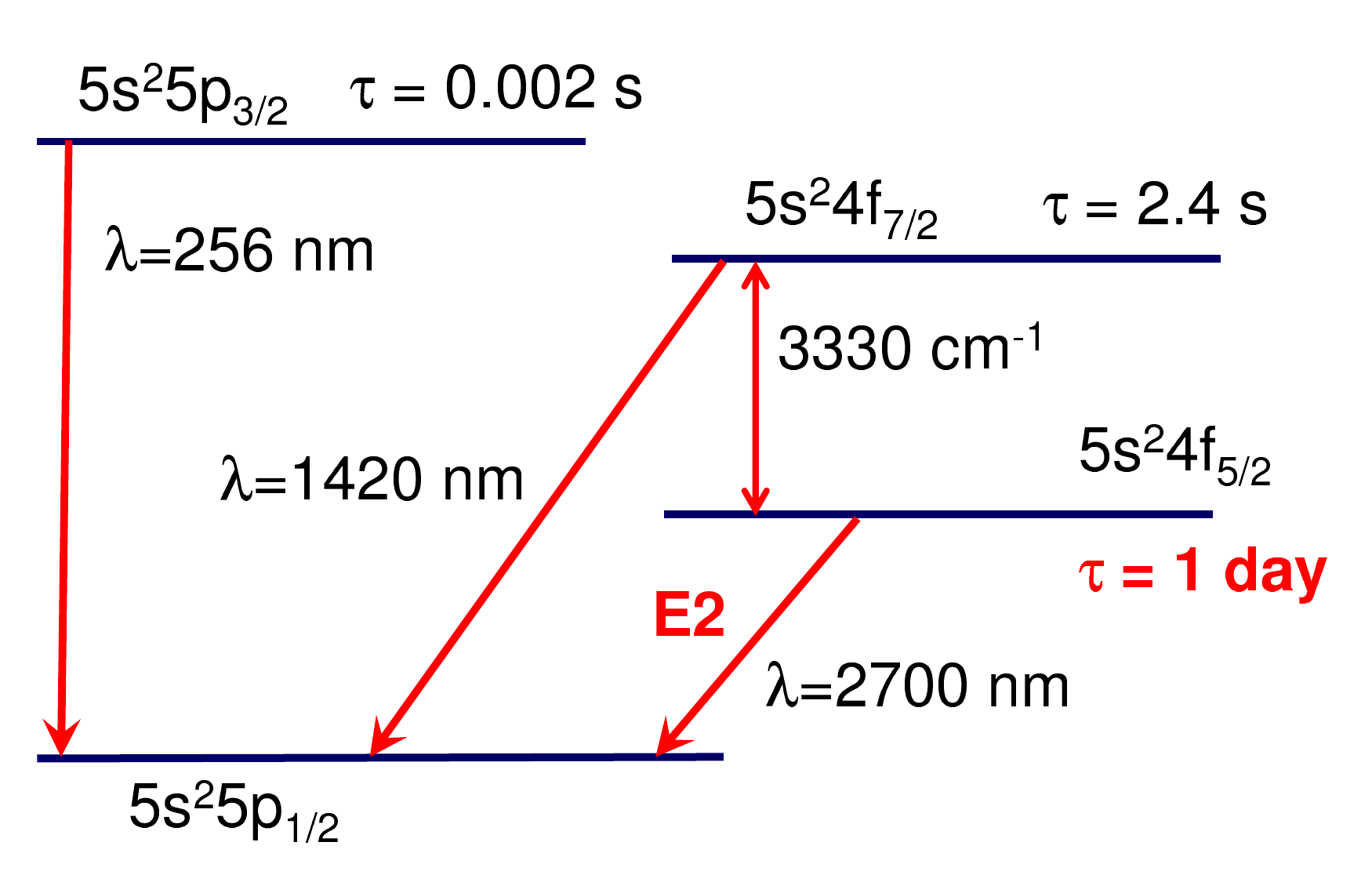}
%\hfill
%\hfill
%\end{center}
\caption{Energy levels and radiative lifetimes of low-lying levels
of In-like Pr$^{10+}$. From \citet{SDFS14}.}
\label{Fig_In-like_Pr}
\end{figure}
%------------------------------------------------------------------

\paragraph{In-like ions.}
There are two level crossings in the In-like isoelectronic sequence
($N=49$). There is a crossing of the $5p$ and $4f$ shells near
$Z=59$ as well as a crossing of the $5s$ and $4f$ shells near
$Z=63$. The ground configuration at the first crossing is $5s^2 nl$,
where $nl=5p$, or $nl=4f$. Near the second crossing the $5s^24f$ and
$4f^25s$ configurations have similar energies. Both crossings can be
adequately studied using a three-electron model with the closed core
$[1s^2\dots 4d^{10}]$.

At the first crossing the order of levels changes from $5p_{1/2}$,
$5p_{3/2}$, $4f_{5/2}$, and $4f_{7/2}$ for Ce$^{9+}$, to $5p_{1/2}$,
$4f_{5/2}$, $4f_{7/2}$, and $5p_{3/2}$ for Pr$^{10+}$, and, finally,
to $4f_{5/2}$, $4f_{7/2}$, $5p_{1/2}$, and $5p_{3/2}$ for
Nd$^{11+}$. The all-order results from \citet{SDFS14,SDFS14a} are
compiled in Table~\ref{tab2}. The theoretical spectrum of the
Pr$^{10+}$ ion is shown in \Fref{Fig_In-like_Pr}. The energies of
the $4f$ levels of Pr$^{10+}$ are very difficult to calculate
accurately as they are very close to the ground state $5p_{1/2}$.
The one-electron binding energies of the $5p_{1/2}$ and $4f_{5/2}$
states are $1.3\times10^6$~cm$^{-1}$, and these values cancel to
99.7\% when two energies are subtracted to obtain a theoretical
prediction for a transition energy, 3700(200)~cm$^{-1}$
\cite{SDFS14a}.

Theory predicts that a second crossing takes place between
Sm$^{13+}$ and Eu$^{14+}$. Calculated energy levels are listed in
Table~\ref{tab2}. For the Sm$^{13+}$ ion, the closest configuration
to the ground fine-structure multiplet ${}^2F_J[5s^24f]$ is
$5s4f^2$. This leads to a very interesting level structure with a
metastable $J=7/2\, [5s4f^2]$ level in the optical transition range
from both levels of the ground multiplet ${}^2F_{5/2,7/2}[5s^24f]$.
For Eu$^{14+}$ the ground state belongs to the configuration
$5s4f^2$ and the first excited level belongs to the configuration
$4f^3$. These levels are very close and their  theoretical
uncertainty is comparable to the energy interval. For example, QED
corrections for the configuration $4f^3$ exceed 1000 \cm\
\cite{TKSSD16} and corrections from the effective three-electron
interactions are of similar size \cite{KSPT16}. It is possible that
missing correlation corrections can change the ground state to
$J=9/2\, [4f^3]$. An experimental measurement of the spectrum of
Eu$^{14+}$ would allow testing of the theory for such difficult
cases.

\begin{figure}[t]
%\begin{center}
%\hfill
\includegraphics[width=3.1in]{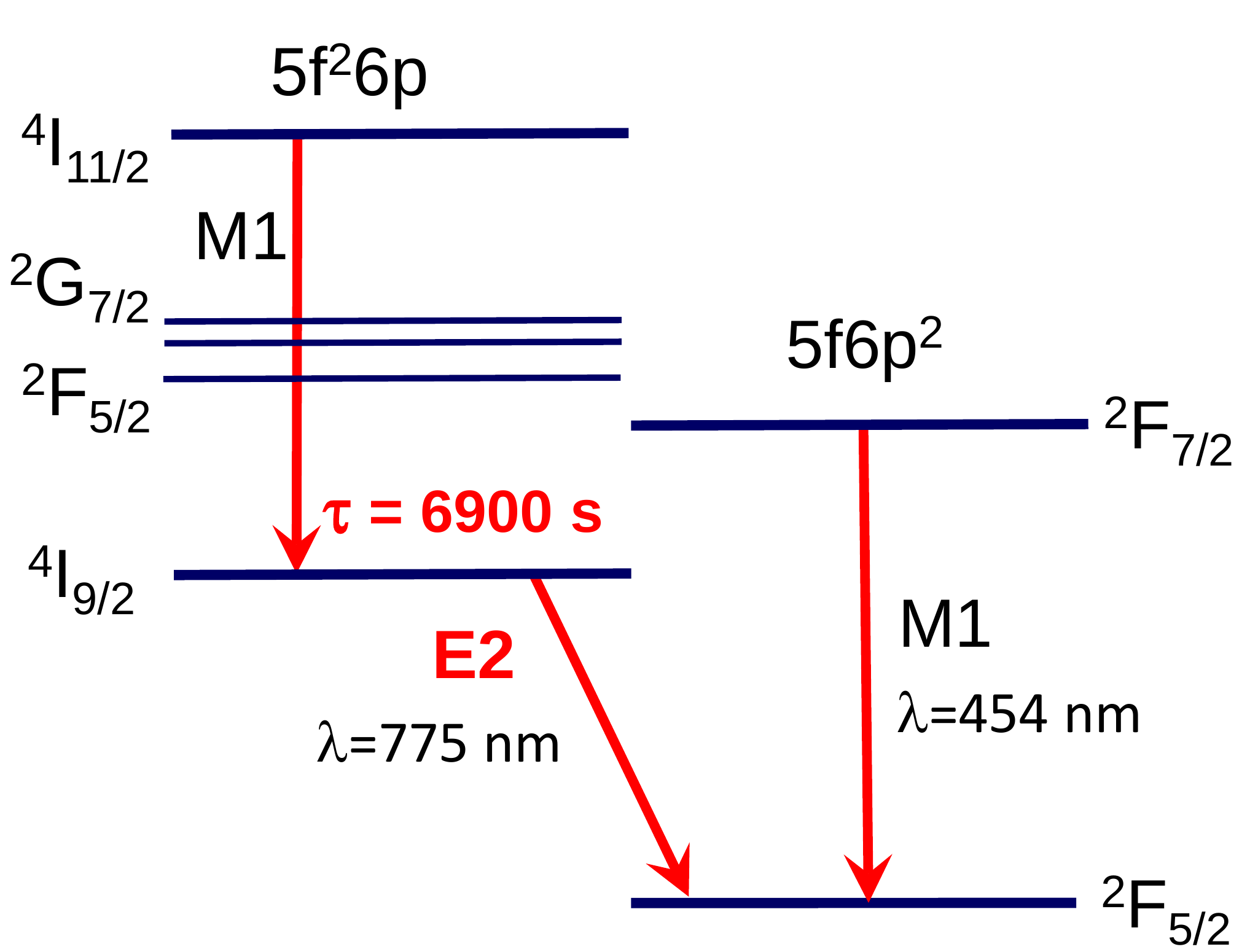}
%\hfill
%\hfill
%\end{center}
\caption{Low-lying energy levels of Cf$^{15+}$. Leading configurations are shown on the
top. E2 clock transitions are between the ground and first excited
states.  } \label{fig_Bi-like}
\end{figure}

\begin{figure}[t]
%\begin{center}
%\hfill
\includegraphics[width=3.1in]{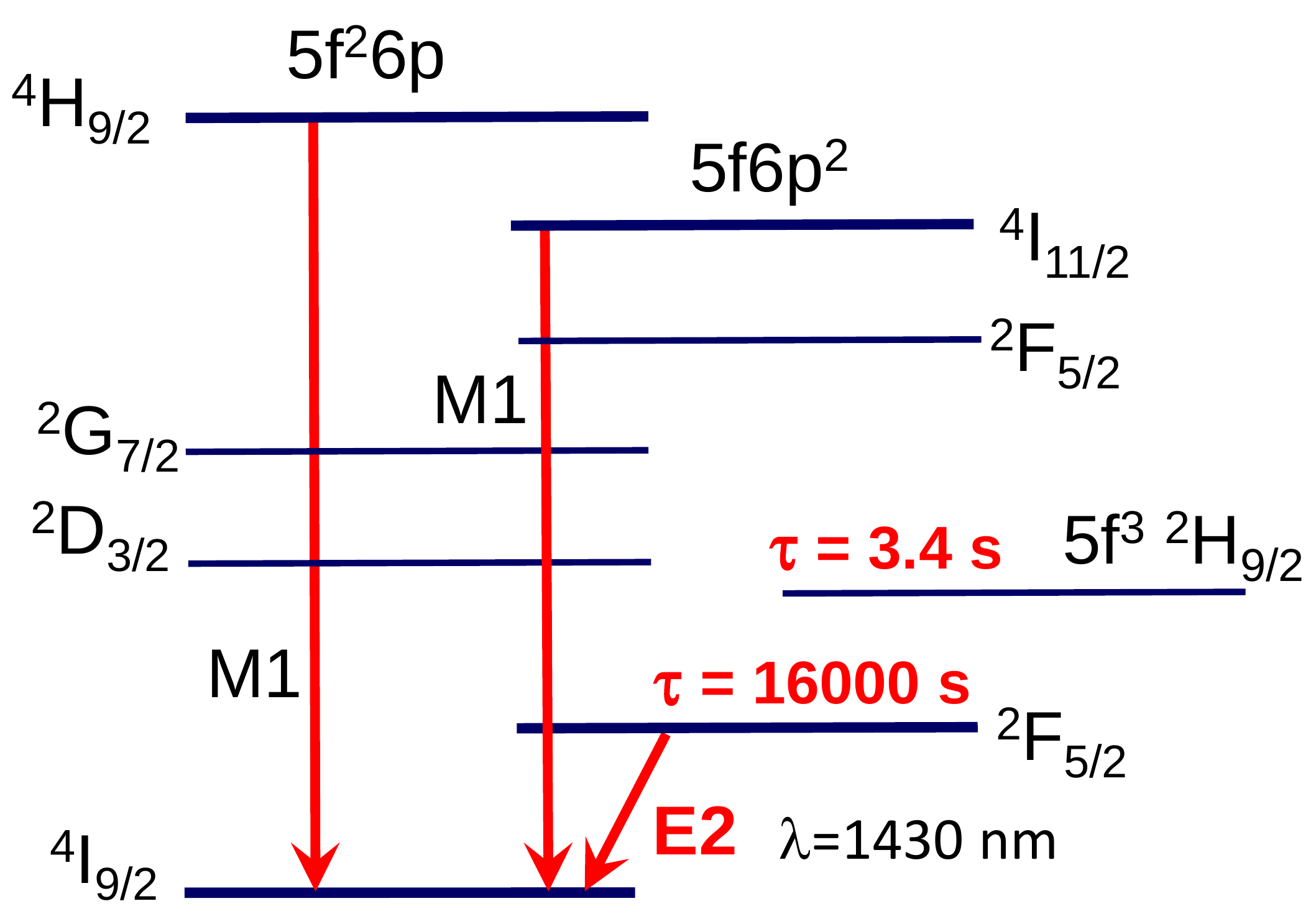}
%\hfill
%\hfill
%\end{center}
\caption{Low-lying energy levels of Es$^{16+}$. Leading configurations are shown on the
top.} \label{fig_Bi-like1}
\end{figure}

\paragraph{Bi-like californium and einsteinium.} Cf$^{15+}$ and
Es$^{16+}$ were studied using the CI+AO method in \cite{DSSF15}.
Theoretical spectra of these ions are shown in \fref{fig_Bi-like}
and \fref{fig_Bi-like1}. In both ions the first excited state is
metastable and linked to the ground state by an E2 transition. For
the Cf$^{15+}$ ion, this transition has a large sensitivity to
$\alpha$-variation, $q=+380000$ \cm. For the Es$^{16+}$ ion both
levels belong to the same $5f^26p$ configuration and the $q$-factor
is smaller, $q=-184000$ \cm. CI+AO results are listed in
Table~\ref{tab2}.

%------------------------------------------------------------------

\subsection{Sn-like ions with four valence electrons}

\begin{figure}[tb]
%\begin{center}
%\hfill
\includegraphics[width=\columnwidth]{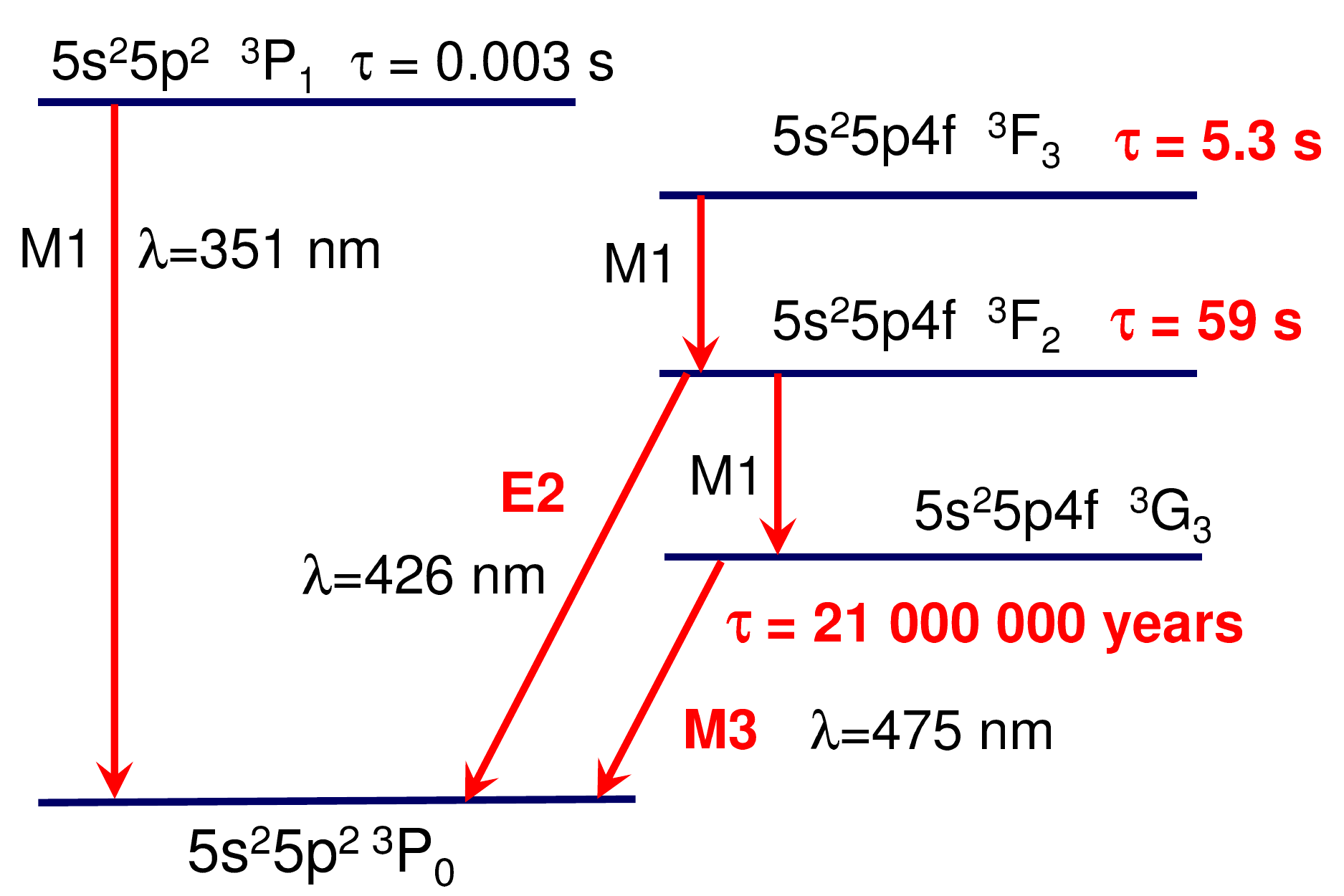}
%\hfill
%\hfill
%\end{center}
\caption{Energy levels and radiative lifetimes of low-lying levels
of Sn-like Pr$^{9+}$. From \citet{SDFS14a}.}
\label{Fig_Sn-like_Pr}
\end{figure}

Pr$^{9+}$ and Nd$^{10+}$ are the ions of interest in the Sn-like
isoelectronic sequence. Pr$^{9+}$ is particularly interesting,
because the lowest metastable state $J=3\,[5p4f]$ is extremely long
lived, with a 495~nm transition to the ground state in the optical
range, see \Fref{Fig_Sn-like_Pr}. The strongest allowed transition
for even isotopes is M3, making this ion a unique system. We expect
that the lifetime of that level will be strongly quenched due to the
E1 hyperfine transition in odd isotopes. The next two levels also
have optical transitions to the ground state, and have lifetimes of
59 and 5.3\,s respectively. In addition, there is a strong M1
transition to the ground state from the ${}^3P_1\,[5p^2]$ level at
351 nm that may be useful for cooling and probing. The first excited
state of Nd$^{10+}$ is so close to the ground state that the
theoretical uncertainty is on the order of the transition energy.
The atomic parameters are listed in Table~\ref{tab2}. Further
details are given by \citet{SDFS14,SDFS14b}.

\subsection{Ions with holes in almost filled $4f$ shell}

\citet{BerDzuFla11b} considered Ir$^{16+}$ and W$^{7+}$ ions with
one hole in the $4f$ shell. The energies of the $4f^{13}5s^2$ and
$4f^{14} 5s$ configurations in Ir$^{16+}$ were predicted to be
sufficiently close for an optical transition. According to the CI
calculation of these authors the ion Ir$^{16+}$ has ground multiplet
$^2F_{7/2,5/2}[4f^{13}5s^2]$ with huge fine-structure splitting
$\Delta_\mathrm{FS}\approx 25900$ \cm\ and excited state
$^2S_{1/2}[4f^{14}5s]$ roughly at 37000 \cm\ (see Table~\ref{tab3}).
Later these levels were recalculated within MBPT method by
\citet{SaFlSa15} and by \citet{NaSa16} using CC approach. Both
calculations gave the same ground doublet and the level
$^2S_{1/2}[4f^{14}5s]$ approximately at 28 and 38 thousand \cm\
respectively. All three calculations gave close values of the fine
structure splitting between 25 and 26 thousand inverse centimeters.

%The ion Ir$^{16+}$ was considered by \citet{BerDzuFla11b} as a
%system with two quasi degenerate ground configurations $4f^{14}5s$
%and $4f^{13}5s^2$. Because of that it should have narrow optical
%transitions, which are highly sensitive to $\alpha$-variation.

%At present, only CI results are available, which predict a ground state multiplet
%$^2F_{7/2,5/2}[4f^{13}5s^2]$ with a huge fine-structure splitting of about 25900
%\cm\ and excited  $^2S_{1/2}[4f^{14}5s]$ state at 37500 \cm. The CI results of
%\citet{BerDzuFla11b} are listed in Table~\ref{tab3}.

%------------------------------------------------------------------

A similar crossing between the one-hole configurations $4f^{13}5p^6$ and
$4f^{14}5p^5$ takes place for W$^{7+}$
\cite{Draganic2003,BerDzuFla11b}. Theory predicts the following order of
levels: $^2F_{7/2} [4f^{13}5p^6]$, $^2P_{3/2} [4f^{14}5p^5]$,
$^2F_{5/2} [4f^{13}5p^6]$, and $^2P_{1/2} [4f^{14}5p^5]$. The fine
splittings for two multiplets are about 18000 and 90000 \cm\
respectively. At present only CI calculations are available and
the accuracy of the theory is not sufficient to reliably predict the
distance between the multiplets.

In addition to one-hole ions \citet{BerDzuFla11b} also considered
the two-hole systems Ir$^{17+}$ and W$^{8+}$. The Ir$^{17+}$ ion has
low lying levels of the $4f^{12}5s^2$, $4f^{13}5s$, and $4f^{14}$
configurations. Spectra of these ions are much denser and, according
to the calculation, include many optical lines. However, present
theory may be rather unreliable for predicting the energy difference
between different configurations. One needs to include more
correlations to reduce theoretical uncertainty. On the other hand,
all such ions are particularly interesting because of the very large
$q$-factors, which determine sensitivity to $\alpha$-variation. For
Ir$^{17+}$ ions, spectra were recently studied by
\citet{WinCreBel15}, along with the isoelectronic, Nd-like sequence
ions of W, Re, Os, and Pt.

Other systems with two-hole $4f^{12}$ configuration were discussed
by \citet{DerDzuFla12} and \citet{DzuDerFla12a}. These authors point
out that there are always optical transitions between levels of this
configuration independently on the degree of ionization, and the
first excited state is always metastable. For the HCI the two holes
form the relativistic configuration $4f_{7/2}^2$. Allowed total
angular momenta for this configuration are: $J = 6,\, 4,\, 2,\, 0$.
According to Hund's rules, the $J = 6$ state is the ground state and
$J = 4$ state is the first excited state. These states are connected
by an electric quadrupole transition. The configuration $5s^2
4f^{12}$ is the ground state configuration for Ce-like ions (Z=58)
beginning from Re$^{17+}$. There are many ions of this type starting
from Re$^{17+}$ to U$^{34+}$. An analysis of the systematic shifts
of transitions within this configuration by \citet{DerDzuFla12}
shows that the largest ones are caused by magnetic fields and the
zero-point-energy motion of the trapped ion. The former effect can
be suppressed by averaging two transitions with different signs of
the projections $M_J$. An estimate of the latter effect is based on
the observation that trapping parameters are similar to those in the
Al$^+$/Be$^+$ clock, but that HCI are about 10 times heavier than
Al. This mass difference leads to a suppression of the time-dilation
effects for heavy HCI, assuming a similarly efficient sympathetic
cooling (see \sref{sec:sympcool}). \citet{DerDzuFla12} concluded
that all investigated systematic effects may be suppressed to a
fractional level of $10^{-19}$ by applying efficient
shift-suppression schemes. We note that the transitions between the
fine-structure states are  generally not sensitive to $\alpha$,
since all multiplet states have similar $q$ coefficients.

The more complex Ho$^{14+}$ ion was proposed for metrology
applications by \citet{DzuFlaKat15}. This ion has as its  ground
state configuration $4f^6 5s$, a first excited configuration $4f^5
5s^2$, and a very rich optical spectrum, which includes a potential
hyperfine-induced clock transition at approximately 400 nm and a
strong cooling and detection transition at 260 nm. The paper
includes estimates of the transition rates and lifetimes. The values
are listed in Table~\ref{tab3}. However, the clock state prediction
has a very large uncertainty, probably as high as 10000~cm$^{-1}$.
Sympathetic cooling of  Ho$^{14+}$ is discussed by \citet{OIW15}.
Their simulations show that at least 10 such ions can be cooled to
sub-milli-Kelvin temperatures by sympathetic cooling with a single
laser-cooled Be$^+$ ion.

%------------------------------------------------------------------
%------------------------------------------------------------------

\begin{table} \caption{\label{tab3} Energies and $\alpha$-variation sensitivity coefficients $q$ s relative to the ground state in cm$^{-1}$; $K=2q/\omega$ is the enhancement factor. Wavelengths $\lambda$ (in nm) for transitions to the ground states is listed. All values are  from the CI calculations and are expected to have large uncertainties, see text. Ir$^{16+}$ and Ir$^{17+}$  values are from \citet{BerDzuFla11b} and Ho$^{14+}$ values are from \citet{DzuFlaKat15}.}
\begin{ruledtabular}
\begin{tabular}{lllrrrc}
\multicolumn{1}{l}{Ion}&
\multicolumn{2}{l}{Level}&
\multicolumn{1}{c}{Energy}&
\multicolumn{1}{c}{$q$}&
\multicolumn{1}{c}{$K$}&
 \multicolumn{1}{c}{$\lambda$}\\
\hline   \\[-0.4pc]
Ir$^{16+}$ & $4f^{13} 5s^2$&$ ^2F_{7/2}$ & 0         &         &      &       \\
           & $4f^{13} 5s^2$&$ ^2F_{5/2}$ &  25898    &   23652 & 1.8  &   386        \\
           & $4f^{14} 5s  $&$^2S_{1/2}$ &  37460    &  367315 & 20   &    267       \\ [0.4pc]
Ir$^{17+}$ & $4f^{13} 5s $&$^3F_4$        & 0         &         &      &              \\
           & $4f^{13} 5s $&$^3F_3$        & 4838      &  2065   & 0.9  &    2067      \\
           & $4f^{13} 5s $&$^3F_2$        & 26272     & 24183   & 1.8  &    381       \\
           & $4f^{14} $&$ ^1S_0$         &  5055     & 367161  & 145  &    1978      \\
           & $4f^{12} 5s^2 $&$ ^3H_6$    & 35285     &-385367  & -22  &   283        \\
           & $4f^{12} 5s^2 $&$ ^3F_4$    & 45214     &-387086  & -17  &   221        \\[0.4pc]
Ho$^{14+}$ & $4f^6 5s $&$^8F_{1/2}$       &           &         &      &               \\
           & $4f^5 5s^2 $&$ ^6H_{5/2}$   &  23823    &-186000  & -16  &   420       \\
             \end{tabular}
\end{ruledtabular}
\end{table}
%MS - not sure where these will go, so commented out
%\section{Loose references, which may be cited somewhere}
 %\begin{itemize}
 %\item Avoided crossing of levels $(2p^{-1}_{1/2}3d_{3/2})_{J=1}$ and
 %$(2s^{-1}_{1/2}3p_{1/2})_{J=1}$ near $Z=68$ is studied in \cite{BSBC16}.
 %\item \citet{KaIv17} discuss the limits on the variation of
 %fundamental constants (no direct link to HCI).
 %\end{itemize}

%%%%%%%%%%%%%%%%%%%%%%%%%%%%%%%%%%
%\input{HCI_v11}
%%%%%%%%%%%%%%%%%%%%%%%%%%%%%%%%%%

%%%%%%%%%%%%%%%%%
\section{Experimental methods for HCI studies}
\label{sec:ExpMethHCI}
\subsection{Early spectral observations}
Experimental observations of forbidden optical transitions in HCI have a surprisingly long history.
Actually, the first detections of forbidden lines were reported in the year 1869 by
Harkness and Young, who saw during a total eclipse how the solar corona emitted green light at 530 nm
from an unknown and extremely light element, as it was believed.
In the 1940s Grotrian \cite{Grotrian1939} and Edl\'en \cite{Edlen1943,Edlen1947} hypothesized the presence of highly ionized atoms in the corona having fine structure transitions that would explain the observed lines.
This implied coronal temperatures in the range of 1\,MK, in contradiction with the then prevailing understanding of the Sun,
and gave a first insight into the very hot universe of modern astrophysics.
Diagnostics of hot astrophysical plasmas was therefore the first application of forbidden optical transitions in HCI.
The first laboratory experiments had to wait until the production of such ions  became feasible in a controlled way.

\subsection{First laboratory methods}
After decades of research with large devices such as theta pinches that used powerful pulsed electrical discharges to generate ions in high charge states, the sliding spark method \cite{Feldman1967} provided a source of comparatively small size and achieving ionization stages as high as Fe\,XVIII.
In the 1970's, the beam-foil method \cite{Bashkin1968,Berry1977Beamfoil},
was introduced. It starts with moderately charged ions that are accelerated to energies in the range 100\,keV-10\,MeV per unit of charge and sent through a submicrometer thin foil which strips further electrons and generates HCI in highly excited states \cite{Berry1977Beamfoil}.
Spectroscopic investigations were carried out by observing the trail of excited ions exiting the foil in a mostly perpendicular direction.
However, with HCI moving at speeds of a few mm/ns, allowed transitions with ps and ns range upper level lifetimes are predominantly detected, while the forbidden optical transitions with ms lifetimes do not produce sufficiently strong signals due to finite detector size.
A severe handicap for precision measurements was the geometry-dependent Doppler shift,
which in spite of various corrections could not be completely canceled.
Another problem was the simultaneous presence of several charge states
in a manifold of multiply excited configurations at the exiting channel.
This hindered line identification and the assignment of electronic levels.
Given those difficulties, the field achieved what was possible at that time,
and produced pioneering systematic spectroscopic data.

Magnetic-fusion research devices such as tokamaks provided for the first time a limited number of forbidden optical observations \cite{Suckewer1982,Edlen1984,Kaufman1983,Finkenthal1984}. Calculations were made for identification and plasma diagnostic purposes \cite{Feldman1985}.
Again, the Doppler width and the difficult control of the plasma composition and conditions hindered dedicated high-resolution  studies. More recently, the analysis of Zeeman splitting and polarization of visible transitions has also been reported \cite{Iwamae2007}. Spectroscopy in general, and forbidden transitions in particular, offer opportunities for temperature and density diagnostics in magnetic fusion plasmas \cite{Beiersdorfer2015Applications}.

\subsection{Early laboratory sources of HCI}

The development of stationary HCI sources solved most of those
problems and made HCI more easily accessible to experimentalists.
Among those, electron-cyclotron resonance ion sources (ECRIS)
\cite{Geller1970,Briand1975,Bechtold1980} based on the electron
heating of a thin plasma by powerful microwaves in a magnetic bottle
yielded microampere beams of HCI in moderate charge states,
typically used to load accelerators. Optical access to the plasma
volume was very constrained by the bulky and complicated magnetic
structures surrounding it. As for the charge states of HCI produced
in ECRIS, they have typically been limited to $q<28+$ by the plasma
conditions. Nonetheless, it was a flexible source and this was soon
recognized. One of the first examples of optical spectroscopy was
the observation by \citet{Prior1987} of fine-structure transitions
in ions with open $p$ and $d$ subshells by using beams from an ECRIS
at the Lawrence Berkeley Laboratory. The very intense currents of
HCI available there (tens of microamperes) made it possible to
observe the decay of a small fraction of the metastable ions passing
in front of a spectrometer equipped with a position-sensitive
microchannel plate detector. In this way, intra-configuration
fine-structure transitions from F-like ions (Ar$^{9+}$, K$^{10+}$,
and Ca$^{11+}$), O-like ions (K$^{11+}$ and Ca$^{12+}$), and B-like
ions (Ar$^{13+}$ and K$^{14+}$) were measured. The transitions
$3d^9\,{}^2D_{5/2\rightarrow 3/2}$  in Nb$^{14+}$,
$3d\,{}^3D_{4\rightarrow 3}$ as well as the
$3d^8\,{}^3F_{2\rightarrow 3}$ in Nb$^{15+}$, and the
$3d^7\,{}^4F_{9/2\rightarrow 7/2}$ in Nb$^{16+}$ were also
investigated. Spectral resolution and wavelength accuracy were both
instrumentally and methodologically limited to a level of 0.1\%.

Another step in the direction of higher charge states and better control was the introduction of electron beam ion sources (EBIS)
by Arianer \cite{Arianer1975,Arianer1976} and Donets \cite{Donets1967,Donets1975,Donets1985,Donets1990}.
An intense, narrow electron beam was generated through magnetic compression that compensated the mutual repulsion of the electrons in the beam.
This development owed much to radar technology, which had made use of such beams to generate microwaves.
Accumulation of multiply charged ions inside the electron beam space charge distribution was recognized as a factor for the beam neutralization.
Unfortunately, due to technical difficulties, spectroscopy on EBIS devices was not strongly pursued
although a few cases of in-source X-ray spectroscopy \cite{Ali1990,Ali1991} were reported.

\subsection{Production of HCI with electron beam ion traps}

It was recognized that ion heating by electron beam instabilities was hindering the production of the highest charge states in EBIS devices \cite{Levine1985}. Correcting this problem, the decisive invention of the electron beam ion trap (EBIT) by Marrs and Levine \cite{Levine1988,Levine1989,Penetrante1992,Marrs1994b} at the Lawrence Livermore National Laboratory (LLNL) prepared the field for many long ranging developments.
To mention some examples, we refer to investigations of QED effects in X-ray emission spectra \cite{Beiersdorfer1993Kalpha,Beiersdorfer1995LilikeTh,Beiersdorfer1998,Beiersdorfer2005TwoLoop},
studies of the dielectronic-recombination process \cite{Knapp1989,Beiersdorfer1992DR,Watanabe2007}
and of quantum interference in photo-recombination processes \cite{Knapp1995,Gonzalez2005,Nakamura2009},
nuclear-size determinations \cite{Elliott1996},
lifetime measurements of forbidden transitions \cite{Wargelin1993,Crespo1998,Crespo2006HelikeSulphur},
laser spectroscopy of the $2s-2p$ He-like transitions \cite{Hosaka2004},
charge exchange \cite{Otranto2006,Beiersdorfer2000CX,Beiersdorfer2003Comets,Wargelin2005,Allen2007},
plasma-polarization spectroscopy \cite{Beiersdorfer1997Polarization,Shlyaptseva1997,Shlyaptseva1998,Nakamura2001,Amaro2017,Shah2018},
effects of Breit interaction in x-ray emission \cite{Hu2012},
soft x-ray laser spectroscopy \cite{Epp2007,Bernitt2012},
photoionization of HCI with soft x-rays \cite{Simon2010},
and mass spectroscopy \cite{Ettenauer2011} as well as in-trap nuclear spectroscopy of radioactive isotopes \cite{Lennarz2014}.

These devices, for reasons we will see below, became true spectroscopy workhorses and produced an enormous scientific harvest.
Experimentally, production of HCI in more or less arbitrary charge states is now routinely performed with EBITs,
and therefore several groups worldwide started utilizing such devices
\cite{Silver1994,Gillaspy1995,Currell1996,Nakamura1997,Biedermann1997,Crespo1999,Ovsyannikov1999,Watanabe2004,Dilling2006,Boehm2007,Fu2010,Xiao2012,Takacs2015}.
In view of their apparent advantages and of the large body of experimental results,
in the following we will  focus on the uses of the versatile EBITs.

\subsubsection{Ionization and trapping mechanism}
The principle of operation is the interaction of an intense (mA to A), strongly focused electron beam with atoms and their ions
(for more details see, e.~g.~, \cite{Currell2005,Currell2000,BeyerKlugeShevelko1997,Gillaspy2001}).
Ionization of neutrals injected as dilute atomic or molecular beams crossing it yields first singly charged positive
ions which stay trapped by, and mostly within, the negative space-charge potential of the beam.
This parameter has its strongest gradient at the beam edge.
Between that point and the beam central axis at tens of micrometers from it a potential difference of tens of volts appears.
From the point of view of a neutral injected from a room-temperature atomic beam with a kinetic energy of 25\,meV,
ionization to the first charge state within the beam means the sudden appearance of a thousand times stronger trapping potential.
This results in instantaneous trapping of the ions produced. Subsequent beam-ion interactions raise the charge state until the physical limit is reached,
namely when the binding energy of the remaining bound electrons is higher than the electron beam energy.
electron-beam energies in the range 40\,eV to 200\,keV, ions from N$^{3+}$ \cite{Simon2010N3+} to U$^{92+}$ \cite{Marrs1994} have been studied with the EBIT method.
Recombination of the HCI is reduced by having an excellent vacuum, suppressing charge exchange with residual gas.
Efficient ionization takes place at energies well above the pertinent thresholds,
but the long trapping times (seconds to hours) allow compensation
for the small electron-impact ionization cross sections close to threshold.

\subsubsection{Photorecombination and charge-exchange processes}
Acting in the opposite direction, photorecombination of free beam electrons with the HCI under emission of a photon (so-called radiative recombination, RR)
is rather weak for ions in relatively low charge states,
but becomes intrinsically strong for ions with open $L$ and $K$ shells, in particular if the charge state is also very high.
This process, RR, is akin to time-reversed photoionization.
Another one is dielectronic photorecombination (DR) involving the resonant excitation of an inner-shell electron during the capture process and the subsequent relaxation of this intermediate state through photon emission.
It is extremely effective at certain discrete energies,
with several orders of magnitude larger cross sections.
While it affects typically only one charge state at each beam energy, its total contribution can be several times larger than that of RR with thermal electron distribution functions.
Charge exchange is the process by which HCI capture electrons from residual gas neutrals by overcoming their ionization potential in collisions at the range of several atomic units. The cross sections for this are rather large (on the order of $10^{-14}$\,cm$^2$) and therefore an excellent UHV is typically needed to store HCI.
The charge-state distribution found in an EBIT is, in general, ruled by a set of coupled rate equations \cite{Penetrante1991Evolution,Penetrante1992}
containing all those ionizing and recombining terms.
Under normal conditions it will be comprised of only a few charge states, and it can be optimized to contain a rather dominant ionic species \cite{Currell2000}.

\subsubsection{Electron-impact excitation of transitions}
An invaluable advantage of an EBIT is the fact that the electron beam also copiously excites the trapped HCI.
This, together with the convenient geometry giving radial optical access to the trap region at few-centimeter distances
allows for spectroscopic observations in all spectral ranges.
The space charge potential forms a narrow (50 to 500 $\mu$m), few-cm long cylindrical ion cloud
which can readily be imaged onto the entrance slit of a spectrograph, or even serve as its substitute in many cases.
Then, spectral dispersion combined with ion-cloud imaging maximizes the signal.
The ion cloud normally contains HCI that compensate a large fraction of the negative space charge of the electron beam.
Taking into account the charge state, this leads to ion densities from $10^9/$cm$^3$ to $10^{11}/$cm$^3$, surrounding or immersed in a negative space charge of $10^{13}/$cm$^3$ electrons.
Since, depending on temperature, the HCI may spend some time outside the electron beam. This reduces the effective electron density for the excitation rate, which may become one hundred times lower than the actual one inside the beam \cite{Liang2009}.

The total HCI count in charge states of interest can be from only a few (e.g. for the 'ultimate' U$^{92+}$) to hundreds of millions, with experiments covering the whole range.
Appropriately designed experiments have been running with count rates of only tens of counts per hour for both visible or x-ray photons, but under favorable conditions tens of kHz have been possible.
Due to the stability of normal EBIT operations, the more time-consuming measurements can be run for days and weeks with little intervention needed.

\subsubsection{Preparation of neutrals for ionization and trapping}
An additional source term  may have to be considered in the charge-state distribution, if a steady-state atomic beam is used.
If neutrals are not constantly injected but introduced in a pulsed mode, the charge state evolution is rather homogeneous.
This is the case if one chooses a starting population of singly charged ions injected along the magnetic field axis.
In the case of a very heavy HCI such as U$^{92+}$, the charge-state distribution reached at 190\,keV may encompass U$^{86+}$ to U$^{92+}$
with its maximum at $q=88$ \cite{Marrs1994}, a small fraction in the hydrogenlike state, and a very small one in the bare state.
In moderate charge states, recombination processes are weaker and the charge-state distribution becomes narrower than in the aforementioned case, and can easily be made to peak at the charge state of interest.
This is the case for most of the ions in the context of the present review.

Concerning the choice of chemical elements, there is no limitation in sight.
EBIT operation has been reported for basically all types of gaseous and solid stable elements, and radioactive isotopes
have also been studied with them.
A common and convenient approach is the use of gases or volatile compounds containing heavier elements.
Very low vapor pressure, e.~g., organometallic substances at the rate of
microgrammes per day have been extensively used to produce molecular beams for injection into the EBIT (e.~g.~ in Ref.\cite{Watanabe2001}.
External ion injection from laser-ion sources \cite{Niles2006,Trinczek2006}, oven-based Knudsen cells \cite{Yamada2007} and vacuum-discharge sources \cite{Brown1986} has also been reported.

Quantities of atoms needed to feed an EBIT are truly microscopic:
operation with nanogramme (and even picogramme) probes electrochemically coated on needles positioned
close to the electron beam was reported for HCI such as $^{233,235}$U$^{88+}$ and $^{248}$Cf$^{96+}$ \cite{Elliott1995,Elliott1996,Beiersdorfer1997}.

More recently, direct injection of radioactive isotopes produced on-line by accelerators, and with lifetimes in the millisecond range, has also been reported.
The experimental cycle lasting only a fraction of a second involved the extraction of the isotope from the target,
generation of a beam of singly charged ions,
transfer into an EBIT, 'charge-breeding' there to HCI state,
extraction from the EBIT, and transfer to a Penning trap \cite{Dilling2006}.
There, precision atomic mass measurements were carried out taking advantage of the increased precision
afforded by the linear scaling of the cyclotron frequency with the HCI charge.
Other experiments involved fundamental studies of $\beta$ decay
and bound-internal conversion processes \cite{Ettenauer2011,Lennarz2014} in charge states relevant for nucleosynthesis in stellar and supernova environments.
In general, these methods have considerably expanded the range of isotopes which can be made available for future optical clocks and precision atomic physics experiments.

%%%%%%%%%%%%%%%%%
\subsubsection{Techniques for HCI delivery}
Since HCI production relies on energetic interactions between neutrals and electrons inside a trap or source, methods for their transfer and re-trapping in Penning or radio-frequency Paul traps, have to be developed.
As with other sources of ions, HCI have to be delivered by means of vacuum beamlines using electrostatic and magnetic guiding fields generated by appropriate electrodes and magnets.
The key limitation is the very small production rate, resulting from the combined effect of minute ionization cross sections and the need to stepwise pass through many successive charge states.
Typical HCI currents from EBITs are therefore in the range from pA to fA for individual charge states in steady-extraction mode; bunches of $10^2$ to $10^7$ HCI are standard in pulsed mode \cite{Gillaspy2001,currell2003,Blessenohl2018,Micke2018}.

%Examples for HCI delivery to external experiments are:
Electron yields caused by the impact of slow HCI on clean surfaces (of up to 280 electrons per ion) were studied using EBIT-extracted $^{136}$Xe$^{21+...51+}$
and $^{232}$Th$^{51+...80+}$ ions \cite{Aumayr1993,McDonald1991}.
For studies of the electronic charge-exchange process in collisions between HCI and neutrals, experiments have used HCI beams extracted from EBITs, and ranging from Ar$^{16+...18+}$ to U$^{88+}$ \cite{Otranto2006,Schneider1994,Xue2014}.
Charge exchange (CX) with residual gas leads to HCI losses, and thus ultra-high vacuum levels ($< 10^{-9}$\,mbar) are required within the apparatus and beamline.

Due to the fact that translational temperature and spatial distribution of the ions at the source (usually parametrized using the so-called emittance in units of mm$\cdot$mrad) can severely reduce the efficiency of the beamline transmission, ion transport is usually performed by applying accelerating potentials on the order of tens of kilovolts,
and ion optics elements such as einzel lenses as well as electrostatic or magnetic quadrupoles are introduced at several places
in order to re-focus the divergent ion beam and keeping it from hitting the guiding elements and vacuum chamber.

Buffer-gas pre-cooling of ions, sometimes applied to improve the emittance of ion sources, is prohibited with HCI due to CX.
However, in spite of the high temperature of the HCI at birth (on the order of MK), EBIS and EBIT show rather good emittance parameters (a few mm$\cdot$mrad), since the ions are produced in a small volume with a well defined magnetic field.
Since the available number of HCI is small, various techniques have been proposed in order to optimize the yield of HCI at the delivery point, e.g., mixing HCI with cold electrons \cite{Poth1991,Beier2005,Kluge2008,Quint2001} or evaporative cooling from a bunch of HCI oscillating inside a Penning trap \cite{Hobein2011}.
One  method successfully applied to HCI is pulsed extraction from an EBIT followed by phase-space cooling of the formed ion bunch by application of a time and position dependent sudden electric pulse \cite{Schmoeger2015RSI} to the moving HCI bunch.
After the pulse, more kinetic energy is removed from the faster ions than the slower ones, and an additional time-focusing effect can be conveniently achieved.
Then, the shorter and more velocity-homogeneous bunches have to overcome a potential barrier before entering a RF quadrupole electrode structure that guides them over a length of several cm having an electrostatic mirror at its end.
Given that the HCI bunch now moves slowly, the entrance electrodes of the RF quadrupole can be quickly switched to a higher positive bias potential.
HCI reflected from the mirror electrode cannot leave the RF quadrupole and perform a linear oscillatory motion back and forth along its axis.
These steps can now be complemented with a dissipative process that removes kinetic energy from the oscillating HCI.
For this purpose, a continuously laser-cooled ensemble of other ions is prepared within the RF quadrupole,
thereby providing stopping power at the beginning, and sympathetic cooling at the end of the procedure as discussed in more detail in \sref{sec:sympcooljose}.

\subsection{Optical spectroscopy of HCI}
\label{Sec:Lab_OS_HCI}
\subsubsection {Measurements of the HFS of hydrogenic ions}

As already mentioned, an important boost to the field of optical
spectroscopy with HCI occurred as both ion storage rings
\cite{Klaft1994,Seelig1998} and EBITs
\cite{Crespo1996,Crespo1998HFS,Beiersdorfer2001} reported HFS
measurements in hydrogenlike ions of heavy elements. It was clear
that scaling laws would shift the 21\,cm microwave transition of
atomic hydrogen into the optical region, and reduce its enormous 11
million years lifetime to milliseconds. At the same time,
relativistic, QED \cite{Persson1996}, and nuclear size effects were
boosted to the level of 50\%, 10\% and 0.5\% of the total ground
state hyperfine transition energy \cite{Shabaev1993,Shabaev1994},
respectively. As detailed in Sec.~\ref{Sec:QED}, tests of QED in
strong fields were the main interest of the community, and
theoretical work aimed at disentangling contributions from two-loop
QED, nuclear recoil effects and nuclear magnetization distribution
in the measured transitions. Lacking sufficient accuracy, models of
nuclear magnetization were deemed far more uncertain than the
dominant first order QED contributions, thus leading to the
application of the HFS data to determining nuclear magnetic radii
\cite{Crespo1998HFS,Beiersdorfer2001}.
\begin{figure}[t]
\includegraphics[width=0.95\columnwidth]{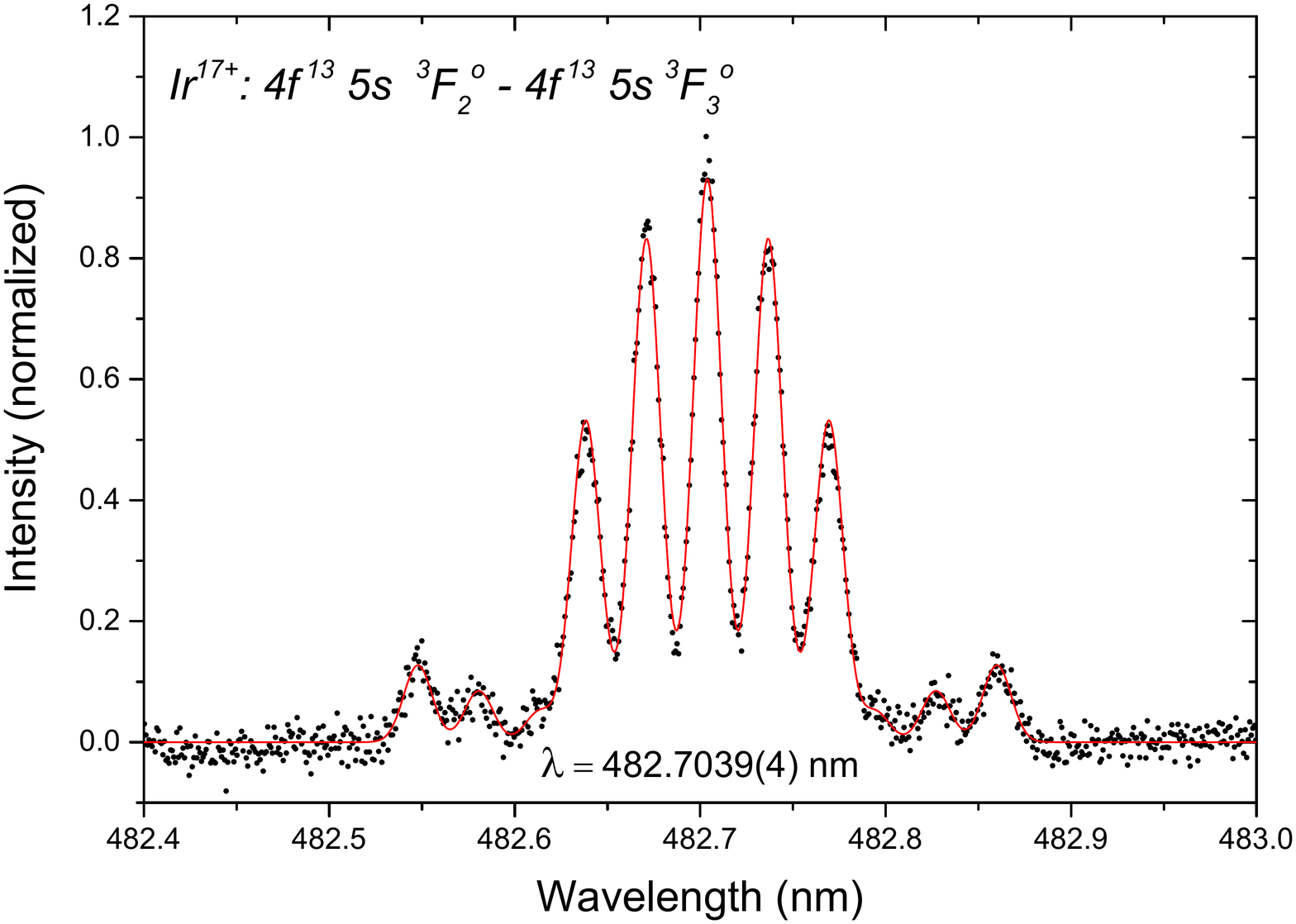}
\caption{Line profile of an optical transition within the $4f^{13}5s$ configuration of Ir$^{17+}$ ion obtained at the Heidelberg EBIT \cite{bekker_private_2017}. The Zeeman splitting due to the 8-T magnetic field of the trap is clearly visible.} \label{IridiumSpectrum}
\end{figure}

\subsubsection{Other optical spectroscopic observations}
A parallel development was the study of specific transitions in
Ti-like ions (starting with Ba$^{34+}$ using optical spectrometers
in the work of the NIST EBIT group \cite{Morgan1995}) that showed an
interesting behavior along that isoelectronic sequence. This was
later also studied by other groups
\cite{Porto2000,Utter2000,Watanabe2001}, namely the very weak
dependence of their wavelengths on the atomic number $Z$ and thus
the charge state of the ion. The reason for this is a competition of
relativistic and correlation effects that results in a near
cancelation of the otherwise very strong scaling of forbidden
transitions to high photon energies with $Z$. Exploratory
experiments showed the existence of many forbidden optical
transitions which could be observed in emission with an EBIT
\cite{Serpa1997b,Bieber1997}. With increased wavelength resolution,
optical measurements in EBITs at the Freiburg (later moved to MPIK
in Heidelberg) EBIT became sensitive to QED contributions,
relativistic nuclear recoil effects and the Zeeman structure of
forbidden lines \cite{Draganic2003,Soria2006,Soria2007}. In order to
explore these possibilities, more and more sophisticated
calculations were produced \cite{Tupitsyn2003,artemyev2007}.
Lifetimes of the metastable levels from which the optical
transitions arise were measured
\cite{Serpa1997a,Traebert2002,Traebert2008}, reaching accuracies at
the 1\% level \cite{Brenner2007,Brenner2009} and beyond, whereas QED
contributions from the electron anomalous magnetic moment could be
resolved \cite{Lapierre2005,Lapierre2006}.

\subsection{Electronic structure determination in HCI}
Since most of the ions of interest have never been investigated,
thorough theoretical and experimental studies of their electronic structure will be required
for clarifying the actual electronic structure and identifying the transitions of interest.
Ritz-Rydberg analysis of optical transitions has been used for the ground state configuration of HCI in a few cases \cite{Windberger2016,Torretti2017} and has helped clarifying the physics of EUV radiation sources for nanolithography, which are based on laser-produced plasmas containing tin HCI \cite{Harilal2006,OSullivan2015}.

Typically, optical measurements in an EBIT are carried out with the help of
grating spectrometers, since the high temperature of trapped HCI
does not support a spectroscopic resolving power $E/\Delta E\leq
30000$. One example of such a measurement is shown in \Fref{IridiumSpectrum}.
A few measurements have been carried out by means of
laser spectroscopy both in storage rings and, as proposed initially by \citet{Back1998}, in EBITs
by \citet{Maeckel2011,Schnorr2013}. Both types of experiments suffer from the wide velocity
distribution of the storage ring (at the level of $E/\Delta E\leq 10000$)
and the aforementioned translational temperature inside EBITs ($E/\Delta E\leq 30000$).
There have also been experiments using free-electron lasers and synchrotrons in combination with EBITs. These studies have progressively expanded the field of laser spectroscopy into the EUV and soft x-ray \cite{Epp2007,Bernitt2012} and x-ray regions \cite{Rudolph2013,RudBerEpp13,Epp2015}, however with the aforementioned limitations in spectral resolution.

%------------------------------------------------------------------
\begin{figure}[tbh]
%\begin{center}
%\hfill
\includegraphics[width=0.8\columnwidth]{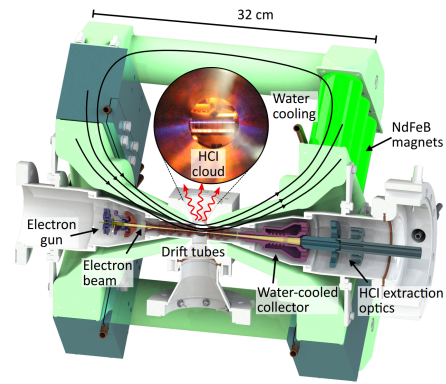}
%\hfill
%\hfill
%\end{center}
\caption{Design of the compact EBIT operating at PTB for the dedicated production of HCI in an optical clock experiment. The insert shows light emission induced by electron impact excitation of the trapped ions (from \textcite{Micke2018}.} \label{PTB-EBIT-Design-3}
\end{figure}

\subsection{Compact EBITs for novel spectroscopic applications}
\label{sec:CompactEBITs}

Laser spectroscopy, metrology and quantum computational studies with
trapped ions are most frequently based on the in-trap production of the
ions using electron impact ionization or photoionization of atoms.
Less often, ions are brought into the system from an external source
by means of an ion transfer beamline. The first approach offers a
more compact setup but can cause certain problems of electrode
contamination which may constitute a hindrance for long-term
stability in metrological work. The second option, generation of
ions in a separate setup, is more complex but has advantages: more
versatility in terms of the types of ions available and lesser
contamination of trap electrodes and optics by the atomic
sources. In principle, HCI would be more amenable to this latter
approach, although one could conceive both electron-beam ionization
(as in, e.g., \citet{Schabinger2012}) and pulsed-laser based
production schemes for moderately charged ions working within the
trap chamber. The most convenient approach, however, is to separate production and
spectroscopy while maintaining an overall compact envelope for the
whole apparatus.

Easy availability of HCI is a prerequisite for these studies.
Production of HCI in accelerator facilities will still remain
restricted to a few groups worldwide. A widespread application of
HCI to atomic physics research and metrology calls for small and
compact sources with minimal setup and maintenance costs, and simple
and reliable operation. There exist already a number of publications
on small HCI sources which can be set up with moderate resources and
effort within a student project, or purchased from a commercial
provider.

%\subsubsection{Current developments on miniaturization}
Most of the spectroscopic and HCI-interaction experiments mentioned above were based on rather powerful cryogenic EBITs using superconducting magnets. Although far smaller than accelerator-based HCI research devices, they still need a dedicated laboratory room and a scientific team for operation.
During the last decades several groups have been working on the development of more compact, and economic, systems.

The first EBIS based on permanent magnets \cite{Khodja1997} were followed by other ``warm-type'' instruments
not requiring superconducting magnets at cryogenic temperatures.
The Tokyo group built an EBIT based on permanent magnets \cite{Motohashi2000} for spectroscopic studies.
Others followed \cite{Kentsch2002,Nakamura2004,Nakamura2008,Sakaue2009,Xiao2012,Takacs2015}.
They can generate and trap ions having ionization potentials of a few keV, that will allow to reach
charge states up to 60+ for heavy elements.
Currently, such devices have typically a footprint of one square meter and
operate unattended for extended periods of time both as sources of
ions or for spectroscopic observation.

At NIST, HCI have been extracted from the EBIT and injected into compact Penning traps \cite{Brewer2013}, where Rydberg states of the HCI shall be prepared for laser spectroscopy. These experiments start with mass selected HCI pulses from the EBIT, extracted at energies of up to $4 \times 10^3 Q\,\times$\,eV, with $Q$ being the charge of the HCI.
The Penning trap captured a significant fraction (up to several thousand) of those ions in a potential well of (4–12)Q\,$\times$\,eV by opening and closing the axial trapping potential.
This fraction has a much narrower kinetic energy distribution (FWHM $ \approx $5.5\,eV) than the HCI inside the EBIT and in the extracted bunch.
Other experiments with extracted ions have been performed there to measure the lifetime of the Kr\,XVIII $3d$ ${}^{2}{D}_{5/2}$ metastable state \cite{Guise2014}. A compact EBIT under construction will be used as a HCI source \cite{Fogwell2014} in combination with the compact Penning trap.

Other compact devices based on the EBIT principle have recently appeared.
A strong point-like magnetic focus generates HCI in charge states as high as Ir$^{55+}$ in a compact and simple device
developed by \citet{Ovsyannikov2016a,Ovsyannikov2016b} with an overall dimension of less than 0.2\,m$\times$0.2\,m$\times$0.2\,m length,
for which HCI extraction in the radial direction was also demonstrated.

For operation of electron beams at lower energies of less than 1\,keV suitable for the production of HCI predominantly in charge states with many optical transitions, and moreover offering good ion beam properties, special care has to be taken in the field design. Recently, novel EBITs using permanent magnets, schematically shown in \Fref{PTB-EBIT-Design-3}, have reached magnetic field strengths of 0.86\,T and demonstrated excellent electron beam energy resolution for X-ray studies, ion extraction and optical spectroscopy \cite{Micke2018}. One of these devices has been installed at the German metrology institute Physikalisch-Technische Bundesanstalt (PTB) to operate an optical clock based on HCI. The magnetic cage of the EBIT has dimensions of 0.3\,m$\times$0.3\,m$\times$0.3\,m. It surrounds a vacuum system that is attached to a transport and deceleration beamline, and has an overall area of approximately 1\,m$^2$. It operates at UHV conditions suitable for its connection with a cryogenic RF-trap, CryPTEx-PTB \cite{Leopold2017} based on the CryPTEx design \cite{Schwarz2012} in which very long HCI storage times will be required for optical clock applications.

One can expect that the rapid development in this field will probably bring even smaller EBIS and EBITs to the market
and substantially reduce the barrier for the use of HCI in the laser spectroscopy community.

\begin{figure*}[bht]
\begin{center}
%\hfill
\includegraphics[width=\textwidth]{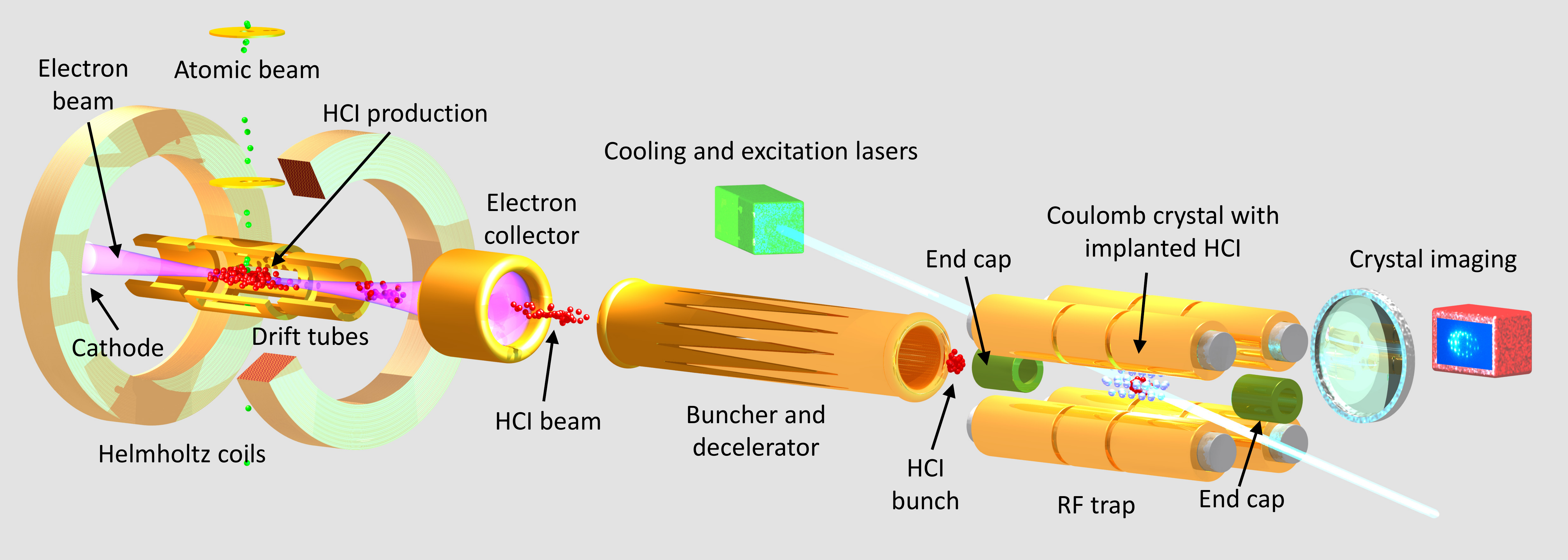}
%\hfill
%\hfill
%
\caption{Schematic representation of an experiment using sympathetic cooling of HCI for optical clock applications \cite{SchVerSch15,Schmoeger2015RSI}. HCI are produced in an electron beam ion trap (left), extracted from it, decelerated and bunched, and implanted into a Coulomb crystal of singly carged ions at mK temperatures. The implanted HCI repel their fluorescing, laser-cooled Be$^+$ neighbors strongly. This provides information about the location and charge of the implanted ions.
} \label{CryPTEx}
\end{center}
\end{figure*}

\section{Preparation of cold highly charged ions}
\label{sec:prep_cold_HCI}
%------------------------------------------------------------------

\subsection {Evaporative cooling of HCI}

A key element for the success of spectroscopy experiments in EBITs was the very early introduction of evaporative cooling to their operation.
The first proposal, modeling and realization of this technique by Penetrante at LLNL in 1990 \cite{Penetrante1991Cooling,Penetrante1991Evolution,Schneider1991,Marrs1994,Marrs1999} preceded its application to Bose-Einstein condensates.
There was an understanding that electron-ion collisions in the deep potential well caused by the negative space charge would lead to heating of the HCI, and to their eventual loss. Therefore, a cooling mechanism would be needed to keep the ions inside the trap for a long time.
This was found in taking advantage of the fact that HCI from lighter elements could not reach the high charge states of co-trapped heavier elements.
Ion-ion collision rates are extremely enhanced between HCI as the product of the squares of the individual colliding charges.
Efficient thermalization accross ion species thus redistributes energy from the heavier ions, subject to strongest electron-impact heating, and the lighter ones.

The relevant trapping parameter is the product of the charge state $Q$ (on the order of $Q\sim 20$) times the potential difference between the trap center and its edge ($\Delta V \approx  50 V$). This gives a trap depth of $\Delta V \times Q \approx$1000\,eV, corresponding to a 'temperature' of 12\,MK.
Therefore, light HCI with lower absolute charge states (say, Ne$^{10+}$) experience a shallow potential well (here of about 200\,eV), while sharing a common temperature with the much more deeply trapped heavier ones reaching far higher charge states under the same electron beam energy conditions (for example, Ba$^{46+}$ would experience a trapping potential of 920\,eV in the present example).
As a consequence, lighter ions in the hot tail of the thermal distribution preferably escape from the trap. Thereby, each evaporating ion in charge state $Q$ removes from the thermal ensemble left behind an energy equivalent to the individual trapping potential, i.~e., $\Delta V \times Q$. A very rough estimate of a typical heating rate for a single heavy HCI (Ba$^{46+}$) yields that the EBIT electron beam transfers a thermal energy of the order of 100\,eV/s through collisions. This heating rate can easily be compensated by the removal of a lighter HCI every couple of seconds.
A steady supply of light atoms, e.g., from an atomic beam to the trap center, will result successively in their ionization, thermalization and evaporation,
accomplishing an efficient removal of the heat constantly generated by the elastic collisions of electrons from the beam with trapped HCI.
Evaporative cooling was the key to achieve in principle unlimited trapping times for HCI from heavy elements,
which are at the same time those really needing long storage times for reaching the highest charge states.
In principle, evaporative cooling also works without the need of a mixture of elements.
Regulation of evaporative cooling is rather unproblematic: The potential barrier which has to be overcome can be tuned by varying the electrostatic potential applied to one of the cylinder-shaped electrodes, or drift tubes in the EBIT.
The use of the magnetic trapping mode \cite{Beiersdorfer1996Magnetic}, in which the electron beam is turned off but the HCI remain trapped in the resulting Penning trap was also combined with evaporative cooling.
In spite of these advantages, trapping-field inhomogeneities, space-charge effects, and voltage noise on the electrodes
have until now limited the efficiency of the method, which in principle should achieve lower temperatures.

Furthermore, compared with other ion traps operating in the mK and $\mu$K regime by means of laser cooling, actual HCI temperatures after evaporative cooling only go down to the level of 0.2\,MK \cite{Beiersdorfer1996Cs,Maeckel2011,Schnorr2013}.
The insufficient cooling causes Doppler broadening and relativistic Doppler shifts, and is the main reason for the lack of data on HCI with a precision better than a few parts-per-million (ppm),
an astounding gap of 12 orders of magnitude to the $10^{-18}$ accuracy of the best optical clock data.
Unfortunately, all other HCI sources are also limited in similar ways.
In some cases, like in accelerators, storage rings and beam-foil methods, the ion beam temperature results from its momentum distribution.
In plasmas, both high ion temperatures and electron-density effects broaden transitions to the same level.
For these reasons, highest-resolution spectroscopy is carried out nowadays in traps under application of laser cooling and related techniques.
This calls for transferring the HCI from their source to another type of trap more amenable to those methods.

\subsection{Sympathetic cooling of HCI}\label{sec:sympcooljose}
Overcoming the difficulty of the high HCI temperatures borne by the violent HCI production processes has been a long-standing aim.
Laser cooling, the tool of choice with atoms and singly charged ions, is impossible with HCI
due to the lack of the necessary fast cycling optical transitions.
%HCI are rich on forbidden lines, but electric dipole (E1) transitions take place in them at much higher energies, typically in the soft x-ray and x-ray domains.

Attempts to solve this shortcoming using sympathetic cooling in Penning traps were already proposed in the 1990's,
and two groups, LLNL and GSI started working on them with the aim of achieving enhanced spectroscopic measurements.
At LLNL, resistive cooling schemes  \cite{Church1999,Gruber2001} for HCI re-trapped in a Penning trap after extraction
from an EBIT were implemented, and sympathetic cooling was achieved \cite{Gruber2001}.
The team at GSI worked at the development of sophisticated deceleration and cooling schemes
\cite{Poth1991,Quint2001,Beier2005,Winters2005,Kluge2008,Rodriguez2010}
needed to bring HCI produced with a relativistic heavy ion accelerator to stillstand in order to load precision ion traps.

\begin{figure*}[htb]
\begin{center}
%\hfill
\includegraphics[width=6in]{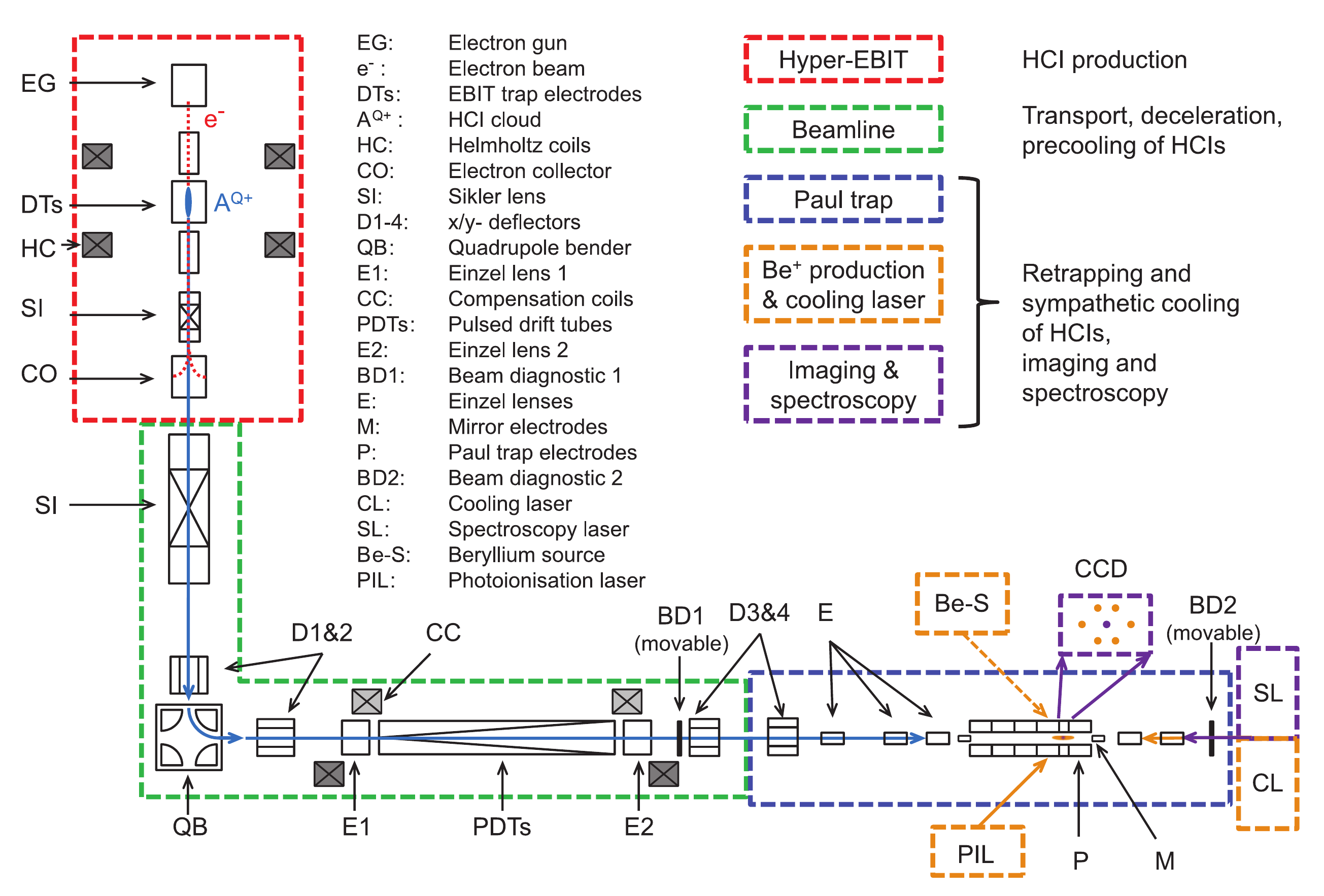}
%\hfill
%\hfill
%
\caption{Scheme of the experiment by \citet{SchVerSch15} in which Ar$^{13+}$ ions were sympathetically cooled by laser-cooled Be$^+$ ions using a laser at $\lambda$\,=\,313 nm.}
 \label{CryPTEx-TotalScheme}
\end{center}
\end{figure*}

\begin{figure}[htb]
\begin{center}
%\hfill
\includegraphics[width=0.8\columnwidth]{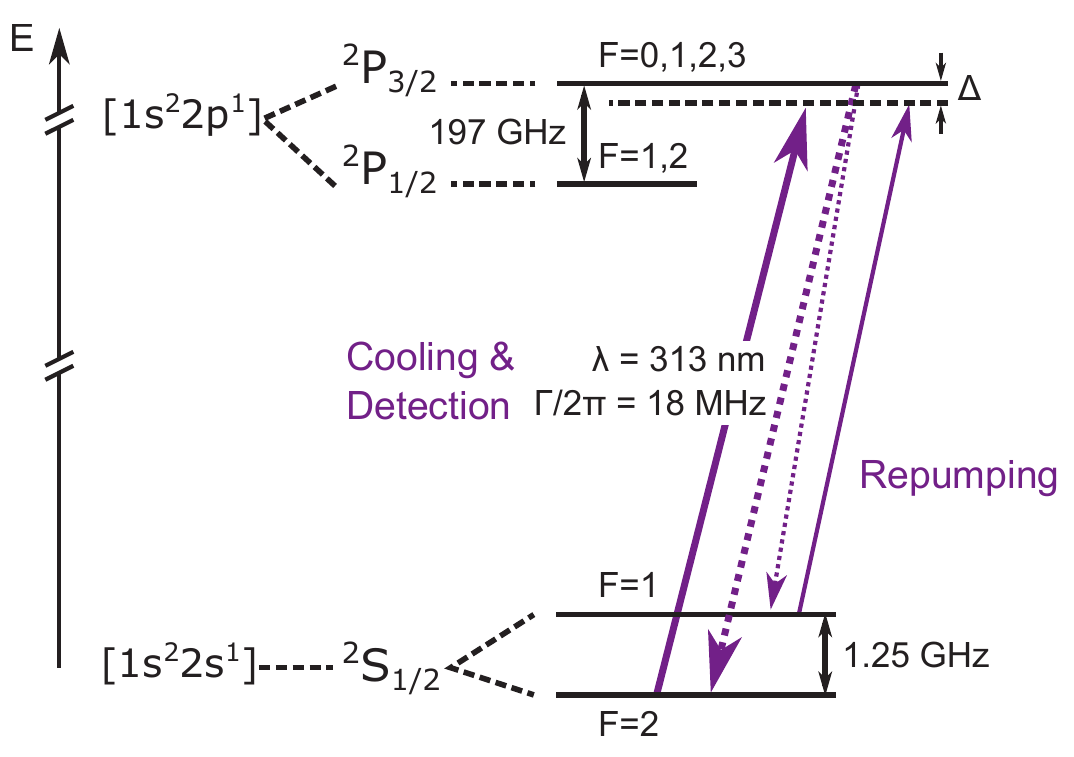}
%\hfill
%\hfill
%
\caption{Doppler cooling of Be$^+$ ions as applied in \cite{SchVerSch15,Schmoeger2015RSI,SchmoegerPhDThesis}. The cooling radiation at $\lambda$\,=\,313 nm and the repumper are produced using the technique proposed by \citet{Wilson2011}.
} \label{Be+CoolingScheme}
\end{center}
\end{figure}

Recently, a combination of HCI production in an EBIT and sympathetic cooling inside of a Coulomb crystal
of laser-cooled Be$^+$ ions contained in a linear RF trap \cite{Schwarz2012,Versolato2013} has demonstrated how to bring the temperature of HCI by 8 orders of magnitude down to the mK regime, and opened the possibilities afforded by RF traps to the study of HCI.
An artist's impression and schematic are shown in \Fref{CryPTEx} and \Fref{CryPTEx-TotalScheme}, respectively. The different components of the system are: an EBIT for the ion production, a transfer beamline for mass selection and deceleration, and a cryogenic RF trap.
\Fref{Be+CoolingScheme} shows the Doppler-cooling scheme applied for preparation of the Coulomb crystals.
Within the work reported by \citet{SchVerSch15,Schmoeger2015RSI}, the HCI transferred to the RF quadrupole were finally embedded in a Coulomb crystal.
Due to the high charge state, HCI tend to occupy the positions along the RF trap axis, where micromotion is minimized.
The laser-cooled Be$^+$ ions are displaced from those positions and surround and sympathetically cool the HCI.

%------------------------------------------------------------------

\begin{figure}[htb]
\begin{center}
%\hfill
\includegraphics[width=\columnwidth]{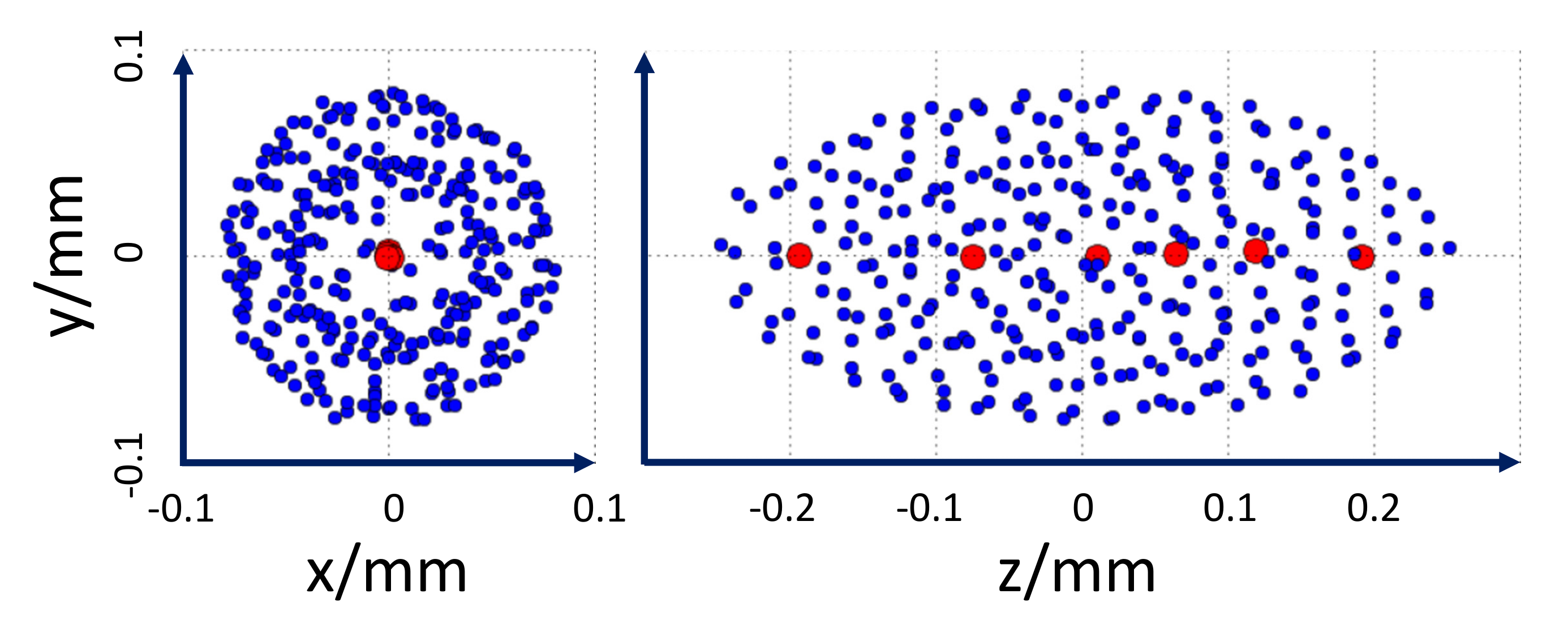}
%\hfill
%\hfill
%
\caption{Simulation \cite{Pedregosa2017} of a Coulomb crystal containing 300 laser-cooled  Be$^+$ ions cooling six Ar$^{5+}$ with a very close charge-to-mass ratio $Q/m$. The thermalization in this case is very good, and the radial and transversal temperatures of all components are about 4\,mK.
} \label{CCSimulation}
\end{center}
\end{figure}

In order to theoretically explore other possible Coulomb crystal configurations,
the full Coulomb interaction and cooling with a thermal bath and laser interaction
molecular dynamics simulations were carried out by J.~Pedregosa \cite{Pedregosa2017}
using a Velocity-Verlet algorithm implemented in Fortran90.
Three phases were analyzed: in the first, no cooling was applied;
in the second, a thermal bath was added, and in the third, laser cooling by two lasers
counterpropagating in the axial direction was turned on.
The simulations probed parameters like $q/m$, the absolute as well as relative number of cooling ions and
sympathetically cooled HCI and the role of the thermal bath temperature.
An example in which Ar$^{5+}$ ions are cooled by 300 Be$^+$ ions is displayed in \Fref{CCSimulation}.
Further simulations were performed for Ar$^{13+}$ ions using the parameters shown in \tref{SimulatedParameters}.
The results indicate for the co-trapped HCI higher radial and axial trap frequencies,
arising from the much steeper potential gradients experienced by those ions.
This could benefit reaching appropriate Lamb-Dicke parameters for the HCI.

\begin{table} \caption{\label{SimulatedParameters} Parameters for a molecular dynamics simulation
of a Coulomb crystal containing sympathetically cooled HCI: Dimensions, radiofrequency, operation potentials, stability parameters $a$ and $q$, axial trap half-length $z_0$, geometric correction factor $\kappa$ and radial/axial trap frequencies $\omega_{x,y/z}$ for the co-trapped Be$^+$ and Ar$^{13+}$. The values are from \citet{Pedregosa2017}.}
\begin{ruledtabular}
\begin{tabular}{lcc}
\multicolumn{1}{l}{Parameter}&
\multicolumn{1}{l}{Value}&
\multicolumn{1}{c}{Unit}\\
\hline \\[-0.4pc]
$r_0$ & 3.5 & mm \\
$\Omega_{RF}$ & 3.96 & MHz \\
$V_{DC} $ & 5.0 & V \\
$V_{RF} $ & 36.25 & V \\
$dt$ & 2.52 & ns\\
$z_0$ & 2.7 & mm \\

$\kappa$            & 0.259 &   \\
$q_{Ar^{13+}}$ & 0.047 &  \\
$q_{Be^+}$      & 0.21 & \\
$a$                     & 0 &  \\

$\omega_{x,y\,Ar^{13+}}$ & 760 & kHz  \\
$\omega_{z\,Ar^{13+}}$ & 531 & kHz  \\
$\omega_{x,y\,Be^+}$ & 189 & kHz  \\
$\omega_{z\,Be^+}$ & 310 & kHz  \\

\end{tabular}
\end{ruledtabular}
\end{table}

These simulations would imply that in a very large trap stable configurations containing 300 Be$^+$ ions
and up to nearly 20 thousand Ar$^{13+}$ sympathetically cooled to 100\,mK are possible .
However, the resulting ensembles would also become rather long and wide (several mm) and thus would suffer from micromotion.

\begin{figure}[htb]
%\begin{center}
%\hfill
\includegraphics[width=0.8\columnwidth]{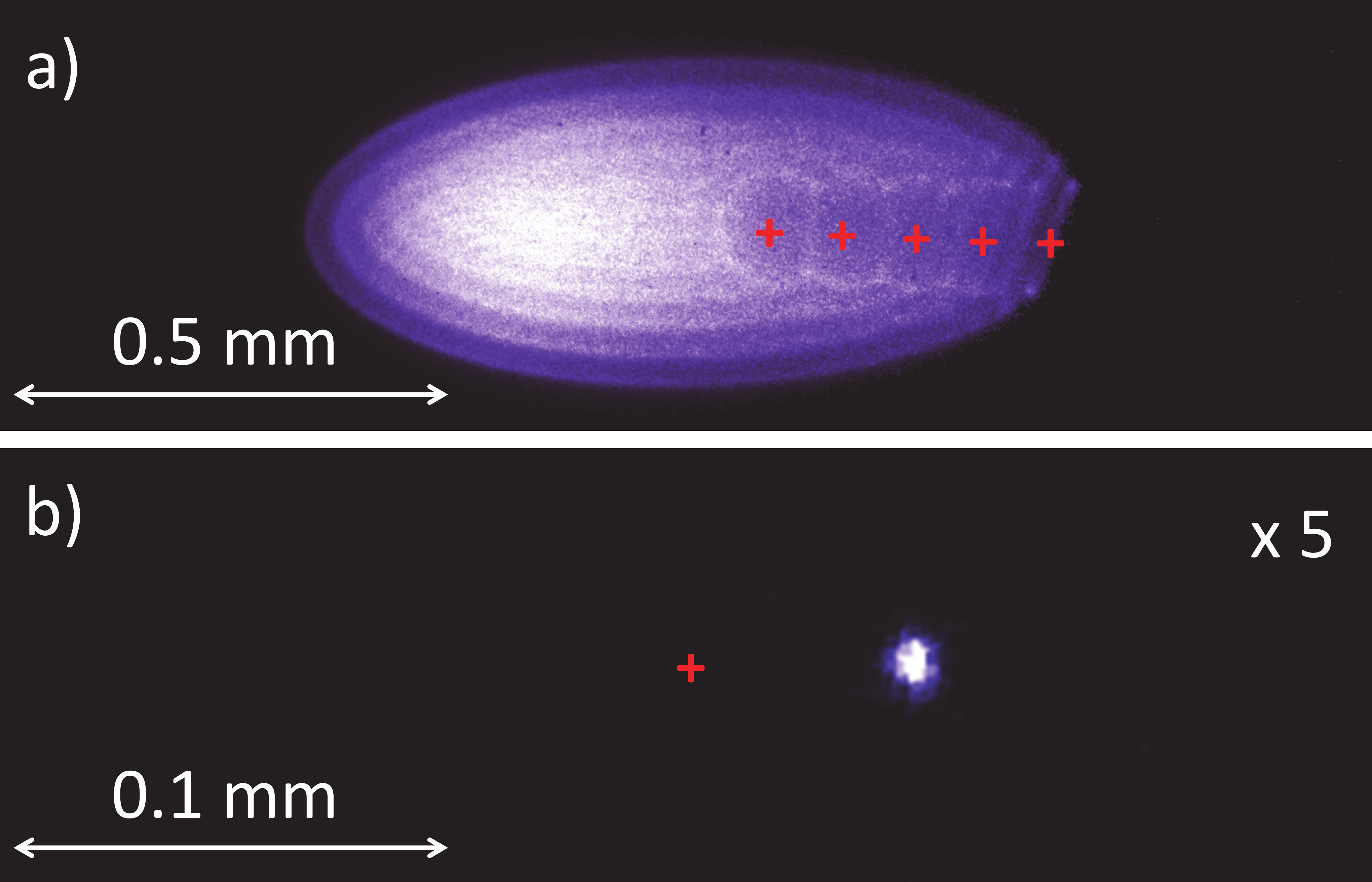}
%\hfill
%\hfill
%
\caption{Images of Be$^{+}$ Coulomb crystals with implanted HCI. a) Ellipsoidal crystal containing 5 Ar$^{13+}$ ions; b) Single Ar$^{13+}$ ion cooled by a single Be$^{+}$ ion \cite{SchVerSch15}. This latter configuration allows for best sympathetic cooling conditions and eventually for ground-state cooling, which is required for quantum-logic detection \cite{schmidt_spectroscopy_2005}.
} \label{CCDImages}
%\end{center}
\end{figure}

Experimental temperature determinations in \cite{SchmoegerPhDThesis} yielded values of 10 mK and below for the HCI.
For faster capture of the HCI in the Coulomb crystals, it was expeditive to use large ensembles of several hundred Be$^+$ ions.
Expelling most of them by means of changes in the trapping parameters leads to the ideal configuration for high resolution quantum logic spectroscopy: a single Be$^+$ cooling a single Ar$^{13+}$, as shown in  \Fref{CCDImages} a) and b), respectively. An effect which has not yet been fully quantified is the partial shielding of the RF field at the HCI positions
by the Be$^+$ ions surrounding them.

In \Fref{CCDImages}a) a Be$^+$ Coulomb crystal containing implanted Ar$^{13+}$ is shown.
\citet{SchmoegerPhDThesis} used these in a first attempt to detect fluorescence from those HCI upon excitation with a low-drift laser system \cite{Leopold2016} developed for this purpose. Due to insufficient knowledge of the exact frequency of the $2p_{3/2}\rightarrow 2p_{1/2}$ transition at approximately 441.255 nm, and the faintness of the expected signal in comparison with technical background noise, the experiment has not yet been successful.
In the near future, an enlarged detection solid angle, noise reduction and better prior knowledge of the scanning range
for the transitions of interests should enable this type of fluorescence studies that can support electronic structure determinations.
Searches for highly forbidden transitions benefit from better knowledge of the intra-configuration transitions of M1 type, as shown by \citet{WinCreBel15}.

For one single Ar$^{13+}$, the fluorescence rate at saturation for this M1 transiton is approximately 100\,Hz \cite{Lapierre2005,Lapierre2006}.
With a detection solid angle of the order of 1\%, and including the photomultiplier detection efficiency and other losses,
a few ions should provide a signal rate of 0.1\,Hz, which in principle could be measured against a PMT dark count rate of 4\,Hz.
%Unfortunately, the search for the exact frequency was hindered by laboratory building and technical problems.
These limitations in signal-to-noise ratio can be overcome by more elaborate quantum state detection systems, e.g. based on quantum logic \cite{schmidt_spectroscopy_2005, wan_precision_2014,hempel_entanglement-enhanced_2013} as discussed in \ref{sec:QLS}. These techniques are crucial for addressing even more-forbidden transition types (E2, M2, E2M3) for optical clocks.

Since trapping of a single HCI with a single cooling Be$^+$ ion has been demonstrated,
the possibility of sideband cooling to the ground state of motion as discussed in \sref{sec:sympcool} will allow for advanced detection schemes addressing very slow clock transitions.
Sympathetic cooling overcomes the main difficulty for high-resolution laser spectroscopy and frequency metrology with HCI.
In this way, the plethora of quantum manipulation techniques available in the
atomic physics community gains a wide class of experimental target
beyond the much-studied alkali-like ions. In essence, a working
technique combining HCI source and RF trap as needed for these
studies is now within reach for high precision frequency-metrology
groups. Optical clocks \cite{Ludlow2015} based on trapped
singly-charged ions have achieved accuracies which in science have
only that of optical lattice clocks as peers. The introduction of HCI
as pacemakers will open up new opportunities for further enhancement
of accuracy and sensitivity for fundamental physics searches.

%------------------------------------------------------------------

%\begin{figure}[htb]
%%\begin{center}
%%\hfill
%\includegraphics[width=\columnwidth]{CCDImagesNew2}
%%\hfill
%%
%\caption{Images of Be$^{+}$ Coulomb crystals with different numbers of implanted implanted HCI. The positions of the Ar$^{13+}$ ions are marked with crosses to guide the eye. The fourth image from the top has been mirrored, since part of the image could not be recorded. Images from \cite{SchmoegerPhDThesis} \cPiet{I suggest removing this figure. It does not add new information compared to the next figure.}.
%} \label{CCDImagesNew}
%%\end{center}
%\end{figure}

%%%%%%%%%%%%%%%%%%%%%%%%%%%%%%%%%%
%\input{clocks_v9}
%%%%%%%%%%%%%%%%%%%%%%%%%%%%%%%%%%

\section{Towards high-resolution spectroscopy}
\label{sec:hires} Several challenges need to be addressed to make
HCI accessible for high precision spectroscopy and optical clocks.
The main goal of an atomic frequency standard is the realization of
the unperturbed frequency of the reference or clock transition. In
this context, two types of uncertainties are important: the
statistical uncertainty (sometimes also called instability or
uncertainty type A) and the systematic uncertainty (sometimes also
called inaccuracy or uncertainty type B). While the systematic
uncertainty quantifies how well we believe we are able to reproduce
the unperturbed transition frequency, statistical uncertainty tells
us for how long we need to average frequency measurements to achieve
a certain resolution. Systematic uncertainties of clocks beyond the
current accuracy of a few parts in $10^{16}$ of the best Cs
frequency standards \cite{guena_first_2017} can only be estimated by
considering all possible shifts to the measured frequency. This
includes changes to the atomic structure due to interaction with
external fields, but also relativistic effects from motion and
gravity. The instability of a frequency standard in which $N$
uncorrelated atoms are probed simultaneously is ultimately limited
by quantum projection noise \cite{itano_quantum_1993}. In the simple
case of Ramsey interrogation \cite{ramsey_molecular_1985} of a
transition with frequency $\nu_0$ using perfect pulses and a probe
time $T_R$, we get a fractional frequency uncertainty expressed in
the form of an Allan deviation
\cite{allan_statistics_1966,riehle_frequency_2004,
riley_handbook_2008} of
\begin{equation}
\sigma_y(t) = \frac{1}{2\pi\nu_0 \sqrt{N T_R t}}\ .
\label{eq:sql}
\end{equation}
From this equation it becomes clear that a high transition
frequency, many atoms and a long probe time reduce the averaging
time $t$ to achieve a certain resolution. The probe time is a
special case, since it can either be limited by the lifetime of the
excited clock state, or by the coherence time of the probe laser
\cite{peik_laser_2006, riis_optimum_2004, leroux_-line_2017}. The
currently best lasers achieve a flicker noise floor-limited
instability of $4\times 10^{-17}$ \cite{matei_1.5_2017} and support
probe times of several seconds. For a single atom, the best possible
statistical uncertainty is achieved for a probe time equal to the
excited state lifetime $\tau$ of the clock transition. It is given
by \cite{peik_laser_2006}
\begin{equation}\label{eq:instability}
\sigma_y(t) = \frac{0.412}{\nu_0\sqrt{\tau t}}.
\end{equation}

In the remainder of this section, we will discuss the technical
issues and systematic frequency shifts, their measurement, and
suppression, specific to HCI. Frequency shifts arise from motion and
from external fields causing a differential shift of the two clock
levels. For evaluating the performance as a frequency reference, we
need to know the atomic parameters (from theoretical atomic
structure calculations or measurements) and the strength of the
external field. In some cases suppression techniques exist to reduce
the influence of external perturbations on the clock transition.
While recent reviews have addressed these issues for single-ion
clocks \cite{Ludlow2015,poli_optical_2013}, we will provide the
scaling of these effects to HCI. Some effects, such as electric and
magnetic field shifts will be reduced, while others (collisional
shift, motion-induced shifts) may be enhanced. The section will end
with an assessment of potential HCI optical clock candidates and
their expected performance.

\subsection{Trapping}\label{sec:trapping}

%José text on this: Radio-frequency linear traps will continue being the workhorse of
%ion-based research. Their size will initially be dictated by the
%needs of external ion transfer and injection, which favour larger
%traps. They are currently also more convenient in terms of spurious
%noise sources, optical access and ultrahigh vacuum (UHV) operation
%than microtraps, but there are efforts to remove such limitations.
%At this moment, cryogenic operation seems mandatory if HCI trapping
%times in the order of hours are envisioned. However, future traps
%could also operate at the necessary UHV levels by simpler means such
%as ion-sputtering and getter pumps if appropriate choices of
%low-outgassing materials and baking procedures are made \cPiet{Is this realistic? If it were possible to achieve such good vacuum conditions outside a cryogenic system, people would have done it already.}. Ideally, a
%future HCI optical clock could be based on liter-volume package with
%a sealed quartz envelope enclosing dispensers for the elements of
%interest and cooling ion of choice, a maintenance-free ion getter
%pump, a miniature EBIT, accessory magnetic shielding, the ion optics
%and trapping electrodes, feedthroughs and optical windows. Such a
%development would be mandatory in order to ease widespread
%applications of such clocks.

\begin{figure}[htb]
%\begin{center}
%\hfill
\includegraphics[width=\columnwidth]{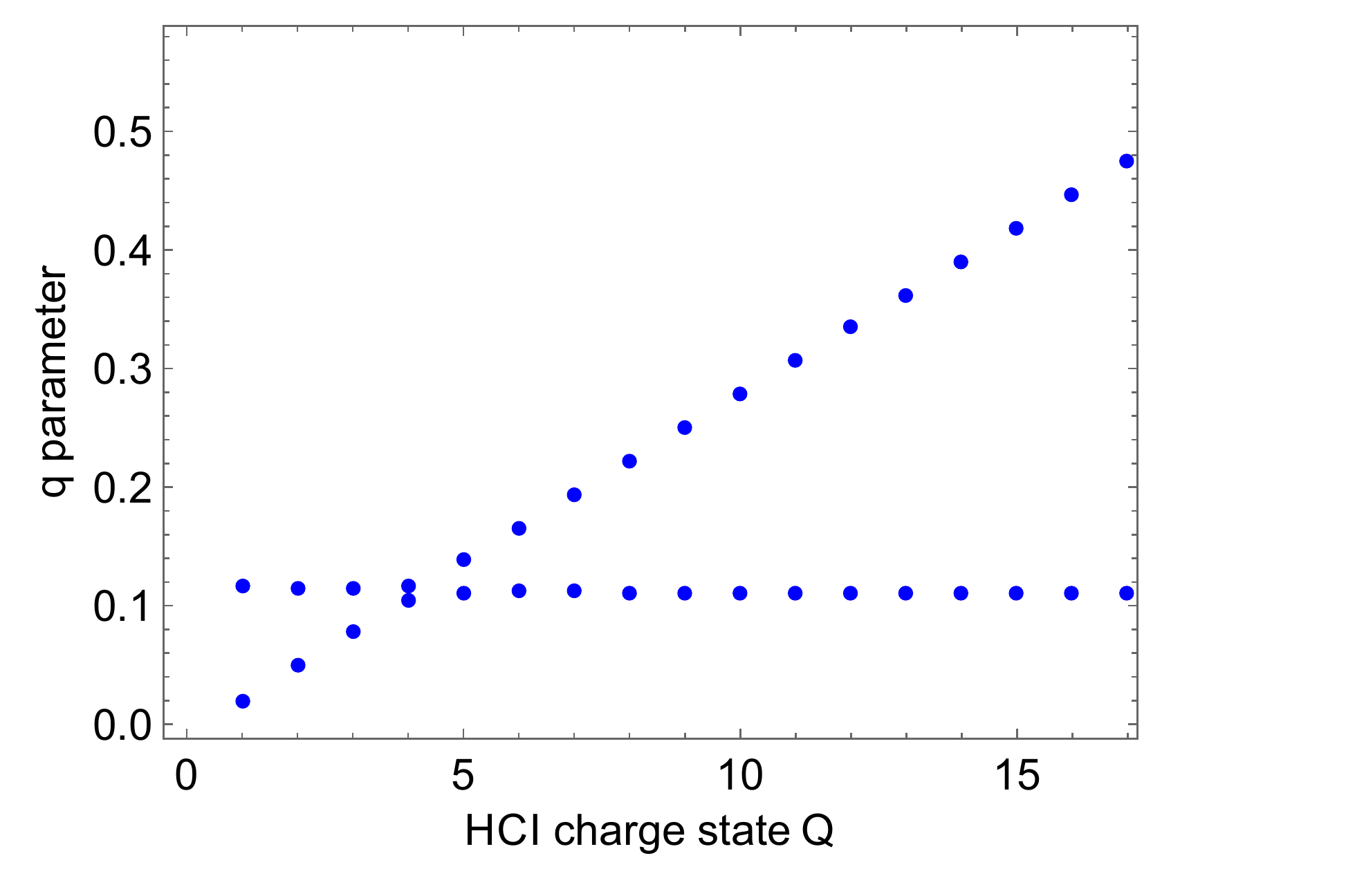}
%\hfill
%\hfill
%\end{center}
\caption{Stability parameter for a HCI in a linear Paul trap. The stability parameter $q$ is plotted for the radial modes of a 2-ion crystal consisting of a singly-charged \Ben and an ${}^{40}$Ar$^{Q+}$ ion as a function of the Ar-ion's charge $Q$ for a single \Be radial trapping frequency of $\omega_r=2\pi\times 2.2$~MHz and $\Orf=2\pi\times 50$~MHz. All charge states satisfy the stability criterion $q<1$.}
\label{fig:qparam}
\end{figure}

Radio-frequency Paul traps have been the workhorse in all areas of
ion-based research that require isolation from external fields, such
as ion optical clocks, quantum information processing, quantum
simulation and quantum metrology \cite{major_charged_2006,
wineland_experimental_1998}. In spherical Paul traps, a 3D
oscillating quadrupole field provides confinement in all directions.
At the center of the quadrupole, a single ion can be stored almost
perturbation free. To trap more ions, linear Paul traps have been
invented in which an oscillating 2D radial electric field combined
with a static axial field provides trapping. In both cases harmonic
trapping at secular frequencies between 100~kHz up to a few MHz can
be achieved with rf drive frequencies $\Orf$ oscillating typically
at least one order of magnitude higher.
%In linear ion traps the axial trapping frequency scales with the charge state $Q$ as $\omega_z\sim\sqrt{Q}$, while the radial trapping frequencies exhibit a linear scaling $\omega_r\sim Q$.
A huge variety of electrode geometries and material choices exist
for ion traps, ranging from microfabricated surface traps providing
a scalable approach with multiple segments
\cite{seidelin_microfabricated_2006, britton_microfabricated_2006,
chiaverini_surface-electrode_2005}, sometimes even in a cryogenic
environment \cite{labaziewicz_suppression_2008} to macroscopic traps
for optical clocks with ion-electrode distances on the order of one
millimeter \cite{dolezal_analysis_2015}. The latter provide large
optical access for efficient detection and probing the ion from
various directions. They feature deep trapping potentials on the
order of a few eV for storage times of many hours up to a few
months. They offer moderately large motional frequencies of up to
several MHz to allow recoil-free spectroscopy in the Lamb-Dicke
regime while maintaining potentially low motional heating rates of
only a few motional quanta per second
\cite{brownnutt_ion-trap_2015}. Recently even linear multi-ion traps
for precision spectroscopy and optical clocks have been developed
\cite{herschbach_linear_2012, pyka_high-precision_2014,
keller_precise_2015, keller_evaluation_2016, dolezal_analysis_2015},
with trap-induced shifts below $10^{-19}$ fractional frequency
uncertainty \cite{keller_optical_2017}. While most single-ion
frequency standards in the past are room temperature systems, with
the notable exception of the Hg$^+$ clock \cite{Ros08}, cryogenic
Paul traps are mandatory for HCI to achieve a sufficiently long
lifetime through excellent vacuum conditions \cite{Schwarz2012}.
Conveniently, at the same time the BBR shift (see
\sref{sec:efields}) becomes negligible. One typically distinguishes
two types of motion of an ion in a Paul trap: fast micromotion at
frequency $\Orf$ and so-called secular motion at lower oscillation
frequencies. Paul traps support only stable trajectories for certain
charge-to-mass ratios for a given geometry and rf trapping field
\cite{major_charged_2006, drakoudis_instabilities_2006}. In the
simplest case, stable trapping in linear Paul traps is achieved if
the radial oscillation frequency, $\omega_r$, is significantly
smaller than the trap drive frequency $\Orf$, or $q\approx
2^{3/2}\omega_r/\Orf<1$. This has to be taken into account when
trapping HCI, in particular when trapping them simultaneously with
singly-charged atomic ions as discussed in the next section. Since
the stability criterium scales with the charge-to-mass ratio, \Be is
a suitable cooling ion species for many medium-charged HCI. As an
example, \fref{fig:qparam} shows the stability parameter for two of
the radial modes of a \Ben-\hci{40}{Ar}{Q+} ion crystal. For
sufficiently small $q$, secular motion decouples from micromotion
and it can be assumed to be harmonic in all three directions
\cite{dehmelt_radiofrequency_1968}, which is what we assume from now
on. Residual effects of micromotion have to be included separately,
as discussed in \sref{sec:micromotion}.

\subsection{Sympathetic cooling}\label{sec:sympcool}
One of the key ingredients for the success of atomic systems over
the past 25~years is the control of their motion through Doppler
laser cooling and other advanced cooling techniques
\cite{chu_manipulation_1998,cohen-tannoudji_nobel_1998,phillips_laser_1998}.
The first laser cooling concepts were in fact developed and
experimentally demonstrated for trapped ions
\cite{wineland_laser_1979,drullinger_high-resolution_1980}. Since
then, laser cooling has become mandatory for all optical ion clocks,
since it allows localization of the ion to well below the wavelength
of the spectroscopy light, which is equivalent to the condition of
recoil-free absorption in the Lamb-Dicke regime
\cite{dicke_effect_1953}. For HCI optical clock candidates that do
not have a sufficiently fast (MHz linewidth) and closed transition
suitable for laser cooling, sympathetic cooling can be provided by
another atomic species that is laser coolable. This is implemented
by trapping the HCI together with the cooling ion into the same
trapping potential and taking advantage of their mutual Coulomb
repulsion
\cite{larson_sympathetic_1986,barrett_sympathetic_2003,wan_efficient_2015}.
While initial stopping and Doppler cooling of HCI is performed in
clouds of laser-cooled atomic ions \cite{SchVerSch15} as described
in \sref{sec:sympcooljose}, precision spectroscopy demands smaller
ion crystals, consisting in the simplest case of two ions.

The strong Coulomb repulsion between two cold ions with charges
$Q_{1,2}$ in a linear Paul trap results in equilibrium positions
$d_{1,2}$ away from the center of the trap according to
\begin{equation} \label{eq:distance}
d_{1,2}=Q_{2,1}/(4\pi\epsilon_0 u_0 (Q_1+Q_2)^2)^{1/3}.
\end{equation}
Here, $u_0=m\omega_z^2/Q$ quantifies the strength of the axial
trapping potential in terms of a single reference ion with mass $m$,
charge $Q$ and oscillation frequency $\omega_z$.

The ions perform a coupled motion around these equilibrium positions
appropriately described in a normal mode framework. For two ions, we
have two modes in each direction, the in-- and out--of--phase mode
(indices $i,o$) with mode frequencies $\omega_{i,o}$. We can thus
write the oscillation $z_{1,2}(t)$ of the two ions along a selected
direction as a superposition of the contributions from the two modes
along this direction \cite{james_quantum_1998,
morigi_two-species_2001, kielpinski_sympathetic_2000,
wubbena_sympathetic_2012}
\begin{eqnarray*}
z_1(t)&=&(\hat{z}_{i}b_{1,i}\sin(\omega_i t+\phi_i)+\hat{z}_{o}b_{1,o}\cos(\omega_o t+\phi_o))/\sqrt{m_1}\\
z_2(t)&=&(\hat{z}_{i}b_{2,i}\sin(\omega_i t+\phi_i)+\hat{z}_{o}b_{2,o}\cos(\omega_o t+\phi_o))/\sqrt{m_2}.
\end{eqnarray*}
Here, $\hat{z}_{i,o}/\sqrt{m_{1,2}}=\sqrt{2E_{i,o}}/\omega_{i,o}$ is
an excitation amplitude that depends on the kinetic energy $E_{i,o}$
in each mode, and $b_{k,\alpha}$ are mode amplitudes normalized to
%$b_{1,\alpha}b_{1,\alpha'}+b_{2,\alpha}b_{2,\alpha'}=\delta_{\alpha,\alpha'}$, which results in
$b_{2,o}=-b_{1,i}$ and $b_{2,i}=b_{1,o}$ with
$b_{1,i}^2+b_{1,o}^2=1$. The modes have phases $\phi_{i,o}$ that
depend on the initial conditions. While identical ions have
(anti-)symmetric mode amplitudes for the in--phase (out--of--phase)
mode in each direction, a difference in mass and/or charge results
in one ion having a large mode amplitude, while the other has a
small amplitude for one mode and vice versa for the other mode along
the same direction.
\begin{figure}[htb]
%\begin{center}
%\hfill
\includegraphics[width=\columnwidth]{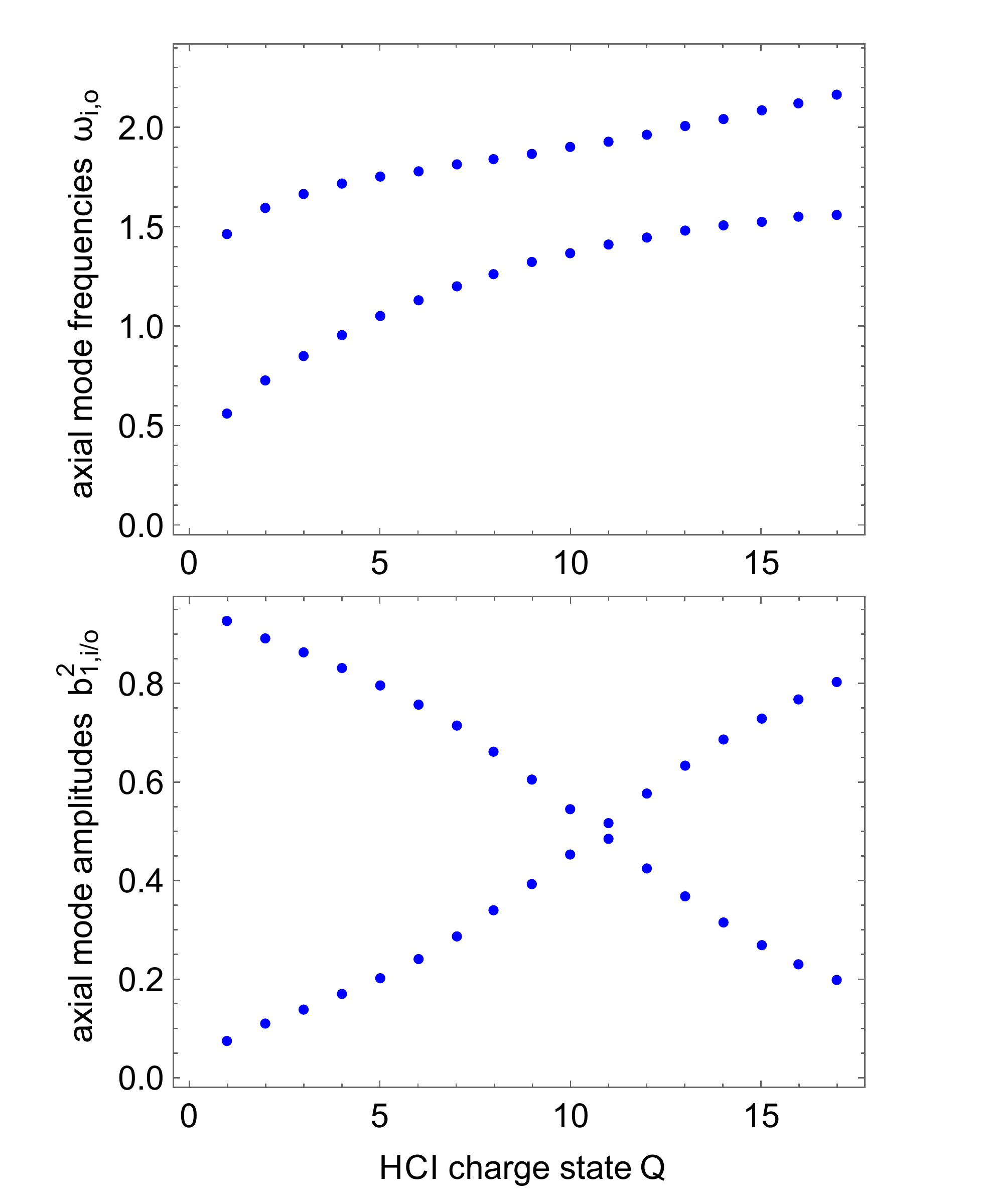}
%\hfill
%\hfill
%\end{center}
\caption{Coupled axial mode parameters for a HCI in a linear Paul trap.}
\label{fig:axial}
\end{figure}

\begin{figure}[htb]
%\begin{center}
%\hfill
\includegraphics[width=\columnwidth]{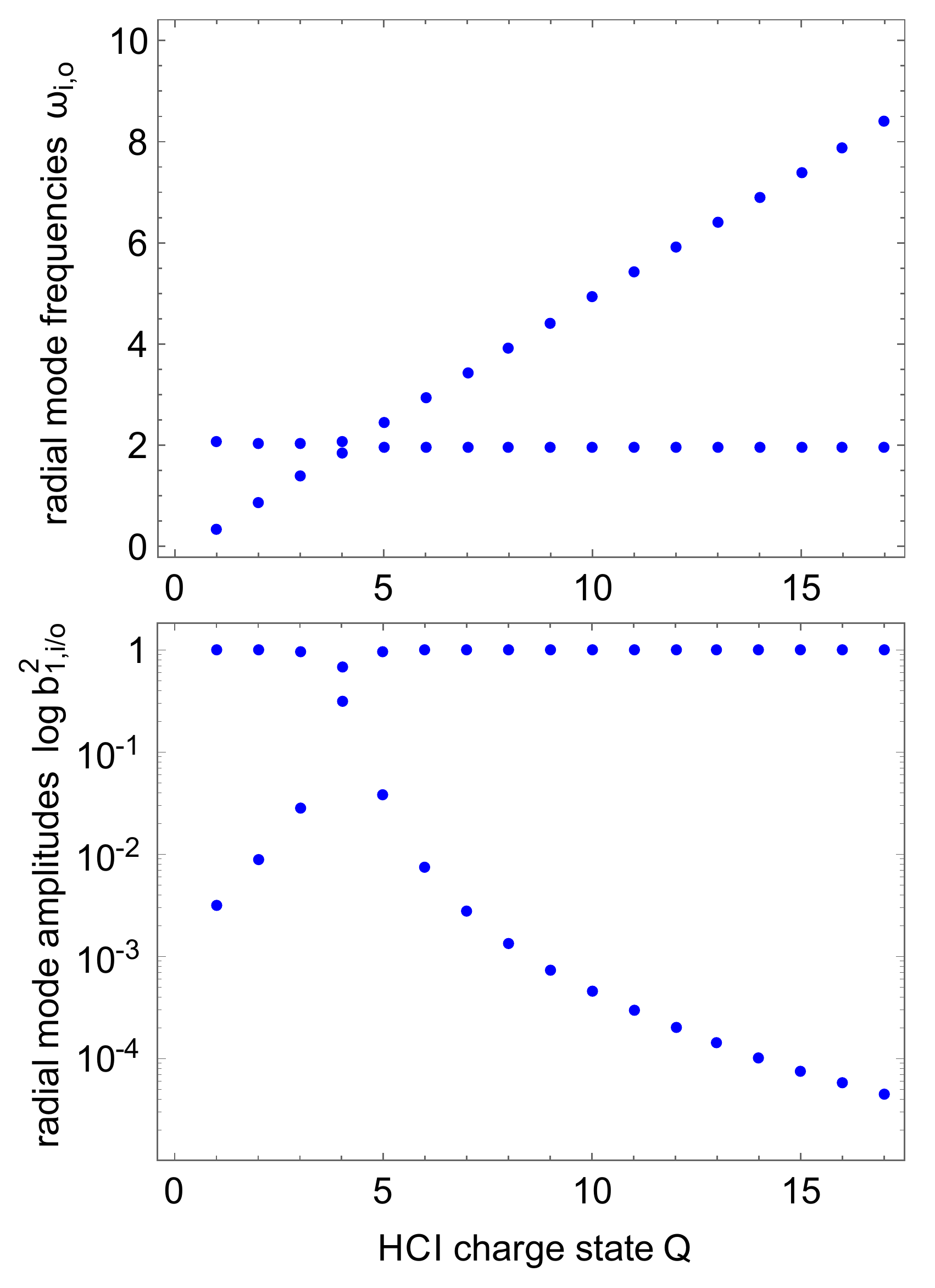}
%\hfill
%\hfill
%\end{center}
\caption{Coupled radial mode parameters for a HCI in a linear Paul trap.}
\label{fig:radial}
\end{figure}
\Fref{fig:axial} shows the axial and \fref{fig:radial} one set of
radial mode frequencies $\omega_{i,o}$ and the square of the
amplitudes $b^2_{1,i}, b^2_{1,o}$ of a 2-ion crystal consisting of a
singly-charged \Ben and an ${}^{40}$Ar$^{Q+}$ ion as a function of
the Ar-ion's charge $Q$. A single \Be ion would have an axial
(radial) mode frequency of $\omega_z=2\pi\times 1$~MHz
($\omega_r=2\pi\times 2.2$~MHz). While the axial modes remain
strongly coupled for all charge states, this is not the case for the
radial modes. The radial mode frequency corresponding closely to the
single \Be ion mode remains almost constant for all charge states,
while the other mode frequency increases as a function of the HCI
charge state. Similarly, one of the mode amplitudes remains almost
constant near a value of one, while the other one is close to zero.
An exception is $Q=4$, for which the charge-to-mass ratio of \Ben
and ${}^{40}$Ar$^{4+}$ is almost equal. Thus, away from this
``resonance'' in the coupling the ions are radially only weakly
coupled. A similar, but less pronounced effect is observed for
singly-charged ions with different masses
\cite{wubbena_sympathetic_2012}. The mode decoupling is more
pronounced for HCI, since (in contrast to singly-charged ions with
different masses) their distance $(d_1+d_2)$ increases with their
charges according to \Eref{eq:distance}, reducing the radial mode
coupling. However, the resonances in the radial mode coupling can be
exploited if a HCI with a suitable cooling transition and matching
$Q/m$ can be found.

The reduction in coupling has two consequences for the two radial
modes for which the Doppler cooling ion has only a small mode
amplitude $b_r$.
%i) cooling to equilibrium takes long ii) in the presence of additional motional heating, the equilibrium temperature is elevated.
Firstly, the Doppler cooling rate scales with the square of the mode
amplitude. Together with typical timescales to reach Doppler cooling
temperature after a background gas collision on the order of
milliseconds for a single ion \cite{wubbena_sympathetic_2012},
cooling a mode with $b^2\sim 10^{-4}$ (see \fref{fig:radial}) would
take up to 10~s, which is unacceptably long. However, cooling these
radial modes can be made more efficient by tilting the ion crystal
out of the axial alignment through application of a static electric
field. This enhances the mode coupling and has already been
successfully demonstrated in the \Al/\Be quantum logic optical clock
\cite[supplementary information]{Ros08}.

%While for typical singly-charged ions the modulus of the ratio of the radial mode amplitudes reaches values as low as $\sim 0.024$ for a \Ben/\Alts ion crystal \cite{schmidt_spectroscopy_2005}, the same trapping potential results in a ratio of $0.0041$ for a \Ben/\Arfztp ion crystal. What are the consequences of these strongly differing mode amplitudes for HCI? The cooling rate scales with the mode amplitude of the cooling ion. Therefore, two of the radial modes are not efficiently cooled and it may take several 10s to 100s of milliseconds to reach Doppler cooling equilibrium temperature, e.g. after a collision with background gas \cite{wubbena_sympathetic_2012}.
Secondly, in the presence of motional heating mechanisms (in
addition to photon scattering), the steady-state temperature will be
larger than the Doppler cooling temperature
\cite{wubbena_sympathetic_2012}. Anomalous motional heating from
fluctuating electric fields is the most common heating mechanism in
ion traps and HCI will be particularly sensitive to it owing to
their high charge state. Although the exact origin of anomalous
heating remains elusive, for most traps the heating rate is
significantly reduced at cryogenic temperatures and for large
ion-electrode distances \cite{brownnutt_ion-trap_2015}. Since for
HCI macroscopic traps at cryogenic temperatures will be employed, it
can be expected that the anomalous heating rate is sufficiently
small to allow efficient sympathetic cooling. Assuming a typical
heating rate of 1~phonon/s for a trap with around 0.7~mm
ion-electrode separation \cite{keller_evaluation_2016}, the
temperature of the weakly coupled modes for the \Be/\Artp example
discussed above would be elevated by less than 10\,\% above the
Doppler cooling limit. The reduction of cooling rates for modes with
a small amplitude of the cooling ion also hold for all other cooling
techniques, such as sideband or electromagnetically-induced
transparency (EIT) cooling
\cite{roos_experimental_2000,lin_sympathetic_2013,
lechner_electromagnetically-induced-transparency_2016,scharnhorst_multi-mode_2017},
that allow reaching the motional ground state.

\subsection{Quantum logic spectroscopy}\label{sec:QLS}
In addition to providing sympathetic cooling, the singly-charged
atomic ion can also be employed for preparation and read out of the
internal state of the HCI during the spectroscopy sequence using
quantum logic spectroscopy \cite{schmidt_spectroscopy_2005,Ros08}.
In the original implementation, state detection is accomplished by
applying a series of laser pulses to the HCI and atomic ion in the
motional ground state that implement so-called sideband pulses
changing the internal state while adding/removing a quantum of
motion \cite{wineland_experimental_1998}. This way, a SWAP operation
between the internal state of the HCI and the motion is implemented,
followed by another SWAP between the motion and the internal state
of the atomic ion. Such a sequence maps the internal state of the
HCI faithfully onto the internal state of the atomic ion, where it
can be detected with high fidelity \cite{schmidt_spectroscopy_2005,
hume_high-fidelity_2007}. Similarly, internal state preparation can
be accomplished by applying sideband pulses to the HCI to drive the
HCI into the target state. Dissipation is provided through
ground-state cooling on the singly-charged atomic ion, which makes
the sideband pulses irreversible \cite{schmidt_spectroscopy_2005,
chou_preparation_2017}. An parallel quantum readout algorithm for
multiple clock ions using as few logic ions as possible has also
been developed \cite{schulte_quantum_2016}. Other forms of quantum
logic spectroscopy are more suitable for fast transitions. For
example, in photon recoil spectroscopy \cite{wan_precision_2014,
hempel_entanglement-enhanced_2013}, recoil upon photon absorption by
the spectroscopy ion manifests changes the motional state, which can
be detected with high efficiency on the logic ion. By employing a
state-dependent optical dipole force, incoherent photon scattering
can be significantly suppressed, which enables spectroscopy of
species with non-closed broad transitions, such as molecules
\cite{wolf_non-destructive_2016}. The same technique can be employed
for an efficient search and identification of previously unknown
lines in HCI. In summary, using a co-trapped singly-charged atomic
ion for sympathetic cooling and quantum logic spectroscopy, any HCI
can be spectroscopically investigated as long as sideband
transitions can be driven with high fidelity on either the
transition of interest or any another suitable transition.
Therefore, the choice of HCI for an optical clock is entirely
dictated by its atomic properties.

\subsection{Systematic frequency shifts}
\label{unc} The evaluation of systematic frequency shifts is one of
the most important tasks when developing a frequency standard. The
most common shifts arise from external fields coupling
differentially to the two clock levels. The shift is characterized
by its magnitude and associated uncertainty. Its magnitude is given
by the properties of the external field (ac/dc, strength,
orientation, polarization, gradient, $\dots$) and the atomic
properties of the clock levels. The field can either be measured
externally or through the atoms, or by using a combination of
measurements and simulation, as is the case e.g.\ for modeling the
thermal environment of the ion to evaluate the black-body radiation
shift \cite{dolezal_analysis_2015,dube_evaluation_2013}. Similarly,
the atomic parameters can either be measured or one can rely on
accurate atomic structure calculations. The uncertainty of a shift
is given by the combined uncertainty of all individual
contributions, i.e. the uncertainty in the atomic properties and the
uncertainty in the properties of the external field. In some cases,
shifts can be suppressed by taking advantage of symmetries in the
shift. Averaging frequency measurements involving different levels
of the Zeeman substructure of the atomic energy levels, allows e.g.\
elimination of the linear and quadratic Zeeman shift and the
electric quadrupole shift. In these cases the uncertainty in the
shift is determined by its variation during the averaging process,
which has to be determined experimentally. Choosing a clock ion
species with advantageous atomic properties can be simpler to
implement and may eventually result in smaller systematic
uncertainties compared to an atom for which cancelation schemes are
required. Following
\cite{Ludlow2015,schmidt_trapped-ion_2015,itano_external-field_2000},
we discuss the dominant shifts and possible cancelation schemes in
trapped ion frequency standards applied to HCI.

\subsubsection{Magnetic field shifts}\label{sec:mfields}
External magnetic fields couple to the magnetic moment of the
electron, $\mu_B$ (order of magnitude 14~MHz/mT/$h$ shifts), and the
nucleus, $\mu_N$ (7.6~kHz/mT/$h$ shifts). The coupling with the
external field is in competition with the internal coupling between
the magnetic moments. This results in non-linear shifts of the
corresponding Zeeman levels as a function of the magnetic field
strength. Of interest for clocks is the differential frequency shift
\DfB between a selected ground and excited state, which can be
expressed as a Taylor expansion according to
\[
\DfB=C_\mathrm{M1}B+C_\mathrm{M2}B^2+C_\mathrm{M3}B^3+\dots.
\]
It is usually sufficient to consider only the first two terms for
typical magnetic fields of a few 100~$\mu$T that are applied to
provide a quantization axis for laser cooling and optical pumping.
Electronic states with a total angular momentum quantum number $J=0$
exhibit only the small nuclear Zeeman effect, whereas states with
$J>0$ have the much larger electronic Zeeman shift. In both cases,
the $C_\mathrm{M1}$ term of \DfB scales linearly with the magnetic
quantum number $m$. The linear Zeeman effect vanishes for
transitions $m=0\rightarrow m'=0$, where the (un)primed magnetic
quantum number denotes the (ground) excited state. In all other
cases, averaging two (or more) transitions with shifts of equal
magnitude, but opposite sign, allows recovering the unshifted
transition frequency. The quadratic Zeeman effect is the next
largest contribution. It arises from the decoupling of nuclear and
electronic magnetic moments as a function of the external magnetic
field strength. For atoms with $J>0$ and hyperfine structure, the
external field mixes states of different hyperfine quantum numbers
$F$. The corresponding quadratic shift can be derived from first
order perturbation theory to be
$C_{M2}\sim(g_J\mu_B-g_I\mu_N)/(h^2A)$, where $g_J$ and $g_I$ are
the electronic and nuclear $g$-factors, respectively, and $A$ is the
hyperfine constant, characterizing the splitting between $F$ states.
For singly-charged ions, typical values for $C_{M2}$ range between a
few to a few ten kHz/mT$^2$. Since the hyperfine splitting scales in
HCI as $Z^3$ \cite{Gillaspy2001}, hyperfine constants become very
large and the quadratic Zeemann shift correspondingly small. In case
the clock transition is a hyperfine transition, the expression for
the shift coefficient simplifies to $C_{M2}=2\mu_B^2/(h^2\nu_0)$
\cite{YTD14}.

In the case of $J=0$ levels, a quadratic Zeeman effect arises
through external field-mediated mixing of fine-structure components.
The shift is then again proportional to the difference of the
involved magnetic moments, divided by the fine-structure splitting.
In singly-charged ions the $C_{M2}$ coefficient is on the order of a
few ten Hz/mT$^2$. Since the fine-structure splitting in HCI scales
with $Z^4$ (even more strongly than the hyperfine splitting), the
quadratic Zeeman shift is further suppressed. To evaluate the shift
precisely, the magnetic field and its variation needs to be
determined with high accuracy. Where available, transitions with
large and calibrated $C_{M1}$ coefficients can be employed for this
task. Alternatively, from the difference of the two transitions
averaged to eliminate the linear Zeeman shift, the magnetic field
can be derived. In fact, by averaging transitions involving all
Zeeman components of a state eliminates linear and quadratic Zeeman
shifts, as well as the electric quadrupole shift discussed below
\cite{dube_evaluation_2013}.

The previous discussion applies to dc magnetic fields. For ac
magnetic fields, the linear Zeeman shift averages to zero. However,
the quadratic term $\Delta f_{M2}=C_{M2}\langle (B-B_0)^2\rangle$
remains and may be significant. In ion traps ac  magnetic fields can
arise from the trap RF drive or from power line noise. In both
cases, the shift needs to be calibrated by either sideband
spectroscopy or by extrapolation to zero field.

\subsubsection{Electric field shifts}\label{sec:efields}
\paragraph{AC Stark shift}
Ions are always located at the position of vanishing electric field.
Therefore, the dominant electric field shifts arise from field
gradients and oscillating electric fields. Oscillating electric
fields couple to the polarizability of the atom's states via the ac
Stark effect, while field gradients couple to electric quadrupole
moments a state might have. The quadratic Stark effect can be
treated as a small perturbation to the linear Zeeman effect,
resulting in a scalar shift for energy levels $\ket{\gamma J}$ with
$J\leq 1/2$ of
\begin{equation}
h\Delta f_S(\gamma,J,\vec{E})=-\alpha_S(\gamma,J)|\vec{E}|^2/2.
\end{equation}
It is characterized by the scalar polarizability $\alpha_S$ which
depends on the atomic state and, in general, on the frequency of the
oscillating electric field $\vec{E}$. Atomic states with $J>1/2$ and
$F>1/2$ have an additional tensor component of the polarizability,
$\alpha_T(\gamma,J)$. The tensorial part depends on the quantum
numbers $F,m_F$ and on the polarization of the electric field with
respect to the quantization axis of the atom. Sources of ac electric
fields are the trap rf drive field, thermal black-body radiation,
the clock interrogation laser or the cooling laser for the atomic
ion applied during interrogation. They all couple the clock states
off-resonantly to other levels, resulting in a differential shift.
For example, the root mean square (rms) electric field associated
with BBR near room temperature is given by $\langle
E^2(T)\rangle=(831.9~\text{V/m})^2(\text{T(K)}/300)^4$. While in
singly-charged ions the shift from BBR of the ion's environment can
be significant, it is negligible for HCI. This has two reasons.
Firstly, HCI traps are operated at cryogenic temperatures near
$T=4$~K where the BBR shift is suppressed by more than seven orders
of magnitude compared to room temperature operation owing to the
$T^4$ scaling of the shift. Secondly, the size of the electron
orbitals and thus the polarizability scale with $1/Z$, contributing
to a further suppression. AC Stark shifts arising from the rf
trapping field of the Paul trap will be discussed in
\sref{sec:micromotion}.

\paragraph{Electric quadrupole shift}\label{sec:QPS}
Atomic states with $J,F > 1/2$ are no longer spherically symmetric,
but exhibit higher order electric multipole moments that couple to
the corresponding electric field components. The largest
contribution is the quadrupole moment which interacts with electric
field gradients according to the Hamiltonian
\cite{itano_external-field_2000}
\begin{equation}
H_Q=\bm{\nabla E^{(2)}}\bm{.}\bm{\Theta^{(2)}},
\end{equation}
where $\bm{\Theta^{(2)}}$ is the electric-quadrupole operator for
the atom and $\bm{\nabla E^{(2)}}$ is a symmetric traceless tensor
of second rank describing the electric field gradient at the
position of the ion. It should be noted that even for states with
$J=0, 1/2$, a small quadrupole moment from mixing of other
electronic states and nuclear quadrupole moments exists, which is
usually negligible at a frequency uncertainty level above $10^{-19}$
\cite{beloy_hyperfine-mediated_2017}. The quadrupole shift depends
on the total angular momentum $F$ and its projection $m_F$ along the
quantization axis \cite{itano_external-field_2000}, according to
\begin{equation}\label{eq:QPS}
\Delta f_Qh\propto \frac{3m_F^2-F(F+1)}{\sqrt{(2F+3)(2F+2)(2F+1)(2F(2F-1))}}.
\end{equation}
Furthermore, it depends on the orientation of the quantization axis
with respect to the electric field gradient.

An electric field gradient is inherent to the axial trapping
mechanism in linear ion traps and can become larger if more than one
ion is trapped. Even in spherical ion traps, that can in principle
be free of electric field gradients, spurious electric fields
typically result in gradients of up to a few V/mm$^2$, resulting in
shifts of a few Hz for a typical atomic quadrupole moment of
$ea_0^2$, where $e$ and $a_0$ are the electric charge and the Bohr
radius, respectively. Quadrupole shift reduction or cancelation
schemes are based on minimizing \Eref{eq:QPS} or averaging frequency
shifts of different transitions to zero. For example, selecting
suitable hyperfine components $F$ for the clock transition, the
quadrupole shift can be made small. The dependence of the shift on
the quantum number $m_F^2$ and the orientation of the gradient with
respect to the quantization axis is identical to other tensorial
shifts, such as the tensor component of the ac Stark effect.
Therefore, averaging suitable pairs of Zeeman transitions cancels
these shifts together with the linear Zeeman shift. Using such
schemes, a suppression of the quadrupole shift by more than four
orders of magnitude has been achieved in a singly-charged ion
\cite{dube_electric_2005,dube_evaluation_2013}. Alternatively, the
quadrupole shift can be canceled by averaging the same transition
over three mutually orthogonal magnetic field directions. The level
of suppression using this technique can reach a factor of 100 if the
magnetic field direction is determined to better than $\pm1^\circ$.
Since the electric quadrupole moment scales with the square of the
size of the electron orbitals $a\sim 1/Z$, the quadrupole shift in
HCI is reduced between one and several orders of magnitude compared
to singly-charged ions.

\subsubsection{Motion-induced shifts}\label{sec:motion}
An atom in motion experiences special relativistic frequency shifts
with respect to an atom in the laboratory frame. Consider the case
of an atom moving with velocity $v_\parallel$ along the direction of
the clock laser with frequency $f$ in the laboratory system probing
the reference transition. According to special relativity, the atom
observes the clock laser with a first order Doppler shift of $\Delta
f_{D1}/f=\langle v_\parallel\rangle/c$, where the average is taken
over typical timescales required for stabilizing the probe laser
frequency to the atomic transition frequency. For an ion oscillating
in an ion trap that is fixed to the laboratory frame, this shift
averages to zero. However, thermal drifts in the relative position
of the trap with respect to the probe laser phase or
probe-synchronous shifts in the ion's position in the trap can
result in fractional frequency drifts of $10^{-17}$ for a relative
velocity of only 3~nm/s. Therefore it may be required to
phase-stabilize the laser to the position of the trap using
interferometric schemes \cite{falke_delivering_2012,
ye_delivery_2003}. Since HCI are particularly sensitive to electric
fields, displacements synchronous to the clock interrogation that
may arise through charging of the ion trap structure via the clock
laser, need to be avoided or measured using counterpropagating probe
laser beams \cite{Ros08}. In addition to the linear Doppler shift,
the atom also experiences a second order Doppler or time dilation
shift
\[
\frac{\Delta f_{D2}}{f}=-\frac{\langle v^2\rangle}{2c^2}=-\frac{\Ekin}{mc^2}
\]
from motion in all directions, which is directly related to the
total kinetic energy \Ekin of the ion in the trap. This shift can be
difficult to quantify, since one has to make assumptions about the
velocity distribution  \cite{chen_sympathetic_2017}. Heating of the
ion during interrogation can increase the shift and result in
additional uncertainty, making this shift one of the largest
contributions to the uncertainty budget, e.g.\ of the \Al and \Yb
clocks \cite{Ros08,huntemann_single-ion_2016, chou_frequency_2010}.
Assuming a constant thermal distribution over the probe time,
characterized by a mean occupation of motional levels, $\bar{n}$,
the kinetic energy in a linear Paul trap is given by the sum over
all modes with frequencies $\omega_j$
\begin{eqnarray}\label{eq:Ekin}
\Ekin&\approx&\left(\frac{1}{2}+\frac{1}{2}\right)\sum_{j\in\mathrm{radial}}\hbar\omega_j\left(\bar n_j+\frac{1}{2}\right)\\
&&+\frac{1}{2}\sum_{j\in\mathrm{axial}}\hbar\omega_j\left(\bar n_j+\frac{1}{2}\right).
\label{eq:Ekintot}
\end{eqnarray}
The first sum is over all radial modes. One of the $1/2$ prefactors
reflects the fact that kinetic energy makes up only half the total
energy in a harmonic oscillator, and the second $\approx 1/2$ is
from intrinsic micromotion of the ion in the trap
\cite{berkeland_minimization_1998}. It will be discussed in more
detail in \sref{sec:micromotion}. The second sum is over axial modes
that ideally are not affected by micromotion. It is interesting to
note that the zero point energy contributes to the time dilation
effect and can be on the order of $10^{-19}$ for typical trap
frequencies of a few MHz and light ion species, such as \Al. For
HCI, motional shifts will depend crucially on the performance of
sympathetic cooling with the singly-charged cooling ion as discussed
in \sref{sec:sympcool}, which in turn will depend on the rate of
collisions with background gas (discussed in \sref{sec:collisions})
and the anomalous motional heating rate in the trap. Since HCI will
be trapped in a cryogenic environment, we expect that both effects
can be made small to not be a limiting factor in clock accuracy.

\subsubsection{Micromotion shifts}\label{sec:micromotion}
The trapping mechanism of Paul traps is based on a periodically
oscillating electric quadrupole field with angular frequency $\Orf$.
This field vanishes in a spherical trap at a single point and in a
linear trap along the (axial) nodal line, on which ideally the ion
is located (see \sref{sec:trapping}). Since an ion in the trap has a
minimal size along each direction corresponding to the zero point
wave function extent $x_0=\sqrt{\hbar/2m\omega_x}$, it is always
subject to an oscillating force leading to intrinsic (and
unavoidable) micromotion in the radial ($x,y$) direction in addition
to any secular motion around its equilibrium position at the much
lower frequency $\omega_{x,y}$ \cite{berkeland_minimization_1998}.
This intrinsic micromotion grows with the oscillation amplitude and
thus the temperature of the ion. Electric dc fields displace the ion
from its equilibrium position and result in additional, so-called
excess micromotion. Similarly, a phase difference between the rf
applied to a pair of electrodes results in excess micromotion. The
amplitude of excess micromotion scales in both cases with the charge
$Q$ of the ion. Therefore, HCI are particularly sensitive to
micromotion and excess micromotion needs to be avoided. This is
typically achieved by probing micromotion using one of several
techniques \cite{berkeland_minimization_1998,keller_precise_2015}
and applying compensation voltages to steer the ion back to the
position of vanishing rf field. Since HCI experience $Q$-times
stronger micromotion, the corresponding signal is also $Q$-times
stronger, allowing micromotion compensation to a level comparable to
singly-charged ions. Therefore, excess micromotion will not pose a
limitation to optical frequency standards based on HCI as long as
the required compensation voltages can be controlled with sufficient
precision and remain constant between micromotion probe cycles.
Micromotion contributes to the second order Doppler shift in the
radial direction as discussed in \sref{sec:motion}, \Eref{eq:Ekin}.
In thermal equilibrium with the sympathetic cooling ion, the second
order Doppler shift of HCI is identical to singly-charged ions.
However, the oscillating electric field also leads to an ac Stark
shift of the clock states as discussed in \sref{sec:efields}. This
shift depends on the differential polarizability $\Delta\alpha_S$
between the two clock states. One can show that both shifts exhibit
the same scaling with the oscillating field \Erf at the position of
the ion. For some ion species, such as \Ca and \Sr with negative
$\Delta\alpha_S$, a so-called ``magic'' rf drive frequency exists,
for which both shifts cancel
\cite{dube_high-accuracy_2014,huang_comparison_2017}. In the case of
HCI, the polarizability of individual states and thus the
differential polarizability is smaller compared to singly-charged
ions owing to the smaller size of electron orbitals. Therefore, the
total micromotion-induced shift in HCI-based optical clocks is
typically smaller compared to singly-charged ion clocks and could
even be made to vanish in case a sufficiently large negative
$\Delta\alpha_S$ is found.

\subsubsection{Collisional shifts}\label{sec:collisions}
Collisions with residual background gas atoms can result in a
transient distortion of the electronic energy levels and thus to a
shift in the clock transition frequency. An upper bound to this
frequency shift $|\Delta\omega|\leq\sqrt{\gamma_g\gamma_e}$ is given
by the total collision rates $\gamma_g$, $\gamma_e$ for the ion in
the ground and excited state, respectively
\cite{vutha_collisional_2017}. These rates can either be calculated
from scattering theory, or measured experimentally by observing
decrystallization and re-ordering of a two-ion crystal \cite{Ros08}.
Since HCI are operated at cryogenic temperatures, background gas
pressure is significantly reduced compared to room temperature
setups, thus minimizing collisional shifts. Furthermore, collisions
of neutral background gas atoms or molecules with a HCI almost
always result in charge-transfer reactions, with the HCI capturing
one electron from its neutral collision partner at a relatively far
distance of several atomic units. The sudden mutual repulsion of the
ionized neutral and the HCI imparts a momentum kick to both
particles, and this usually removes the down-charged HCI from the
trap. In this case the systematic uncertainty from collisional
shifts vanishes.
%%%%%%%%%%%%%%%%%%%%%%%%%%%%%%%%%%%%%%%%%%%%%%%%%%%%%%%
%%%%%%%%%%%%%%%%%%%%%%%%%%%%%%%%%%%%%%%%%%%%%%%%%%%%%%%

\subsection{Evaluation of HCI clock candidates}
In this section, we will assess HCI candidates proposed for optical clocks and frequency references.
%For this we have defined a number of primary criteria a clock ion species should meet to outperform current clocks in terms of systematic shifts and statistical uncertainty.
%This includes in particular that relevant atomic properties should be superior to the currently two most advanced singly-charged clocks based on the \Al \ssztpz and the \Yb octupole transitions. This is a somewhat arbitrary choice, since efficient mitigation strategies are known for many of the systematic shifts that are also applicable for HCI.
Ideally, HCI clock candidates would outperform their neutral and
singly-charged counterparts in the atomic properties that are
responsible for systematic frequency shifts and the statistical
uncertainty of the clock. Unfortunately, for many of the HCI
candidates proposed as optical clock references not enough atomic
data for a proper evaluation are available. This illustrates the
strong demand for more detailed and more accurate atomic structure
calculations and measurements. Therefore, we provide a list of
primary clock ion selection criteria and order of magnitude values
for the most important atomic parameters, inspired by \Al and \Yb,
two of the currently most advanced ion clock candidates:
\begin{itemize}
\item clock transition in a range accessible by current laser technology, i.e. 200~nm$\dots$2~$\mu$m
\item transition with a large quality factor, i.e. high transition frequency and a long-lived excited state with a lifetime time $\gtrsim 1$~s to provide high stability
\item small linear Zeeman shift coefficient, i.e. $|C_{M1}|\lesssim 100$~kHz/mT
\item small quadratic Zeeman shift coefficient, i.e. $|C_{M2}|\lesssim 100$~Hz/mT$^2$
\item small electric quadrupole moment $\Theta$, i.e. $|\Theta|\lesssim 0.1 ea_0^2$
\item small differential polarizability, i.e. $|\Delta\alpha_S|\lesssim 10^{-41}$~Jm$^2$/V$^2$
\item sparse level structure to simplify intial state preparation and laser complexity
\item long-lived isotopes with a lifetime exceeding years
\end{itemize}
The second requirement is often neglected in the literature.
Assuming a transition wavelength of 500~nm (transition frequency
$\nu_0=600$~THz) and an excited state lifetime of $\tau=1$~s, we get
according to \Eref{eq:instability} an instability of
$\sigma_y(T)\approx 7\times 10^{-16}/\sqrt{t/\mathrm{s}}$. This
already corresponds to an averaging time of 5.5~days to achieve a
relative frequency uncertainty of $10^{-18}$ for the measurement.

Despite the lack of atomic structure data, some general guidelines
for suitable clock transitions are at hand and can be applied with
caution, since exceptions may exist:
\begin{itemize}
\item vanishing electronic spin or availability of a $m=0\rightarrow m'=0$ transition to eliminate the linear Zeeman shift
\item small total angular momentum to reduce the number of hyperfine levels and magnetic substructure which simplifies initial state preparation and minimizes second order Zeeman and tensorial shifts
\item large fine- and hyperfine-structure splittings to reduce second order Zeeman and ac Stark shifts
\end{itemize}

In addition to the primary criteria, one might apply secondary
criteria such as the availability of a cooling transition, known
level structure from experiment or accurate atomic structure
calculations. Particularly important for possible applications is
the sensitivity of the clock transition to a change in fundamental
constants, QED tests or other physics beyond the Standard Model.
This will result in a trade-off between achievable accuracy and the
sensitivity to such effects.

In the following, we will discuss some selected species
representative of a whole class of HCI with slightly different
properties. In this assessment, we have taken into account the
atomic properties and will discuss possible cancelation techniques
that will further reduce systematic frequency shifts.

%\begin{table}
%\caption{\label{tab:assessment} Assessment of atomic properties of HCI candidates for optical clocks.}
%\begin{ruledtabular}
%\begin{tabular}{lcc}
%\end{tabular}
%\end{ruledtabular}
%\includegraphics[width=\columnwidth]{assessment}
%\end{table}

\subsubsection{Hyperfine transitions}
\label{HFS1} Optical clocks based on $M1$ hyperfine transitions in
HCI have been discussed by \cite{Schiller2007, YTD14}. All
investigated species with transitions between the
$F=0\leftrightarrow F=1$ hyperfine components of the \dsoh
electronic ground state with nuclear spin $I=1/2$ have very
favorable atomic properties concerning systematic frequency shifts.
They have no electric quadrupole moment, feature a vanishing first
order Zeeman shift by employing a $m=0\rightarrow m'=0$ transition
and have a very small second order Zeeman coefficient $C_{M2}$,
since the hyperfine splitting is large compared to any Zeeman shift
(see Sec.~\ref{sec:mfields}). Since all dipole-allowed transitions
are in the XUV wavelength regime, only a small differential
polarizability arises from M1 couplings between hyperfine components
\cite{itano_shift_1982}. As a consequence, all electric and magnetic
field shifts are extremely small with a fractional frequency
uncertainty below $10^{-20}$, rendering HCI based on hyperfine
transitions ideal candidates for high-accuracy clocks. The only drawback in these systems is the
achievable statistical uncertainty, since either the transition
frequency is low with a long excited state lifetime, or vice versa.
The longest investigated excited state lifetime for a transition
near the optical regime is that of \hci{171}{Yb}{69+} with a
transition wavelength of $2160\,$nm and an excited state lifetime of
0.37\,s \cite{YTD14}. The achievable instability of $\sigma_y(\tau)\approx
4.9\times 10^{-15}/\sqrt{t/\mathrm{s}}$ \cite{peik_laser_2006}
%requiring an impractically long averaging time $T$ of 285~days to reach a fractional uncertainty of $10^{-18}$. However, this instability
is better by more than an order of magnitude compared to the best Cs
fountain clocks. Given the high sensitivity of hyperfine transitions
to changes in $\mu$, $X_q$, and $\alpha$, as discussed in
\sref{sec:variation}, HCI clocks based on these transitions could
help to significantly improve bounds on a possible variation of
these quantities compared to what is currently possible with
measurements involving Cs clocks. Since only motional and
collisional frequency shifts have to be evaluated, they represent a
promising starting point for HCI-based clocks.

\subsubsection{Fine-structure transitions}
At more moderate charge states, fine-structure transitions in the
optical regime can be found and optical clocks based on one-valence
electron (OVE) $\dpoh\leftrightarrow\dpth$ ($I=3/2$) and two-valence
electrons (TVE) $\tpz\leftrightarrow\tpo$ ($I=1/2$) transitions have
been investigated in \cite{YTD14}. The sensitivity to frequency
shifting effects are similar, but not quite as small compared to the
hyperfine clocks discussed in the previous section. The electric
quadrupole shift is non-zero for the excited clock state of the OVE
systems. However, it can be made small by a proper choice of
hyperfine components (see Eq.~\ref{eq:QPS}). Currently no estimates
on the actual value of the quadrupole moment exist that would allow
a proper evaluation of this shift. While for the OVE species
$m_F=0\rightarrow m_F'=0$ transitions are available, the large
electronic linear Zeeman shift needs to be canceled by averaging
suitable combinations of $m_F$ transitions for the TVE systems. As a
consequence of the smaller hyperfine splitting, the second order
Zeeman effect and M1 transition-induced differential polarizability
are somewhat larger compared to the hyperfine transition clocks, but
still significantly smaller compared to clocks based on neutral or
singly-charged atoms. Transitions between the ${}^2$F$_{5/2}\leftrightarrow ^2$F$_{7/2}$ fine-structure in 
\hci{184}{W}{13+} ($I=1/2$) and \hci{191}{Ir}{16+} ($I=3/2$) have been investigated in
\cite{NaSa16}. These transitions share the properties of the OVE systems discussed above and exhibit non-zero electric quadrupole moments on the order
of $0.015ea_0^2$ that are smaller by a factor of 4 compared to the
\Yb excited clock state due to common mode suppression between ground and excited state. While the accuracy of clocks based on
fine-structure transitions is very promising, their achievable
statistical uncertainty is a limiting factor. The best instability
of $\sigma_y(t)\approx 3.2\times 10^{-15}/\sqrt{t/\mathrm{s}}$ for
the OVE species is found for \hci{79,81}{Br}{4+} with an excited
state lifetime of $\tau\approx 0.5$~s and a transition wavelength of
$\lambda\approx 1642$~nm. A similar instability of
$\sigma_y(t)\approx 4\times 10^{-15}/\sqrt{t/\mathrm{s}}$ is
achieved e.g.\ by the TVE system \hci{123,125}{Te}{2+} with an
excited state lifetime of $\tau\approx 0.51$~s and a transition
wavelength of $\lambda\approx 2105$~nm.
%In contrast to hyperfine clocks, fine-structure clocks are only sensitive to the fine-structure constant with a rather small sensitivity coefficient. \cPiet{Do we have a number?}

\subsubsection{Level crossing transitions}
The largest investigated group of HCI optical clock candidates is
based on level crossing transitions \cite{BerDzuFla12}. Many of them
feature a large sensitivity to a change in the fine-structure
constant \cite{BerDzuFla12a,BDF10,BerDzuFla11b}, which was the
original motivation to study them and is discussed in more detail in
\sref{sec:variation}. This group can be divided into one-valence
electron systems for which atomic structure calculations can provide
estimates of the atomic properties required for a proper evaluation
of the clock candidates, and systems with a more complicated
electronic structure for which accurate data is currently
unavailable. HCI clock candidates belonging to the former group
include \hci{\xspace}{Nd}{13+} and \hci{\xspace}{Sm}{15+} (Ag-like
isoelectronic sequence) that have optical transitions between the
$5s_{1/2}\leftrightarrow 4f_{7/2}$ electronic states at wavelengths
of 170~nm and 180~nm, respectively and excited state lifetimes of
several days with zero nuclear spin ($I=0$) \cite{DzuDerFla12} (see
also Fig.~\ref{Fig_Ag-like_Nd}). A partial systematic frequency
shift evaluation reveals \cite{DzuDerFla12} that the differential
polarizability is with $\Delta\alpha_S\sim 10^{-41}$~Jm$^2$/V$^2$
comparable to the \Al polarizability, the large linear electronic
Zeeman shift needs to be canceled by averaging suitable transitions,
whereas the second order Zeeman shift is extremely small with
$C_{M2}\sim 10$~mHz/mT$^2$. The electric quadrupole moments of the
$4f_{5/2}$ states have not been calculated, thus cancelation
techniques as discussed in Sec.~\ref{sec:efields} need to be
applied. While the transition itself has only a mild sensitivity to
a change in the fine-structure constant, the $4f_{5/2}$ ground state
in \hci{}{Sm}{15+} may exhibit a large sensitivity to a violation of
local Lorentz invariance (LLI) as discussed in Sec.~\ref{sec:LLI}.
In \cite{NaSa16}, the same transition has been investigated in \hci{195}{Pt}{17+} ($I=1/2$) at a wavelength of around 400~nm. This transition exhibits a long lifetime of 128~years, a strong linear Zeeman effect, which can be mitigated by probing $m_F=0\rightarrow m_F'=0$ states, an electric quadrupole moment of $-0.081ea_0^2$, which is comparable to the Yb$^+$ octupole transition, and small second order Zeeman ($\Delta f/f\sim 1.6\times 10^{-24}$) and polarizability ($\Delta f/f\sim 2\times 10^{-18}$) shifts.

The larger group of HCI clocks based on level-crossings contains HCI
for which estimates of the atomic properties are missing. Therefore,
only a qualitative assessment for selected cases is possible. All
investigated species have a transition in the optical with a
sufficiently long excited state lifetime, typically well-exceeding
1~s. HCI with an optical level-crossing transition for the
isoelectronic sequences of Ag-like, Cd-like, In-like, and Sn-like
ions have been identified \cite{SDFS14, SDFS14a, SDFS14b}. The
Cd-like sequence, such as \hci{}{Nd}{12+} with $I=0$ offers no
transitions satisfying either the specified wavelength range nor
excited state lifetime. Examples of the Ag-like sequence with $I=0$
have already been discussed earlier in this section. Optical clocks
based the In-like sequence with $I=0$ use level crossings between
similar states, such as $s_{1/2}$/$p_{1/2}$ and $f_{5/2}$/$f_{7/2}$,
and therefore will share similar properties: a large linear and
small quadratic Zeeman shift, a non-zero electric quadrupole moment
and small polarizabilities. Species with hyperfine structure will in
general have worse properties, such as an increased quadratic Zeeman
effect, while featuring an $m_F=0\rightarrow m_F'=0$ transition free
of the linear Zeeman shift. A representative of the Sn-like sequence
is \hci{141}{Pr}{9+} with $I=5/2$ shown in
Fig.~\ref{Fig_Sn-like_Pr}. It features several long-lived excited
states at laser accessible wavelengths. The \tpz ground state is
free of a first order electric quadrupole shift, and exhibits only
small linear and quadratic Zeeman shifts. However, the excited
$^3$G$_3$ and $^3$F$_2$ excited clock states have non-vanishing
electric quadrupole moments, strong linear ($J>0$) and medium
quadratic (hyperfine splitting) Zeeman shifts.

Electron-hole transitions at the $4f$-$5s$ level crossing have been
proposed \cite{BerDzuFla11b} and experimentally investigated
\cite{WinCreBel15,bekker_private_2017} for the case of
\hci{}{Ir}{17+}. The single-hole system \hci{}{Ir}{16+} has a
transition of unknown linewidth at 267~nm between the
$^2$F$_{7/2}\leftrightarrow ^2$S$_{1/2}$ states
\cite{bekker_private_2017} with non-vanishing electric quadrupole
moment. Hyperfine structure ($I=3/2$) results in a linear
Zeeman-free $m_F=0\rightarrow m_F'=0$ transition with medium second
order Zeeman shift. The double-hole system \hci{}{Ir}{17+} features
several transitions of unknown linewidth in the optical, the lowest
being at 1415.6~nm between the $^3$F$_4^0\leftrightarrow^3$H$_6$
states \cite{bekker_private_2017} with non-vanishing electric
quadrupole moment. All transitions exhibit a large electronic linear
Zeeman shift and a medium second order Zeeman shift. However, of all
considered stable atomic systems, the transitions in \hci{}{Ir}{17+}
feature the largest sensitivity to a change in the fine-structure
constant.

A HCI with even more complex electronic structure investigated as an
optical clock candidate is \hci{\xspace}{Ho}{14+}
\cite{DzuFlaKat15}. A transition at around 400~nm connects the
$^8$F$_{1/2}\leftrightarrow ^6$H$_{5/2}$ states with non-vanishing
electric quadrupole moment. The excited state has an estimated
lifetime of around 37\,s, mostly from E1 decay into other states.
While the nuclear spin of $I=7/2$ enables a $m_F=0\rightarrow
m_F'=0$ transition free of the linear Zeeman effect, the large
number of hyperfine levels and small hyperfine splitting will result
in a large second order Zeeman shift and complicates state
initialization.

%\cite{DzuDerFla12}
%\cite{SDFS14}
%\cite{SDFS14a}
%\cite{SDFS14b}
%\cite{BerDzuFla12}
%Ho14+ \cite{DzuFlaKat15}
%Cf15+ \cite{DSSF15}
\subsubsection{Intra-configuration transitions}
Another category of HCI for optical clocks are based on optical
intra-configuration transitions in the $4f^{12}$ shell that have
been investigated with \cite{DerDzuFla12} and without
\cite{DzuDerFla12a} hyperfine structure. One example with hyperfine
structure is \hci{209}{Bi}{25+} with $I=9/2$ and a transition
between the $^3$H$_6\leftrightarrow ^3$F$_4$ states with an excited
state lifetime of about 3~h. This transition features a very small
differential polarizability, large linear and quadratic Zeeman
shifts and a quadrupole moment up to four times larger compared to
the \Yb F-state. Through a proper choice of transition, the
effective quadrupole moment can be reduced by two orders of
magnitude, at the expense of complicated state initialization. While
the transition is not particularly sensitive to a change in
fundamental constants, the large angular momentum may exhibit a high
sensitivity to a violation of LLI as discussed in
Sec.~\ref{sec:LLI}. HCI species of this kind without hyperfine
structure ($I=0$) have a much simpler level structure and
consequently smaller quadratic Zeeman shift \cite{DzuDerFla12a}. All
other properties are similar compared to the case with hyperfine
structure, except that a more conventional electric quadrupole
suppression technique (see \sref{sec:efields}) has to be employed.

%(1) "Ion clock and search for the variation of the fine-structure constant using optical transitions in Nd13+ and Sm15+".  PHYSICAL REVIEW A 86, 054502 (2012) Category: ?Normal? cases where theory can give relatively good predictions.
%
%(2) "High-precision atomic clocks with highly charged ions: Nuclear-spin-zero f12-shell ions", PHYSICAL REVIEW A 86, 054501 (2012). Category: fine-structure cases, clock only, no alpha-sensitivity. Some of such type are in the next paper too.
%
%(3) "Magnetic-dipole transitions in highly-charged ions as a basis of ultra-precise optical clocks", PRL 113, 233003 (2014). Also 2007 paper Category: H-like hyperfine. These are actually very important, as the only way to improve on mu and mq/LQCD with atomic clocks. In the application for dark matter searches, this accesses limits on different dark matter/SM couplings, comparing to ?alpha-only? searches which can only get de constraints, see Mina?s paper on dilatons, PHYSICAL REVIEW D 91, 015015 (2015). I just had a long conversation with Junwu, who is a co-author, and this DM scenario now has good theory background behind it.
%
%($) Optical clock sensitive to variation of the fine structure constant based on the Ho14+ ion. PHYSICAL REVIEW A 91, 022119 (2015). Category: complicated ions where theory has limited predictive ability, Ir17+ is in here too.

\subsection{Evaluation summary}
All discussed HCI clock candidates have rich features and partially
fulfill the primary criteria listed above. However, it remains an
open challenge to obtain sufficient information about the atomic
properties of most of the species that would allow a full assessment
of their systematic frequency shifts and enable identification of a
superior candidate. By employing systematic shift cancelation
schemes, such as magic-drive frequency operation and averaging the
frequencies of all Zeeman components, all magnetic- and
electric-field shifts can be significantly suppressed and very
likely brought to a level below the current best singly-charged ion
optical clocks. This opens up exciting prospects for testing
fundamental physics with HCI. As an example, transitions in
Ir$^{17}$ \cite{BerDzuFla11b} and Cf$^{15+}$/Cf$^{17+}$ have more
than an a factor of 20 higher sensitivity to a change in the
fine-structure constant $\alpha$ compared to the Yb$^+$ E3
transition, which is the most sensitive system currently employed.
Similarly, HCI clocks based on hyperfine transitions are sensitive
to a change in the electron-to-proton mass ratio, providing more
than an order of magnitude better statistical uncertainty compared
to the currently employed Cs clocks. Other optical transitions in
HCI may not be well-suited for clocks with ultimate performance, but
rather are sensitive to other New Physics effects as discussed in
the next section. Improving the bounds on such effects using HCI
with their high sensitivity might turn out to be a much more
efficient approach compared to improving the systematic uncertainty
of conventional neutral or singly-charged atom clocks and performing
frequency comparisons at a challenging level of below $10^{-18}$.

\section{Other applications and future developments}
\label{sec:OtherApps}

%%%%%%%%%%%%%%%%%%%%%%%%%%%%%%%%%%
%\input{LV_v8}
%%%%%%%%%%%%%%%%%%%%%%%%%%%%%%%%%%

\subsection{Tests of local Lorentz invariance}\label{sec:LLI}

Local Lorentz invariance (LLI) is one of the cornerstones of modern
physics: the outcome of any local non-gravitational experiment is
independent of the velocity  and the orientation of the
(freely-falling) apparatus. The recent interest in tests of Lorentz
symmetries is motivated by theoretical developments in quantum
gravity  suggesting that Lorentz symmetry may be violated at some
energies,  tremendous progress in experimental precision, and
development of a theoretical framework to analyze different classes
of experiments. Separate  violations of LLI are possible for each
type of particle, and the experiments include searches for Lorentz
violation (LV) in the matter, photon, neutrino, and gravity sectors.
In this section, we limit the discussion to specific LLI tests
relevant to HCI applications.

Lorentz violation tests are analyzed in the context of an effective
field theory  known as the Standard Model extension (SME). The
\textit{Data Tables for Lorentz and $CPT$ Violation} by
\citet{KosRus11,KosRus17} gives tables of the measured and derived
values of coefficients for Lorentz and $CPT$ violation in the SME.
In minimal SME, the  Standard Model Lagrangian is augmented with
every possible combination of the SM fields that are not
term-by-term Lorentz invariant, while maintaining gauge invariance,
energy--momentum conservation, and Lorentz invariance of the total
action \cite{ColKos98}. A general expression for the quadratic
Hermitian Lagrangian density describing a single spin-$1/2$ Dirac
fermion of mass $m$  (electron, proton, or neutron) in the presence
of Lorentz violation is given by \cite{KosLan99}
\begin{equation}
{\cal{L}}=\frac{1}{2} i c \overline{\psi} \Gamma_\nu \overleftrightarrow{\partial^\nu} \psi - M c^2 \overline{\psi} \psi,
\end{equation}
where $\psi$ is a four-component Dirac spinor, $c$ is the speed of
light in a vacuum, $f \overleftrightarrow{\partial^\nu}g = f
\partial^\nu g - g \partial^\nu f$,
\begin{equation}
\label{LV1}
M =  m + a_{\mu}\gamma^{\mu} + b_{\mu} \gamma_5 \gamma^{\mu} + \frac{1}{2} H_{\mu \nu} \sigma^{\mu \nu}
\end{equation}
and
\begin{equation}
\label{LV2}
\Gamma_{\nu} = \gamma_{\nu} + c_{\mu \nu} \gamma_{\nu}+d_{\mu \nu} \gamma_5 \gamma_{\nu} + e_\nu +  i \gamma_5 f_\nu +
\frac{1}{2} g_{\lambda \mu \nu} \sigma_{\lambda \mu}.
\end{equation}

The $\gamma _\mu$ are Dirac matrices, $\mu=0,1,2,3$, $\gamma _5$ is
a Dirac matrix associated with pseudoscalars, and
 $\sigma^{\mu \nu}=\frac{i}{2}\left( \gamma^\mu \gamma^\nu - \gamma^\nu \gamma^\mu \right)$.
The first terms in the expressions for $M$ and $\Gamma_{\nu}$ give
the usual SM Lagrangian. Lorentz violation is quantified by the
parameters $a_{\mu}$,  $b_{\mu}$, $c_{\mu \nu}$, $d_{\mu \nu}$,
$e_\mu$, $f_\mu$, $ g_{\lambda \mu \nu}$, and $H_{\mu \nu}$. The
coefficients in Eq.~(\ref{LV1}) have dimensions of mass; the
coefficients in Eq.~(\ref{LV2}) are dimensionless. The framework of
interpreting the laboratory experiments involving monitoring atomic
or nuclear frequencies in terms of the SME coefficients is described
in detail by \citet{KosLan99,KosMew02}.

Violations of Lorentz invariance  in bound electronic states
result in a perturbation of the Hamiltonian that can be described by \cite{KosLan99,Hoh13}
\begin{equation}
\delta H=-\left(  C_{0}^{(0)}-\frac{2U}{3c^{2}}c_{00}\right)
\frac{\mathbf{p}^{2}}{2m_e}-\frac{1}{6m_e}C_{0}^{(2)}T^{(2)}_{0},\label{LV3}%
\end{equation}
where $\mathbf{p}$ is the momentum of a bound electron. The second
term in the parentheses  gives the leading order gravitational
redshift anomaly in terms of the Newtonian potential $U$. The
parameters $C_0^{(0)}$ and $C_{0}^{(2)}$ are elements of the $c_{\mu
\nu}$ tensor in the laboratory frame introduced by Eq.~(\ref{LV2}):
 \begin{eqnarray}
 C^{(0)}_0 &=& c_{00}+(2/3)c_{jj},\\
 C^{(2)}_0 &=& c_{jj}+(2/3)c_{33},
 \end{eqnarray}
where $j={1,2,3}$.

The non-relativistic form of the $T^{(2)}_{0}$ operator is
$T^{(2)}_{0}=\mathbf{p}^{2}-3p_{z}^{2}$. Predicting the energy shift
due to LV involves the  calculation of the expectation value of the
above Hamiltonian for the atomic states of interest. The larger the
matrix elements, the more sensitive is this atomic state. In atomic
experiments aimed at the LLI tests in the electron-photon sector
\cite{Hoh13,PruRamPor15}, one searches for variations of the atomic
energy levels when the orientation of the electronic wave function
rotates with respect to a standard reference frame, such as the Sun
centered celestial-equatorial frame (SCCEF). The rotation is simply
supplied by Earth 24h or year long periodic motion.

\citet{PruRamPor15}  performed a test of Lorentz symmetry using an
electronic analogue of a Michelson- Morley experiment using the
$^2$D$_{5/2}$  atomic states of $^{40}$Ca$^+$  ion with anisotropic
electron momentum distributions. A pair of $^{40}$Ca$^+$ ions was
trapped in a linear Paul trap, with a static magnetic field applied,
defining the eigenstates of the system. The direction of this
magnetic field changes with respect to the Sun as the Earth rotates.
The Lorentz violation effects depend on the magnetic quantum number
$m_J$. A test of LLI can be performed by monitoring the frequency
difference  between the LV shifts of the $m_J=5/2$ and $m_J=1/2$
substates of the $3d~^2D_{5/2}$ manifold
\begin{equation}
\frac{1}{h}\left( E_{m_J=5/2} -E_{m_J=1/2}\right) = \left[-4.45(9)\times10^{15}~ \textrm{Hz} \right]\times \,C_{0}^{(2)}
\end{equation}
as the Earth rotates. The Ca$^{+}$ experiment improved the limits to
the $c_{JK}$ coefficients ($J,K=1,2,3$) of the LV-violation in the
electron-photon sector to the 10$^{-18}$ level.

Further significant improvement of the LV constraints with such
experiments requires another system with a long-lived or ground
state that has a large $\langle j|T^{(2)}_0|j \rangle$ matrix
element. \citet{DzuFlaSaf16,ShaOzeSaf17} calculated this matrix
element in a variety of systems and identified the $4f^{13}
6s^2$~$^2F_{7/2}$ state of Yb$^+$ and HCI as systems with large
sensitivities to LLI. HCI ions of interest are among those already
proposed for the tests of the $\alpha$-variation. The advantage of
HCI for LLI tests is the possibility to use a ground rather than an
excited state as the LLI probe state and larger matrix elements
$\langle j|T^{(2)}_0|j \rangle$. An experimental scheme for a search
of Lorentz violation with HCI is described in \citet{ShaOzeSaf17}.

%%%%%%%%%%%%%%%%%%%%%%%%%%%%%%%%%%
%\input{IS_v8}
%%%%%%%%%%%%%%%%%%%%%%%%%%%%%%%%%%

\subsection{Probing for new forces}
Motivated by the failure of the Standard Model to e.g.\ describe
dark matter or dark energy, searches for possible candidate fields
or other, yet unidentified fields and their non-gravitational
effects on atoms and molecules have commenced
\cite{ClockMasters2017}. Precision spectroscopy measurements allow
searches for new light scalar fields and constraining their
couplings to ordinary matter (see e.g. \cite{FJKL17}). Recently, a
technique based on isotope shift spectroscopy has been proposed to
probe for such fields that mediate forces e.g.\ between electrons
and neutrons \cite{frugiuele_atomic_2016, delaunay_probing_2016,
berengut_probing_2017, flambaum_isotope_2017, delaunay_probing_2017,
delaunay_probing_2017-1}. The idea is based on the observation of
King \cite{king_comments_1963} that appropriately scaled isotope
shifts of two transitions exhibit a linear dependence. An additional
force between neutrons and electrons would break this linearity. Two
major effects result in a change of the transition frequency of a
selected transition with the neutron number: the field and mass
shifts. The field shift arises from a difference in the nuclear
charge radius for the two isotopes. This results in a change in the
overlap of the wavefunctions of the involved electronic states with
the nuclear charge distribution and thus a change in their binding
energy. The mass shift takes into account the change in recoil upon
photon absorption by the electrons bound to the nucleus. Neglecting
higher order effects, the isotope shift of transition $i$ between
isotopes $A$ and $A'$ can thus be written as
\cite{heilig_changes_1974}
\begin{equation}\label{eq:IS}
\delta\nu_i^{A,A'}=F_i\delta\langle r^2\rangle_{A,A'}+k_i\frac{A-A'}{AA'},
\end{equation}
where $F_i$ and $k_i$ are the field and mass shift constants,
respectively, and $\delta \langle r^2\rangle$ is the change in the
nuclear charge radius. While optical spectroscopy can achieve very
high fractional resolutions of better than $10^{-17}$, the change in
nuclear charge radius is much less well known. By measuring at least
two transitions $i, i'$ and introducing the modified isotope shift
$m\delta\nu_i=\delta\nu_i^{A,A'}\frac{AA'}{A+A'}$, the dependence on
$\delta\langle r^2\rangle$ can be eliminated:
\begin{equation}\label{eq:King}
m\delta\nu_i=\frac{F_i}{F_i'}m\delta \nu_i'+k_ik_i'\frac{F_i}{F_i'}.
\end{equation}
Plotting the two modified isotope shifts in one graph has become
known as a King plot \cite{king_comments_1963}. A hypothetical new
force carrier represented e.g.\ by a scalar field $\phi$ mediated by
an unknown particle with mass $m_\phi$ that couples to neutrons and
electrons adds an additional term to \Eref{eq:IS}
\cite{berengut_probing_2017}
\begin{equation}\label{eq:nonlinearIS}
\delta\nu_i^{A,A'}=F_i\delta\langle r^2\rangle_{A,A'}+k_i\frac{A-A'}{AA'}+ \alpha_\mathrm{NP}X_i\gamma_{AA'}.
\end{equation}
In this equation, $\alpha_\mathrm{NP}$ is the new physics coupling
constant, $X_i$ depends on the form of the potential of the scalar
field $\phi$, while $\gamma_{AA'}$ depends only on nuclear
properties. The additional frequency shift for a single transition
will in general result in non-linearity in \Eref{eq:King} that could
be bounded experimentally by measuring at least two transitions on
at least four even isotopes (without hyperfine structure). The
highest sensitivity is obtained when comparing transitions for which
two of the electronic states have very different orbitals to
maximize the differential effect. To be able to distinguish between
higher-order nuclear effects and $\alpha_\mathrm{NP}$, the range of
the scalar field should be larger than the size of the nucleus. The
sensitivity for massive fields is enhanced for electron orbitals
that are concentrated just outside the nucleus. As a consequence,
narrow optical transitions between electronic state of different
character in HCI are ideal for this purpose. However, even fine- or
hyperfine transitions can serve as narrow reference lines when
combined with transitions in neutral or singly-charged atoms.
Possible candidates for which narrow lines and at least four stable
isotopes exist are Yb and in particular Ca, since experimental data
for \Ca is already available \cite{gebert_precision_2015-1,
shi_unexpectedly_2017, solaro_direct_2017}. Once nonlinearities in
the King plot have been observed, the challenge remains to isolate
$\alpha_\mathrm{NP}$ from all other standard physics higher order
effects neglected in \Eref{eq:IS} \cite{flambaum_isotope_2017}.

%%%%%%%%%%%%%%%%%%%%%%%%%%%%%%%%%%
%\input{higher_frequency_v10}
%%%%%%%%%%%%%%%%%%%%%%%%%%%%%%%%%%

\subsection{Towards higher transition frequencies}
\label{sec:HigherTransitions}
Science based on frequency metrology experienced a tremendous boost
with the development of optical frequency combs which enabled the counting and comparison of optical frequencies \cite{hall_nobel_2006,hansch_nobel_2006}. 
Moving from microwave to optical frequencies implies a four order of magnitude gain in statistical uncertainty according to \Eref{eq:sql} for the same probe time. 
One may ask if a similar jump can be envisioned by again going higher in photon energy, say into the
vacuum-ultraviolet (VUV) or soft x-ray range. 
The main obstacle for this in the realm of atoms and singly charged ions
\cite{Ros08,Wolf2009,wubbena_sympathetic_2012} as x-ray clocks, is the ionization of such systems under irradiation. 
Ions in higher charge states overcome this limitation and offer both allowed and forbidden
transitions, the latter with suitably long lifetimes. Recent
developments of VUV frequency combs based on high-harmonic sources
\cite{Gohle2005,Jones2005,Yost2008} have paved the way for the
extension of the current photon-metrology methods by at least one
order of magnitude. In view of the very rapid developments in this
field, one can expect that within few years the same quality
of combs becomes available at wavelengths of a few nm. By then, the
methods of trapping and cooling of HCI have to be sufficiently
improved to take full advantage of the exquisite frequency control
at the $10^{18}$\,Hz range.

The main advantage of x-ray clocks would be their improved statistical uncertainty according to \Eref{eq:sql} for the same probe times. However, this will put stringent requirements to the phase coherence of the probe laser and its delivery all the way to the ion. To take full advantage of the improved statistical uncertainty, systematic shifts need to be suppressed using advanced schemes developed in the context of quantum-information processing. 
Examples are decoherence-free subspaces \cite{roos_designer_2006,PruRamPor15} or other correlation-based measurement schemes \cite{chwalla_precision_2007}.
A long term perspective for the development of the field can be drawn along the following lines: The development of VUV and soft x-ray frequency combs will enable HCI frequency standards as well as HCI probes for fundamental physics in this wavelength regime, supported by quantum-computing based control schemes for systematic shift suppression and advanced sensing schemes.

\section{Conclusion}

Stimulating the further development of ideas for new applications of highly charged ions, enabled by rapidly improving control of these systems is the central aim of this work. We hope that this review will assist in promoting further rapid experimental progress, and that it will also serve to bring together the HCI and laser-cooled trapped ion communities, which have been previously somewhat disjoint due to very distinct experimental approaches. On the background of earlier theoretical and experimental work, we discuss novel developments and their implications for the future. Based on recent developments, exciting new avenues of research are opened by the use of cold HCI in such diverse applications as  tests of fundamental physics, metrology, development of frequency combs, and quantum information.

\section*{Acknowledgements}
M.~G.~K.~acknowledges
support from Russian Foundation for Basic Research under Grant No.~17-02-00216.
M.~S.~S.~acknowledges the support of the Office of Naval Research, USA, under award number N00014-17-1-2252.
J.~R.~C.~L.~-U.~acknowledges support by the DFG Collaborative Research Centre SFB 1225 (ISOQUANT).
P.~O.~S.~acknowledges support from PTB and DFG through SCHM2678/5-1 and the Collaborative Research Centre SFB 1227 DQ-\textit{mat}, project B03. We thank Steven King for helpful comments on the manuscript.
%\bibliographystyle{apsrev}%
%\bibliography{hci-final_v14} % this should be v13 !!
%\bibliography{BigBIBList8}
%merlin.mbs apsrmp4-1.bst 2010-07-25 4.21a (PWD, AO, DPC) hacked
%Control: author (3) reversed first dotless
\end{document}